\documentclass[structabstract]{aa}
\usepackage{txfonts}
\usepackage{natbib}
\usepackage{graphicx}
\bibpunct{(}{)}{;}{a}{}{,}
\usepackage[normalem]{ulem}

\begin{document}

\title{Signatures of internal rotation discovered in the \textit{Kepler} data\\of five slowly pulsating B stars\thanks{Based on observations made with the Mercator telescope, operated by the Flemish Community, and with the William Herschel Telescope operated by the Isaac Newton Group, both on the island of La Palma at the Spanish Observatorio del Roque de los Muchachos of the Instituto de Astrof\'{i}sica de Canarias.}\fnmsep\thanks{Based on observations obtained with the \textsc{Hermes} spectrograph, which is supported by the Fund for Scientific Research of Flanders (FWO), Belgium, the Research Council of KU Leuven, Belgium, the Fonds National Recherches Scientific (FNRS), Belgium, the Royal Observatory of Belgium, the Observatoire de Gen\`{e}ve, Switzerland and the Th\"{u}ringer Landessternwarte Tautenburg, Germany.}}
%\subtitle{}

\author{P.~I.~P\'{a}pics\inst{\ref{inst1}}\thanks{Postdoctoral Fellow of The Research Foundation -- Flanders (FWO), Belgium}
\and A.~Tkachenko\inst{\ref{inst1}}\thanks{Postdoctoral Fellow of The Research Foundation -- Flanders (FWO), Belgium} %spectroscopy
\and T.~Van Reeth\inst{\ref{inst1}} %10spectra, initial pattern fitting
\and C.~Aerts\inst{\ref{inst1},\ref{inst2}} %help
\and E.~Moravveji\inst{\ref{inst1}}\thanks{Marie Curie Postdoctoral Fellow} %models
\and M.~Van de Sande\inst{\ref{inst1}} %4spectra
\and K.~De Smedt\inst{\ref{inst1}} %3 spectra
\and S.~Bloemen\inst{\ref{inst1},\ref{inst2}} %3 spectra
\and J.~Southworth\inst{\ref{inst3}} %3 spectra (PI)
\and J.~Debosscher\inst{\ref{inst1}} %sample selection
\and E.~Niemczura\inst{\ref{inst11}} %1 spectra (PI)
\and J.~F.~Gameiro\inst{\ref{inst9}}} %1 spectra

\institute{Instituut voor Sterrenkunde, KU Leuven, Celestijnenlaan 200D, B-3001 Leuven, Belgium\\ \email{Peter.Papics@ster.kuleuven.be}\label{inst1}
\and Department of Astrophysics, IMAPP, Radboud University Nijmegen, PO Box 9010, 6500 GL Nijmegen, The Netherlands\label{inst2}
\and Astrophysics Group, Keele University, Staffordshire, ST5 5BG, United Kingdom\label{inst3}
\and Instytut Astronomiczny, Uniwersytet Wroc\l{}awski, Kopernika 11, 51-622 Wroc\l{}aw, Poland\label{inst11}
\and Instituto de Astrof\'{i}sica e Ci\^{e}ncias Espaciais and Faculdade de Ci\^{e}ncias, Universidade do Porto, Rua das Estrelas, PT4150-762 Porto, Portugal\label{inst9}}

\date{Received Day Month Year / Accepted Day Month Year}

\abstract{Massive stars are important building blocks of the Universe, and their stellar structure and evolution models are fundamental cornerstones of various fields in modern astrophysics. The precision of these models is strongly limited by our lack of understanding of various internal mixing processes that significantly influence the lifetime of these objects, such as core overshoot, chemical mixing, or the internal differential rotation.}{Our goal is to calibrate models by extending the sample of available seismic studies of slowly pulsating B (SPB) stars, providing input for theoretical modelling efforts that will deliver precise constraints on the parameters describing the internal mixing processes in these objects.}{We used spectral synthesis and disentangling techniques to derive fundamental parameters and to determine precise orbital parameters from high-resolution spectra. We employed custom masks to construct light curves from the virtually uninterrupted four year long \textit{Kepler} pixel data and used standard time-series analysis tools to construct a set of significant frequencies for each target. These sets were first filtered from combination frequencies, and then screened for period spacing patterns.}{We detect gravity mode period series of modes, of the same degree $\ell$ with consecutive radial order $n,$ in four new and one revisited SPB star. These series (covering typically 10 to 40 radial orders) are clearly influenced by moderate to fast rotation and carry signatures of chemical mixing processes. Furthermore, they are predominantly prograde dipole series. Our spectroscopic analysis, in addition to placing each object inside the SPB instability strip and identifying KIC\,4930889 as an SB2 binary, reveals that KIC\,11971405 is a fast rotator that shows very weak Be signatures. Together with the observed photometric outbursts this illustrates that this Be star is a fast rotating SPB star. We hypothesise that the outbursts might be connected to its very densely compressed oscillation spectrum.}{}


\titlerunning{Signatures of internal rotation discovered in the \textit{Kepler} data of five SPB stars}
\maketitle

%%%%%%%%%%%%%%%%%%
%%%Introduction%%%
%%%%%%%%%%%%%%%%%%

\section{Introduction}\label{intro}

Slowly pulsating B (SPB) stars are main-sequence stars of between B3 and B9 in spectral type, $11\,000$ and $22\,000$\,K in effective temperature, and $2.5\,\mathcal{M}_{\sun}$ and $8\,\mathcal{M}_{\sun}$ in mass, which show non-radial, heat-driven, gravity-mode oscillations \citep[see e.g.][Chapter 2]{2010aste.book.....A}. Although it has been a quarter of a century since their discovery by \citet{1991A&A...246..453W}, in-depth seismic modelling based on high-order $g$ modes was only achieved recently for KIC\,10526294 by \citet{2014A&A...570A...8P}. Even though SPB stars are not massive according to the classical definition of supernova progenitors, they share the same internal structure on the main sequence by having a convective core and a radiative envelope. This means that SPB stars are ideal asteroseismic probes of ill-understood internal mixing processes that have a significant influence on the lifetime of the metal factories of the Universe, such as core overshooting, diffusive mixing, or internal differential rotation. Therefore these stars hold the key to a precise calibration of stellar structure and evolution models of massive stars. Such calibrations are much needed because the observed distribution of massive stars does not agree with theoretical predictions, as shown by \citet{2014A&A...570L..13C}, and internal mixing is likely to be one of the possible solutions.

Before the space revolution of asteroseismology began with the launch of the MOST \citep[Microvariablity and Oscillations of STars;][]{2003PASP..115.1023W} and CoRoT \citep[Convection Rotation and planetary Transits;][]{2009A&A...506..411A} missions, it was very difficult to collect sufficient photometric data of SPB stars, mainly because their dominant oscillation modes have typical periods of 0.5--3\,days \citep[see\ e.g.][]{2002A&A...393..965D}, and also their mode amplitudes are relatively low (e.g. compared to the slightly hotter $\beta$\,Cephei stars that pulsate in $p$ modes). While some extremely time- and resource-intensive multisite campaigns and the CoRoT observations have resulted in a few in-depth observational studies showing an unexpected richness in the pulsation behaviour of OB stars in general \citep[for a historical overview see e.g.][]{papics_phd}, the number of stars with actual constraints on the extent of the core overshoot and/or the internal rotation profile remained very limited \citep{2013EAS....64..323A}. The real breakthrough arrived with the \textit{Kepler} Mission, thanks to its monitoring over four years.

The primary goal of \textit{Kepler} was the detection of Earth-like exoplanets \citep{2010Sci...327..977B}, but in the process the initial mission also provided us with four years of virtually uninterrupted photometric time series of more than $150\,000$ stars with micromagnitude precision in a 105-square-degree field of view (FOV) fixed between the constellations of Cygnus and Lyra, which are a goldmine for asteroseismology \citep{2010PASP..122..131G}. While the positioning of the FOV above the Galactic plane was not optimal in terms of the number of observable OB stars \citep{2011A&A...529A..89D,2011MNRAS.413.2403B,2012AJ....143..101M}, the long time base and high precision meant that previously inaccessible seismic diagnostic methods could be applied with success for the first time.

Forward seismic modelling is not possible without mode identification. Earlier this could be only achieved for a few selected modes from ground-based, follow-up, multicolour photometry and/or line-profile variation studies \citep[e.g.][]{2005A&A...432.1013D}; but now the full process can be reduced to a relatively simple pattern search in the periodogram combined with the approximate knowledge of fundamental parameters from spectroscopy. It has been known for a long time that there is structure in the dense frequency spectrum of SPB stars \citep{1993MNRAS.265..588D,1993MNRAS.262..213G}. A series of $g$ modes of the same degree $\ell$ with consecutive radial orders $n$ is not only predicted to be nearly equally spaced in period, but the periodic deviations from such an equal spacing carry information about the physical properties in the near-core region \citep{2008MNRAS.386.1487M}. The first detections of $g$-mode period series came from CoRoT data of the hybrid SPB/$\beta$\,Cep pulsators HD\,50230 \citep{2010Natur.464..259D,2012A&A..542A..88D} and HD\,43317 \citep{2012A&A...542A..55P}, but these were not precise enough to provide input for in-depth modelling.

This changed with \textit{Kepler} and the case of the extremely slow rotator KIC\,10526294. Aided especially by the tenfold increase in frequency resolution compared to CoRoT, \citet{2014A&A...570A...8P} managed to identify a clear series formed by the rotationally split multiplets of the 19 consecutive (in radial order) dipole modes of the star that are quasi-equally spaced. Based on the unprecedented observational data, \citet{2015A&A...580A..27M} found that a diffusive core-overshooting prescription that is exponentially decaying fits better than a step function formulation, and these authors also put stringent constraints on the value of the core overshoot; they derived $f_\mathrm{ov}\in[0.017,0.018]$ reaching a $\sim5\%$ precision in $f_\mathrm{ov}$ compared to the typical $\sim50\%$ achieved before. They showed that the inclusion of extra diffusive mixing ($\log{D}\in[1.75,2.00]\,\mathrm{cm}^2\mathrm{s}^{-1}$) in the radiative envelope is needed to explain the structure in the period spacing pattern. The relative difference between the observed and their model frequencies is better than 0.5\%. As a result of having such a good seismic model, \citet{2015ApJ...810...16T} have shown that, by using different inversion techniques, it is possible to derive the internal rotation profile of an SPB star using rotationally-split multiplets of different modes that penetrate to different depths of the star. In this particular case, while the seismic data are consistent with rigid rotation, a profile with counterrotation within the envelope has a statistical advantage over constant rotation. This finding was immediately validated by the 2D numerical simulations of \citet{2015ApJ...815L..30R}. Such an inversion has never been performed before based only on gravity modes, and even overall there were only a few earlier studies of hydrogen-burning stars (besides the Sun) that resulted in constraints on the internal rotation profile \citep[see][]{2003Sci...300.1926A,2004MNRAS.350.1022P,2007MNRAS.381.1482B,2014MNRAS.444..102K}. In the meantime a faster rotator analogue of this star was also found. KIC\,7760680 exhibits a rotationally affected series, consisting of 36 consecutive high-order, sectoral dipole gravity modes \citep{2015ApJ...803L..25P}, which carries clear signatures of chemical mixing and rotation. This star was modelled thoroughly by \citet{2016arXiv160402680M}, who reconfirmed that the exponential overshoot prescription on top of the receding convective core is superior  to the step function formulation. They constrained the overshooting parameter to $f_\mathrm{ov}=0.024\pm0.001$, which is slightly larger than the value in its virtually non-rotating counterpart KIC\,10526294. Furthermore, an extra diffusive mixing of $\log{D}=0.75\pm0.25\,\mathrm{cm}^2\mathrm{s}^{-1}$ was necessary to explain the deviations of the period spacing pattern from the asymptotic value.

In this paper we present our discovery of five additional SPB stars from the initial \textit{Kepler} Mission that show rotationally affected period series. Besides analysing the remaining four targets of our original Guest Observer (GO) sample of eight stars, we also revisit one of the primary components of the binary stars already presented in \citet{2013A&A...553A.127P}. We refer to the latter for details on the target selection. Future modelling based on this observational data will significantly extend the list of SPB stars that have well-constrained internal mixing parameters \citep{2015IAUS..307..154A}, and bring us closer to the seismic calibration of stellar structure and evolution models of massive stars.

%%%%%%%%%%%%%%%%%%%%%%%%%%%%
%%%Fundamental parameters%%%
%%%%%%%%%%%%%%%%%%%%%%%%%%%%

\section{Spectroscopy}\label{spectroscopy}

\subsection{Spectroscopic data}
 We gathered high-resolution spectra of our targets between 2010 and 2015 to confirm the initial classification of the SPB stars, obtain precise
fundamental parameters, test for their binary nature, and cover the binary
orbit if necessary. Depending on the apparent brightness, we used the \textsc{Hermes} spectrograph \citep{2011A&A...526A..69R} installed on the 1.2 metre Mercator telescope or the \textsc{Isis} spectrograph mounted on the 4.2 metre William Herschel Telescope, which are both located on La Palma (Spain). We refer to Table\,\ref{specsummary} for a summary of the spectroscopic observations of the new SPB stars.

\begin{table*}
\caption{Logbook of the spectroscopic observations -- grouped by stars and observing runs.}
\label{specsummary}
\centering
\renewcommand{\arraystretch}{1.1}
\begin{tabular}{c c c c c c c c c}
\hline\hline
Instrument & N &BJD first & BJD last & $\langle\mathrm{S/N}\rangle$ & S/N$_\mathrm{i}$ & $\mathrm{T}_{\mathrm{exp}}$ & R & Observer (PI)\\
\hline
\multicolumn{9}{l}{KIC\,3459297}\\
\textsc{Isis}     & 1 &\multicolumn{2}{c}{2456088.66015} &  90 & [76, 104] &   1500        &$22\,000\,\&\,13\,750$\tablefootmark{a} & SB (JS)\\
\hline
\multicolumn{9}{l}{KIC\,4930889}\\
\textsc{Hermes}   & 1 &\multicolumn{2}{c}{2455484.47403} &  90 &  90       &   1500        &$85\,000$              & PIP\\
\textsc{Hermes}   & 1 &\multicolumn{2}{c}{2455821.53738} & 100 & 100       &   1900        &$85\,000$              & JFG (EN)\\
\textsc{Hermes}   & 3 & 2456903.46202 & 2456907.45660    &  93 & [87,  97] &  [900, 1200]  &$85\,000$              & KDS\\
\textsc{Hermes}   &10 & 2457129.71807 & 2457139.62990    &  80 & [73,  86] &   1200        &$85\,000$              & TVR\\
\textsc{Hermes}   & 3 & 2457189.56028 & 2457197.63215    &  94 & [93,  97] & [1320, 1500]  &$85\,000$              & MVS\\
\textsc{Hermes}   & 8 & 2457201.68442 & 2457208.47316    &  96 & [83, 104] & [1200, 1800]  &$85\,000$              & PIP\\
\hline
\multicolumn{9}{l}{KIC\,9020774}\\
\textsc{Isis}     & 2 & 2456090.57710 & 2456090.59895    &  27 & [19,  35] &   1800        &$22\,000\,\&\,13\,750$\tablefootmark{a} & SB (JS)\\
\hline
\multicolumn{9}{l}{KIC\,11971405}\\
\textsc{Hermes}   & 2 & 2455487.34427 & 2455487.36568    &  79 & [75,  84] &  1800         &$85\,000$              & PIP\\
\textsc{Hermes}   & 1 &\multicolumn{2}{c}{2457190.64447} &  88 &  88       &  1800         &$85\,000$              & MVS\\
\textsc{Hermes}   & 2 & 2457201.65336 & 2457203.61089    &  56 & [44,  68] &  1200         &$85\,000$              & PIP\\
\hline
\end{tabular}
\tablefoot{For each observing run, the instrument, number of spectra N, BJD of the midpoint of the first and last exposure, average S/N level, range of S/N values (where the S/N of one spectrum is calculated as the average of the S/N measurements in several broad line free regions), typical exposure times (in seconds), resolving power of the spectrograph, and initials of the observer (and the PI if different) are given. \tablefoottext{a}{The resolution for the blue and red arms of \textsc{Isis} is different.}}
\end{table*}

The raw \textsc{Hermes} exposures were reduced using the dedicated pipeline of the instrument, resulting in the extracted, cosmic-removed, log-resampled (which includes barycentric correction), order-merged spectra that we used afterwards. The \textsc{Isis} frames were reduced using standard STARLINK routines following the optimal extraction described by \citet{1989PASP..101.1032M}. All spectra were rectified using the method described in \citet{2013A&A...553A.127P}.

We used the method of spectral disentangling \citep{1994A&A...281..286S} formulated in Fourier space \citep{1995A&AS..114..393H} and implemented in the FDBinary code \citep{2004ASPC..318..111I} to separate the individual spectral contributions for the KIC\,4930889 binary star (see below for details). The method bypasses the step of radial velocity (RV) determination but simultaneously optimises orbital elements of the system and individual spectra of the binary components instead. We start with the disentangling of several (typically 4 or 5) spectral regions where the contributions from both binary components can be clearly visible and determine orbital elements from each of the considered regions separately. The final set of the orbital parameters is then computed by taking average values from the above-mentioned individual solution, and errors bars are computed as the standard deviations of the mean. Once computed, the orbital elements are fixed and the disentangling is performed in the entire wavelength range, including the Balmer lines.

The spectra of all stars in this paper were analysed using the Grid Search in Stellar Parameters \citep[GSSP;][]{2015A&A...581A.129T} software package\footnote{https://fys.kuleuven.be/ster/meetings/binary-2015/gssp-software-package}. The method is based on a comparison of the observed spectra in an arbitrary wavelength range with a grid of synthetic spectra in six fundamental parameters ($T_\mathrm{eff}$, $\log g$, micro- and macroturbulent velocities, $v \sin i$, and metallicity) and optionally light dilution factor (specifically for the disentangled spectra of binary stars). The latter is assumed to be constant over considered wavelength interval, which is justified in our case, given similar spectral types of the individual binary components. Synthetic spectra are computed using the SynthV spectrum synthesis code \citep{1996ASPC..108..198T} based on the grid of atmosphere models pre-computed with the most recent version of the LLmodels code \citep{2004A&A...428..993S}. Both codes operate in LTE mode and make use of the atomic data extracted from the Vienna Atomic Line Database \citep[VALD;][]{2000BaltA...9..590K}. Chi-square merit function is used to evaluate the quality of the fit between the observed spectrum and each of the synthetic spectra in a grid; the error bars on the fundamental parameters are computed from chi-square statistics taking all possible correlations between the parameters in question into account. \citet{2015A&A...581A.129T} provides a detailed description of the method and full capabilities of the GSSP code.

\begin{figure*}
\resizebox{\hsize}{!}{\includegraphics{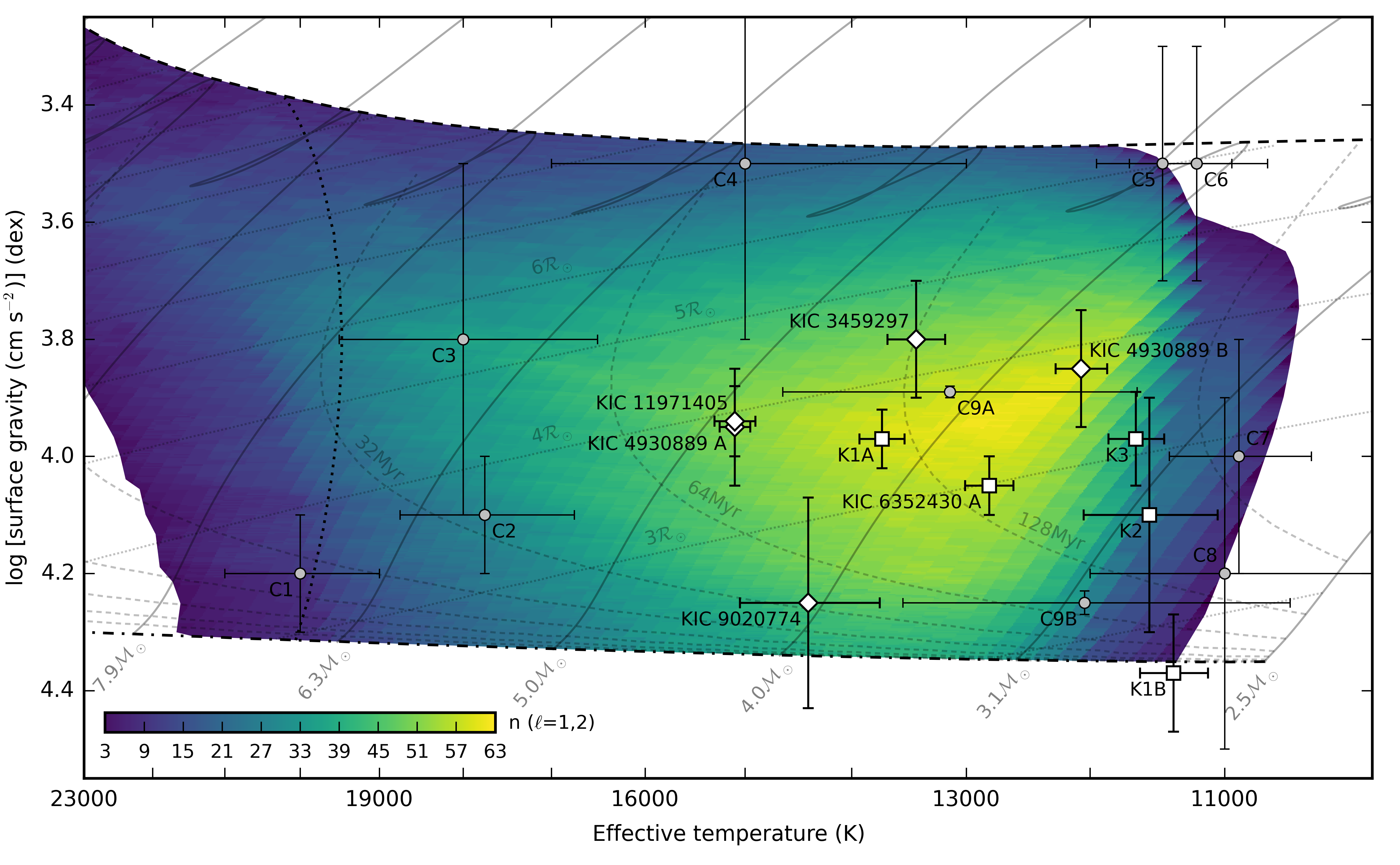}} 
\caption{Kiel diagram of a sample of B-type stars near the main sequence, for which an in-depth seismic data analysis is available, observed by CoRoT (circles) and \textit{Kepler} (squares). The new SPBs from this paper are plotted with diamonds. Stars analysed or revisited in this paper are labelled using their full names, while for the others we use the following annotations: C1 -- HD\,48977, C2 -- HD\,43317, C3 -- HD\,50230\,A, C4 -- HD\,50846\,A, C5 -- HD\,182198, C6 -- HD\,181440, C7 -- HD\,46179, C8 -- HD\,174648, C9 -- HD\,174884\,AB, K1 -- KIC\,4931738\,AB, K2 -- KIC\,10526294, and K3 -- KIC\,7760680. The dot-dashed line represents the zero-age main sequence (ZAMS), while the dashed line indicates the terminal-age main sequence (TAMS). The cool edge of the $\beta$\,Cep instability strip is plotted with a dotted line (for the $\ell=0$ modes). The thin grey solid lines denote evolutionary tracks for selected masses. Isochrones from $2^0$ to $2^8$\,Myr and isoradii corresponding to integer multiples of the Solar radius are plotted using thin dashed and dotted grey lines, respectively. The instability strip of SPB stars is defined as the region where the sum of excited $\ell=1$ plus $\ell=2$ gravity modes is at least three, and it is shown in colours ranging from purple (3 excited modes) to bright yellow (63 excited modes). All model data were taken from \citet{2016MNRAS.455L..67M} for $Z=0.014$, the \citet{2009ARA&A..47..481A} mixture, an exponential core overshoot of $f_\mathrm{ov}=0.02$, and opacity enhancement factors for iron and nickel of $\beta_\mathrm{Fe}=1.75$ and $\beta_\mathrm{Ni}=1.75$. Different error bars reflect differences in data quality and methodological approach.}
\label{hrd}
\end{figure*}

In the following subsections we give the basic observable parameters and discuss the results from the spectral analyses for all targets individually. \citet[Table 5,][]{2013A&A...553A.127P} provide the results for KIC\,6352430\,A. A general overview of the position of the stars in the SPB instability strip among other CoRoT and \textit{Kepler} targets is given in Fig.\,\ref{hrd}.

\subsubsection{KIC\,3459297}

\begin{figure*}
\resizebox{\hsize}{!}{\includegraphics{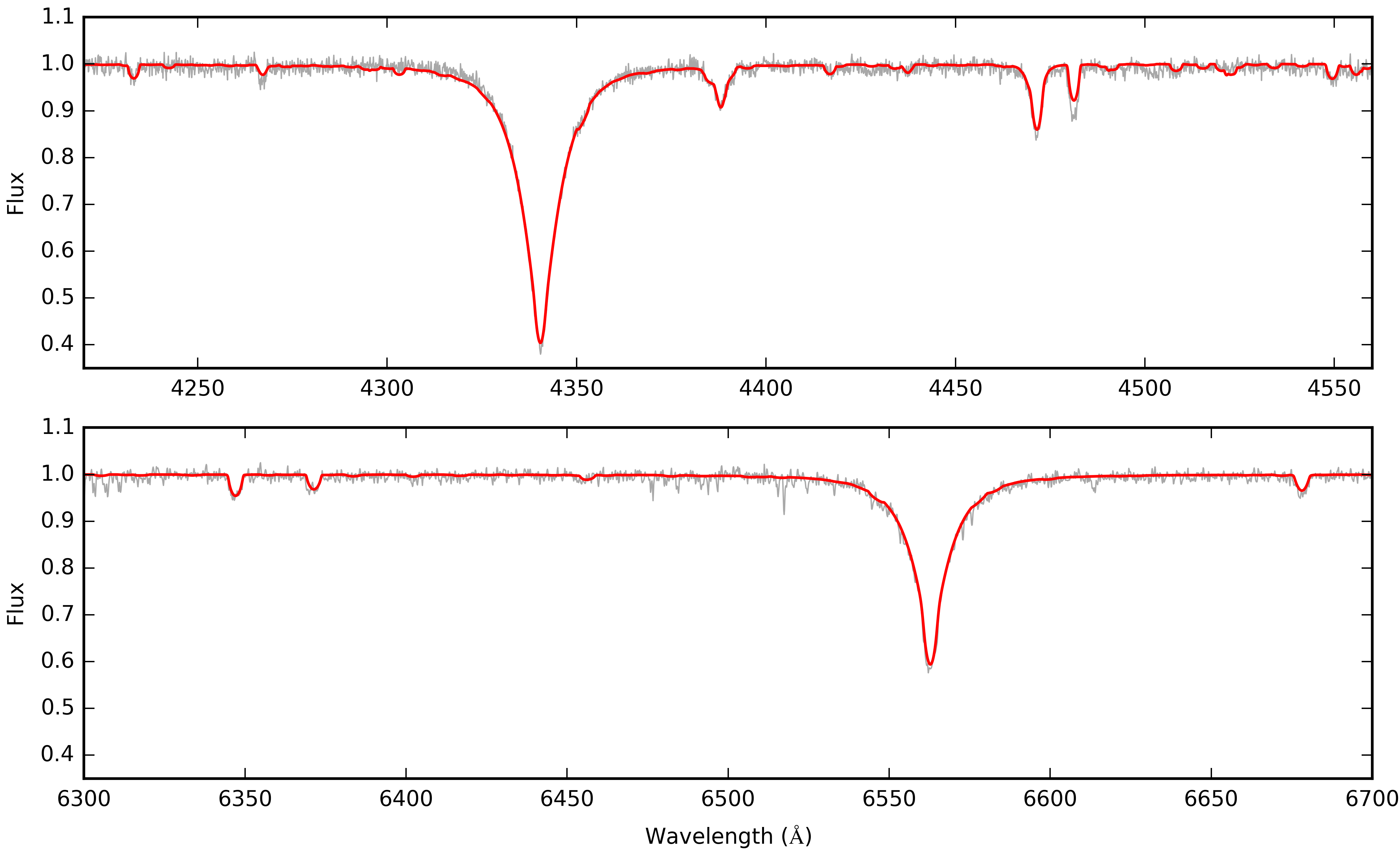}} 
\caption{Comparison of the rectified observed spectrum and the synthetic spectrum for the two observed wavelength regions of KIC\,3459297. In each panel, the \textsc{Isis} spectrum is plotted with a grey solid line and the synthetic spectrum is plotted with a red solid line.}
\label{spectralfitKIC3459297}
\end{figure*}

The fundamental parameters from the spectral synthesis of KIC\,3459297 are listed in Table\,\ref{fundparamsKIC3459297}, while we show the quality of the fit between the synthetic and observed spectra in Fig.\,\ref{spectralfitKIC3459297}. The derived parameters suggest that KIC\,3459297 is the most evolved of the four new SPBs of this paper.

\begin{table}
\caption{Fundamental parameters and basic observable properties of KIC\,3459297.}
\label{fundparamsKIC3459297}
\centering
\renewcommand{\arraystretch}{1.25}
\setlength{\tabcolsep}{1pt}
\begin{tabular}{l l r r c l c r c l}
\hline\hline
\multicolumn{2}{l}{Parameter} && \multicolumn{3}{c}{KIC\,3459297}\\
\hline
\multicolumn{2}{l}{$T_\mathrm{eff}\,(\mathrm{K})$} && $13430$&$\pm$&$250$\\
\multicolumn{2}{l}{$\log g\,\mathrm{(cgs)}$} && $3.8$&$\pm$&$0.1$\\
\multicolumn{2}{l}{$[M/H]$} && $-0.14$&$\pm$&$0.21$\\
\multicolumn{2}{l}{$v \sin i$\,$(\mathrm{km\,s}^{-1})$} && $109$&$\pm$&$14$\\
\multicolumn{2}{l}{$\xi_\mathrm{t}\,(\mathrm{km\,s}^{-1})$}&& $2$&\multicolumn{2}{l}{(fixed)}\\
\multicolumn{2}{l}{Spectral type}&&\multicolumn{3}{c}{B6.5\,IV}\\
\hline
\multicolumn{2}{l}{$\alpha_{2000}$}&&\multicolumn{3}{l}{$19^\mathrm{h}40^\mathrm{m}54\fs458$}\\
\multicolumn{2}{l}{$\delta_{2000}$}&+&\multicolumn{3}{l}{$38\degr34\arcmin50\farcs38$}\\
\multicolumn{2}{l}{\textit{Kepler} mag.}&&$12.554$\\
\hline
\end{tabular}
\tablefoot{Spectral type were determined based on $T_\mathrm{eff}$ and $\log g$ values using an interpolation in the tables given by \citet{1982SchmidtKalerBook}. Observable properties are from the \citet{2009yCat.5133....0K}.}
\end{table}

\subsubsection{KIC\,4930889}

\begin{figure}
\resizebox{\hsize}{!}{\includegraphics{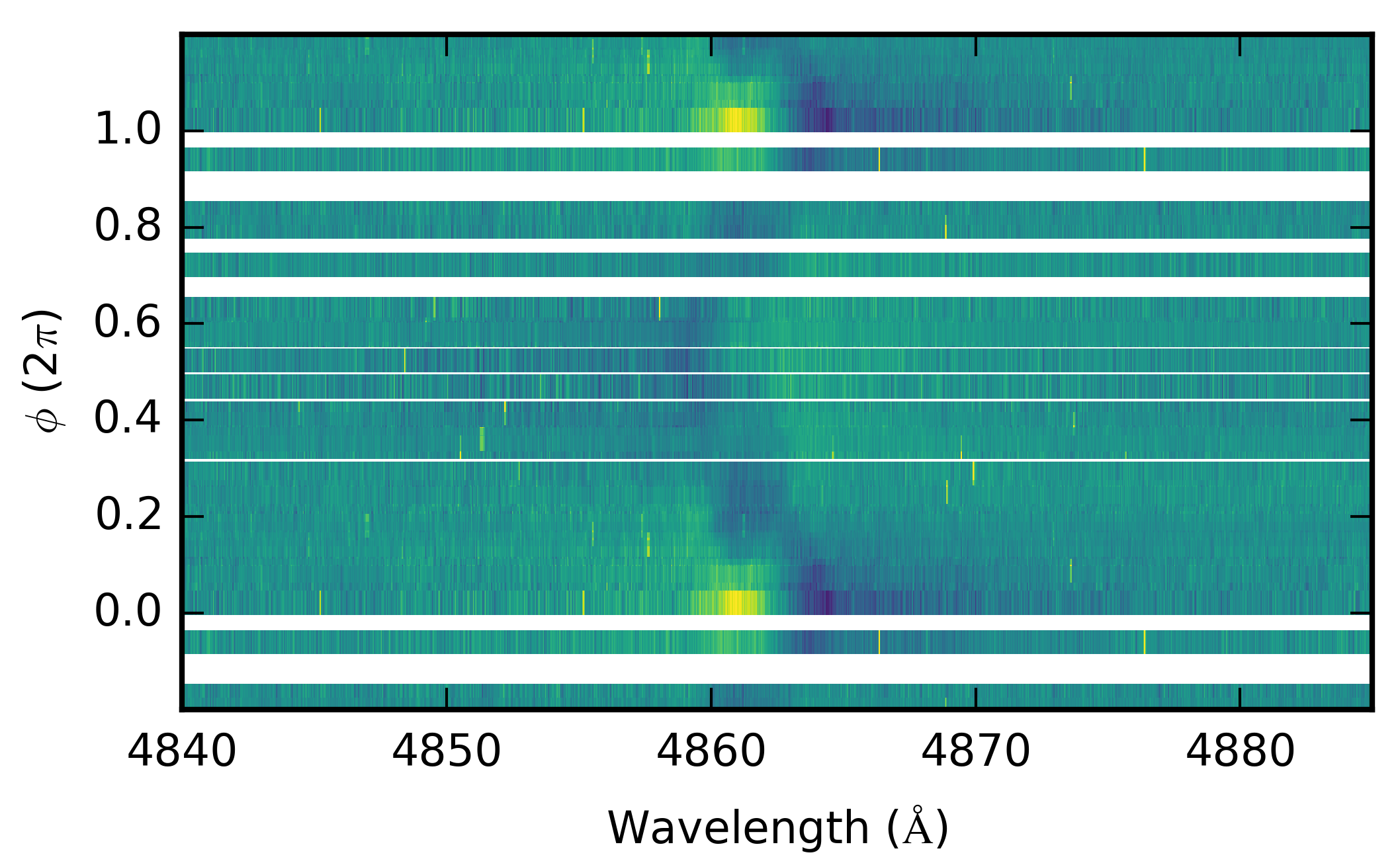}} 
\caption{Phase-folded residual spectra of KIC\,4930889 around the \ion{H}{i} line at 4861\AA\  obtained after subtraction of the average. The colour scale represents the residual intensity at each wavelength point. White horizontal regions denote phases that are more than $0.05\pi$ away from an observed spectrum.}
\label{binaryspectrachangesKIC4930889}
\end{figure}

\begin{figure}
\resizebox{\hsize}{!}{\includegraphics{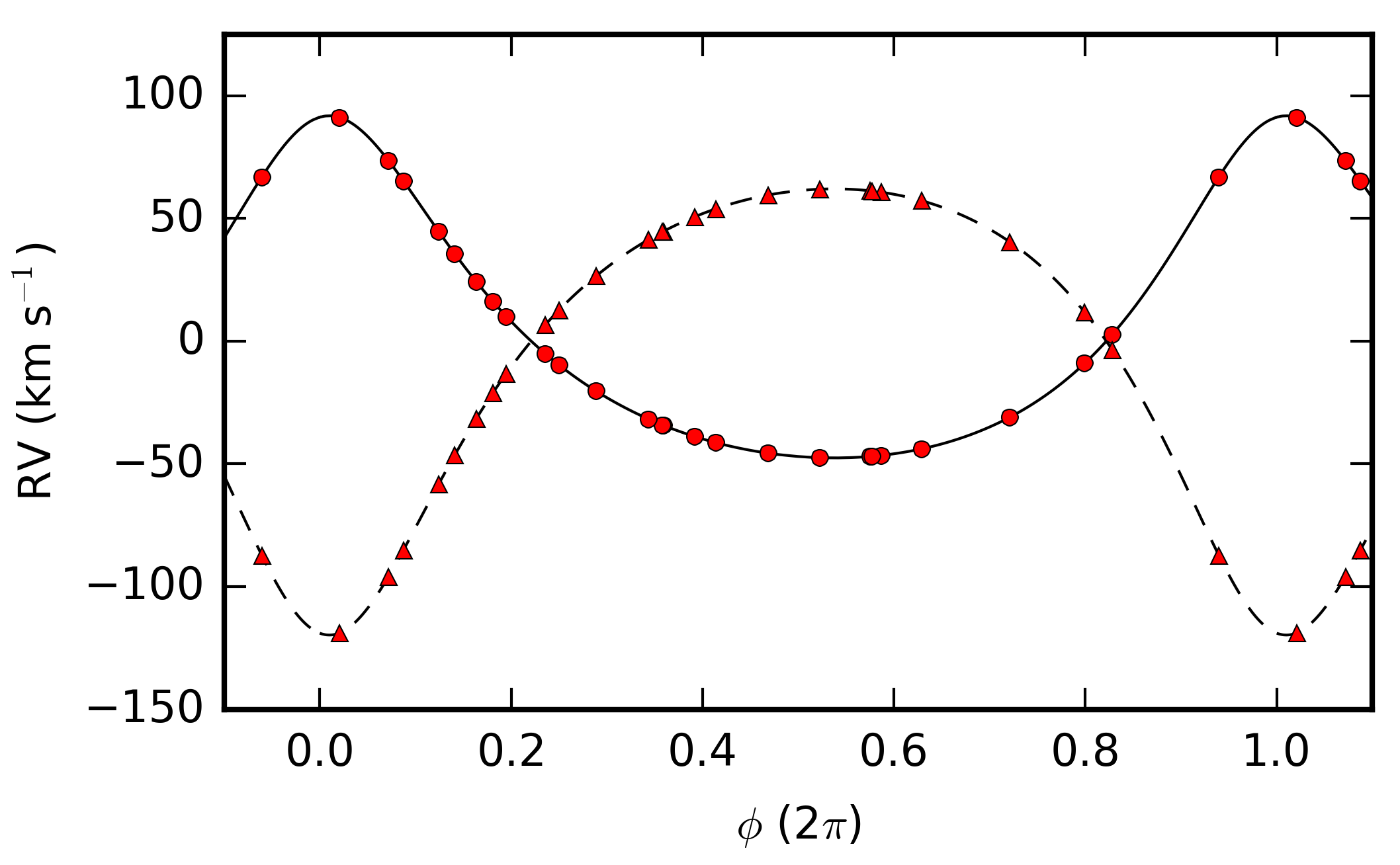}} 
\caption{Spectroscopic orbital solution for KIC\,4930889 (solid line: primary; dashed line: secondary) and the radial velocities from the disentangling process. Filled circles indicate RVs of the primary; filled triangles denote RVs of the secondary component.}
\label{binaryorbitKIC4930889}
\end{figure}

\begin{figure*}
\resizebox{\hsize}{!}{\includegraphics{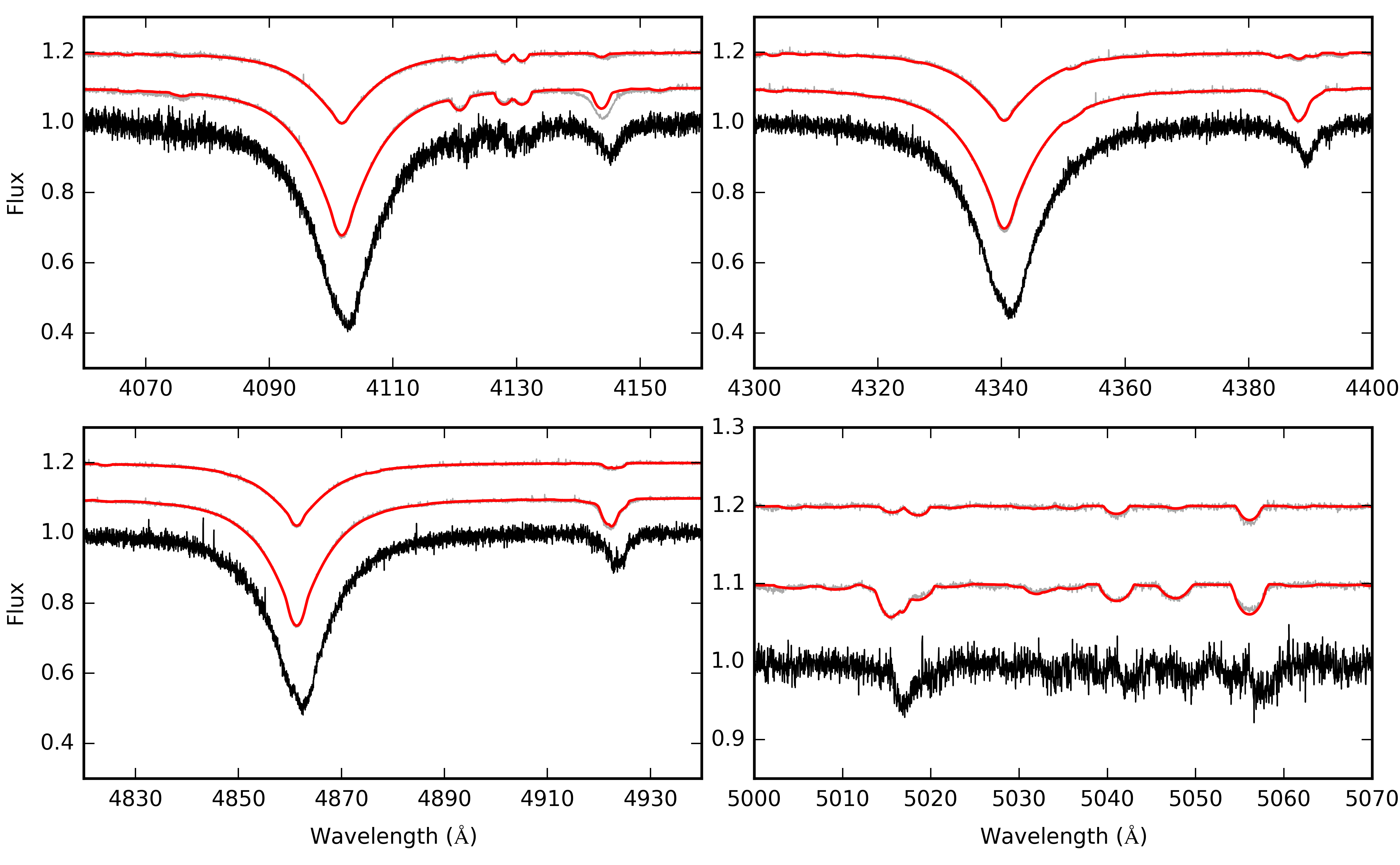}} 
\caption{Comparison of the \textsc{Hermes} spectrum at the phase of the largest observed radial velocity difference, disentangled component spectra, and synthetic spectra for each component for selected wavelength regions of KIC\,4930889. In each panel, the \textsc{Hermes} spectrum is plotted with a black solid line, the disentangled spectra are plotted with grey solid lines, and synthetic spectra (scaled to the disentangled spectra) are plotted with red solid lines. For clarity, the primary and secondary spectra are plotted with a vertical offset of $+0.1$ and $+0.2$ flux units, respectively.}
\label{spectralfitKIC4930889}
\end{figure*}

KIC\,4930889 turned out to be a double-lined spectroscopic binary (SB2), thus we took a total of 26 exposures with \textsc{Hermes} to cover the binary orbit. The SB2 nature of the original spectra is visible in Fig.\,\ref{binaryspectrachangesKIC4930889}. The derived orbital parameters from the disentangling process are listed in Table\,\ref{fundparamsKIC4930889}, while the reconstructed orbit and radial velocities are shown in Fig.\,\ref{binaryorbitKIC4930889}. The fundamental parameters, obtained via spectral synthesis from the disentangled spectra place both components in the SPB instability strip, are listed in Table\,\ref{fundparamsKIC4930889}, while the fits of the disentangled spectra are shown in Fig.\,\ref{spectralfitKIC4930889}.

\begin{table}
\caption{Fundamental parameters, orbital parameters, and basic observable properties of KIC\,4930889.}
\label{fundparamsKIC4930889}
\centering
\renewcommand{\arraystretch}{1.25}
\setlength{\tabcolsep}{1pt}
\begin{tabular}{l l r r c l c r c l}
\hline\hline
\multicolumn{2}{l}{Parameter} && \multicolumn{3}{c}{KIC\,4930889 A} && \multicolumn{3}{c}{KIC\,4930889 B} \\
\hline
\multicolumn{2}{l}{$T_\mathrm{eff}\,(\mathrm{K})$} && $15100$&$\pm$&$150$ && $12070$&$\pm$&$200$\\
\multicolumn{2}{l}{$\log g\,\mathrm{(cgs)}$} && $3.95$&$\pm$&$0.1$ && $3.85$&$\pm$&$0.1$ \\
\multicolumn{2}{l}{$[M/H]$} && $-0.08$&$\pm$&$0.1$ && $-0.09$&$\pm$&$0.1$ \\
\multicolumn{2}{l}{$v \sin i_\mathrm{rot}\,(\mathrm{km\,s}^{-1})$} && $116$&$\pm$&$6$ && $85$&$\pm$&$5$ \\
\multicolumn{2}{l}{$\xi_\mathrm{t}\,(\mathrm{km\,s}^{-1})$}&& $1.85$&$\pm$&$0.8$ && $2$&$\pm$&$1$\\
\multicolumn{2}{l}{Light factor}&& $0.71$&$\pm$&$0.01$ && $0.29$&$\pm$&$0.01$\\
\multicolumn{2}{l}{Spectral type\tablefootmark{a}}&&\multicolumn{3}{c}{B5\,IV-V}&&\multicolumn{3}{c}{B8\,IV-V}\\
\hline
\multicolumn{2}{l}{$P\,(\mathrm{days})$}&&\multicolumn{3}{r}{$18.296$}&$\pm$&\multicolumn{3}{l}{$0.002$}\\
\multicolumn{2}{l}{$T_0\,(\mathrm{BJD})$}&&\multicolumn{3}{r}{$2\,455\,481.48$}&$\pm$&\multicolumn{3}{l}{$0.17$}\\
\multicolumn{2}{l}{$e$}&&\multicolumn{3}{r}{$0.32$}&$\pm$&\multicolumn{3}{l}{$0.02$}\\
\multicolumn{2}{l}{$\omega\,({\degr})$}&&\multicolumn{3}{r}{352.7}&$\pm$&\multicolumn{3}{l}{4.9}\\
\multicolumn{2}{l}{$K\,(\mathrm{km\,s}^{-1})$}&&$69.7$&$\pm$&$1.0$&&$90.9$&$\pm$&$0.8$\\
\multicolumn{2}{l}{$\mathcal{M}\sin^3 i_\mathrm{orb}\,(\mathcal{M}_{\sun})$}&&$3.8$&$\pm$&$0.2$&&$2.9$&$\pm$&$0.2$\\
\multicolumn{2}{l}{$a\sin i_\mathrm{orb}\,(R_{\sun})$}&&$23.9$&$\pm$&$0.5$&&$31.1$&$\pm$&$0.5$\\
\multicolumn{2}{l}{$q$}&&\multicolumn{3}{r}{$0.77$}&$\pm$&\multicolumn{3}{l}{$0.09$}\\
\hline
\multicolumn{2}{l}{$\alpha_{2000}$}&&\multicolumn{3}{l}{$19^\mathrm{h}35^\mathrm{m}07\fs831$}\\
\multicolumn{2}{l}{$\delta_{2000}$}&+&\multicolumn{3}{l}{$40\degr01\arcmin28\farcs88$}\\
\multicolumn{2}{l}{\textit{Kepler} mag.}&&$8.883$\\
\hline
\end{tabular}
\tablefoot{See Table\,\ref{fundparamsKIC3459297}.}
\end{table}

\subsubsection{KIC\,9020774}

\begin{figure*}
\resizebox{\hsize}{!}{\includegraphics{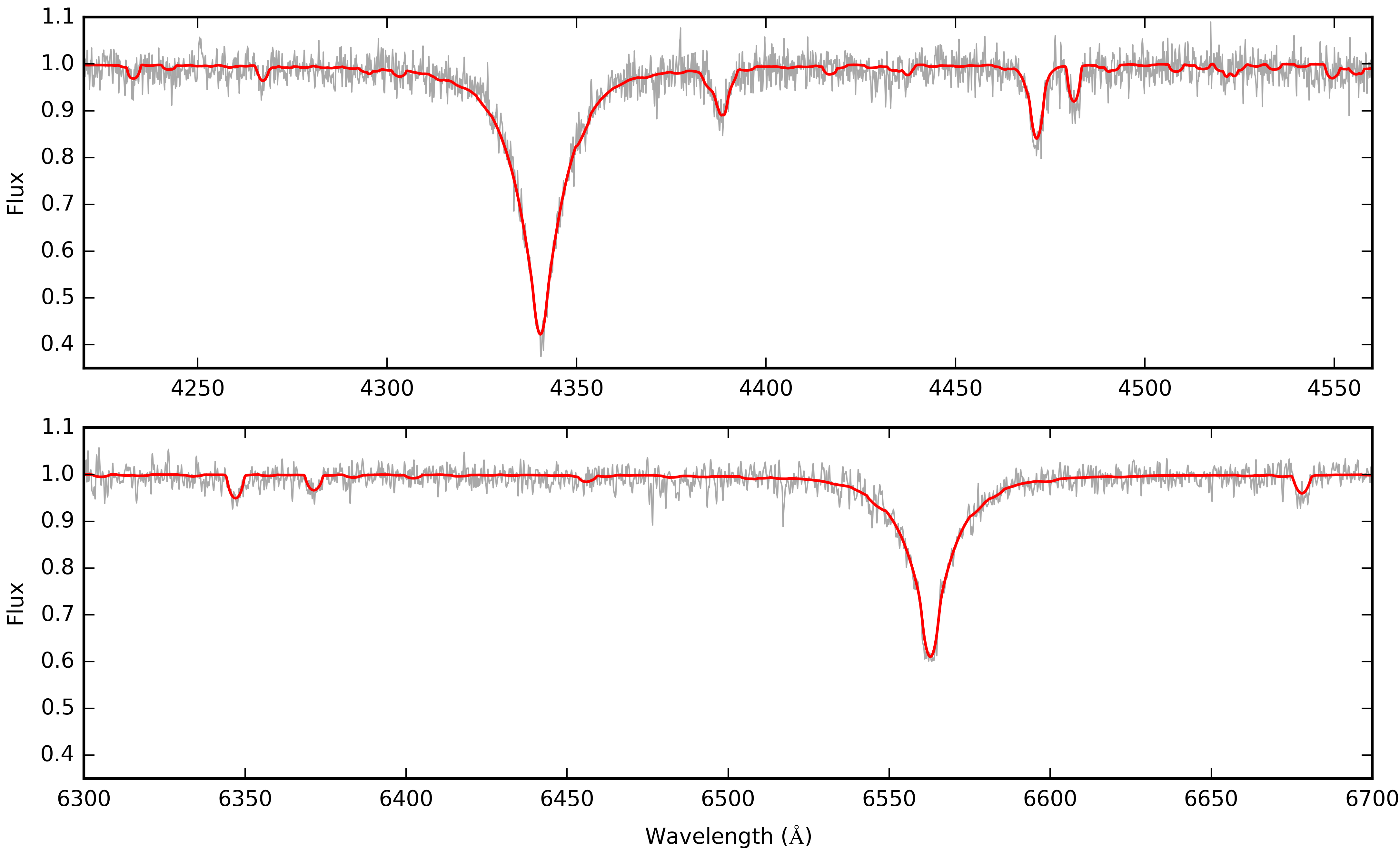}} 
\caption{Comparison of the synthetic spectrum and the average of the rectified observed spectra for the two observed wavelength regions of KIC\,9020774. In each panel, the \textsc{Isis} spectrum is plotted with a grey solid line, and the synthetic spectrum is plotted with a red solid line.}
\label{spectralfitKIC9020774}
\end{figure*}

Its low brightness (with \textit{Kepler} mag. = 15.091, the dimmest source in our GO sample) limited the precision of our analysis, but the fundamental parameters derived from the spectral synthesis (see Table\,\ref{fundparamsKIC9020774}, and Fig.\,\ref{spectralfitKIC9020774}) strongly suggest that KIC\,9020774 is a young B star.

\begin{table}
\caption{Fundamental parameters and basic observable properties of KIC\,9020774.}
\label{fundparamsKIC9020774}
\centering
\renewcommand{\arraystretch}{1.25}
\setlength{\tabcolsep}{1pt}
\begin{tabular}{l l r r c l c r c l}
\hline\hline
\multicolumn{2}{l}{Parameter} && \multicolumn{3}{c}{KIC\,9020774}\\
\hline
\multicolumn{2}{l}{$T_\mathrm{eff}\,(\mathrm{K})$} && $14400$&$\pm$&$650$\\
\multicolumn{2}{l}{$\log g\,\mathrm{(cgs)}$} && $4.25$&$\pm$&$0.18$\\
\multicolumn{2}{l}{$[M/H]$} && $0.20$&$\pm$&$0.35$\\
\multicolumn{2}{l}{$v \sin i$\,$(\mathrm{km\,s}^{-1})$} && $129$&$\pm$&$28$\\
\multicolumn{2}{l}{$\xi_\mathrm{t}\,(\mathrm{km\,s}^{-1})$}&& $2$&\multicolumn{2}{l}{(fixed)}\\
\multicolumn{2}{l}{Spectral type}&&\multicolumn{3}{c}{B5.7\,V}\\
\hline
\multicolumn{2}{l}{$\alpha_{2000}$}&&\multicolumn{3}{l}{$19^\mathrm{h}25^\mathrm{m}36\fs998$}\\
\multicolumn{2}{l}{$\delta_{2000}$}&+&\multicolumn{3}{l}{$45\degr23\arcmin44\farcs81$}\\
\multicolumn{2}{l}{\textit{Kepler} mag.}&&$15.091$\\
\hline
\end{tabular}
\tablefoot{See Table\,\ref{fundparamsKIC3459297}.}
\end{table}

\subsubsection{KIC\,11971405}\label{KIC11971405spectroscopy}

\begin{figure*}
\resizebox{\hsize}{!}{\includegraphics{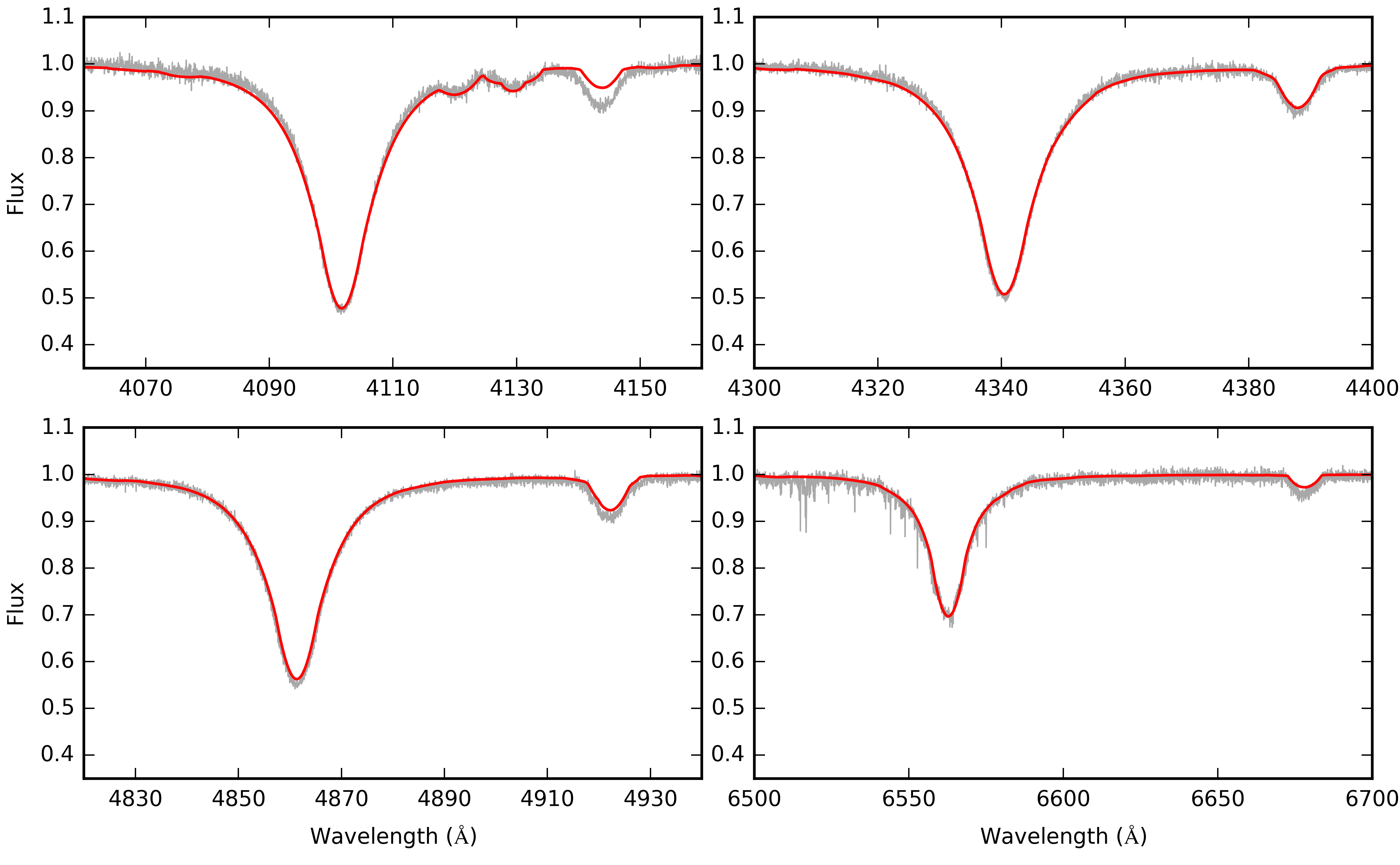}} 
\caption{Comparison of the median of the rectified observed spectra, and the synthetic spectrum for selected wavelength regions of KIC\,11971405. In each panel, the \textsc{Hermes} spectrum is plotted with a grey solid line, and the synthetic spectrum is plotted with a red solid line.}
\label{spectralfitKIC11971405}
\end{figure*}

\begin{figure}
\resizebox{\hsize}{!}{\includegraphics{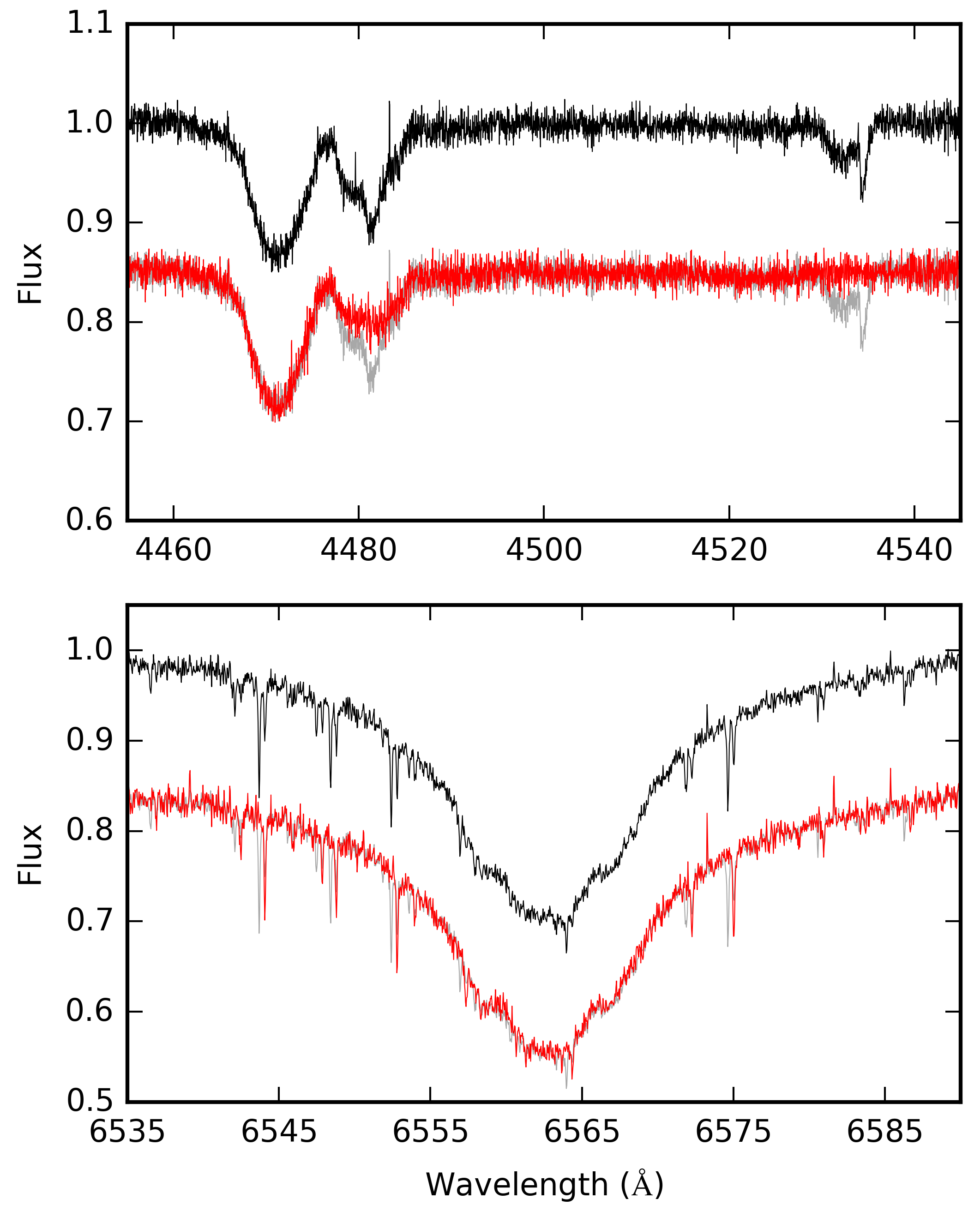}} 
\caption{Temporal changes in the spectrum of KIC\,11971405. The average of the two consecutive exposures from October 2010 is plotted on the top using a black solid line, while the average spectrum from June 2015 is shown below with a red solid line with an offset of $-0.15$ in flux units. For clarity, the average from 2010 is repeated in the background using a grey solid line. The sharp absorption features in the bottom panel are telluric lines.}
\label{spectrachangesKIC11971405}
\end{figure}

Spectroscopy of KIC\,11971405 revealed that the star is a fast rotator near the middle of the SPB instability strip (see Table\,\ref{fundparamsKIC11971405}, and Fig.\,\ref{spectralfitKIC11971405}) with a projected rotational velocity of $\sim240\,\mathrm{km\,s}^{-1}$, which is the highest from the stars in our GO sample. Furthermore, the stellar spectrum shows very weak emission features near the core of the $\mathrm{H}\alpha$ line ($6563\AA$), and changes over time (see Fig.\,\ref{spectrachangesKIC11971405}). The emission component in $\mathrm{H}\alpha$ is extremely weak; it appears only as two small bumps inside the absorption line around the core and it is not visible in any other Balmer line. Nevertheless, this grants the spectral classification of Be \citep[for a review, see][]{2013A&ARv..21...69R}. We present, in Sect.\,\ref{KIC11971405photometry}, that the star showed small photometric outbursts during the nominal \textit{Kepler} mission, which strengthens this classification. The exposures taken in October 2010 (before the start of the photometric outbursts) show excess absorption around the \ion{Mg}{ii} lines at $4481\AA$ and $4534\AA$ compared to the observations from 2015 (after the start of the outbursts). These features consist of a broader and narrower absorption component. We checked whether these features could be from external contamination (e.g. of telluric origin), but the examination of other exposures from the same and neighbouring nights did not show anything unexpected at these wavelengths. Similar sharp red-shifted transient absorption features were observed in the Be star $\omega$\,CMa \citep{2003A&A...402..253S}, but in the outburst phase. In addition, the emission features in $\mathrm{H}\alpha$ also show some minor changes between the two epochs.

\begin{table}
\caption{Fundamental parameters and basic observable properties of KIC\,11971405.}
\label{fundparamsKIC11971405}
\centering
\renewcommand{\arraystretch}{1.25}
\setlength{\tabcolsep}{1pt}
\begin{tabular}{l l r r c l c r c l}
\hline\hline
\multicolumn{2}{l}{Parameter} && \multicolumn{3}{c}{KIC\,11971405}\\
\hline
\multicolumn{2}{l}{$T_\mathrm{eff}\,(\mathrm{K})$} && $15100$&$\pm$&$200$\\
\multicolumn{2}{l}{$\log g\,\mathrm{(cgs)}$} && $3.94$&$\pm$&$0.06$\\
\multicolumn{2}{l}{$[M/H]$} && $-0.15$&$\pm$&$0.14$\\
\multicolumn{2}{l}{$v \sin i$\,$(\mathrm{km\,s}^{-1})$} && $242$&$\pm$&$14$\\
\multicolumn{2}{l}{$\xi_\mathrm{t}\,(\mathrm{km\,s}^{-1})$}&& $2$&\multicolumn{2}{l}{(fixed)}\\
\multicolumn{2}{l}{Spectral type}&&\multicolumn{3}{c}{B5\,IV-Ve}\\
\hline
\multicolumn{2}{l}{$\alpha_{2000}$}&&\multicolumn{3}{l}{$19^\mathrm{h}42^\mathrm{m}48\fs842$}\\
\multicolumn{2}{l}{$\delta_{2000}$}&+&\multicolumn{3}{l}{$50\degr21\arcmin39\farcs85$}\\
\multicolumn{2}{l}{\textit{Kepler} mag.}&&$9.315$\\
\hline
\end{tabular}
\tablefoot{See Table\,\ref{fundparamsKIC3459297}.}
\end{table}

%%%%%%%%%%%%%%%%
%%%Photometry%%%
%%%%%%%%%%%%%%%%

\section{The \textit{Kepler} photometry}\label{photometry}

\subsection{Description and reduction of the data}

\begin{figure}
\resizebox{\hsize}{!}{\includegraphics{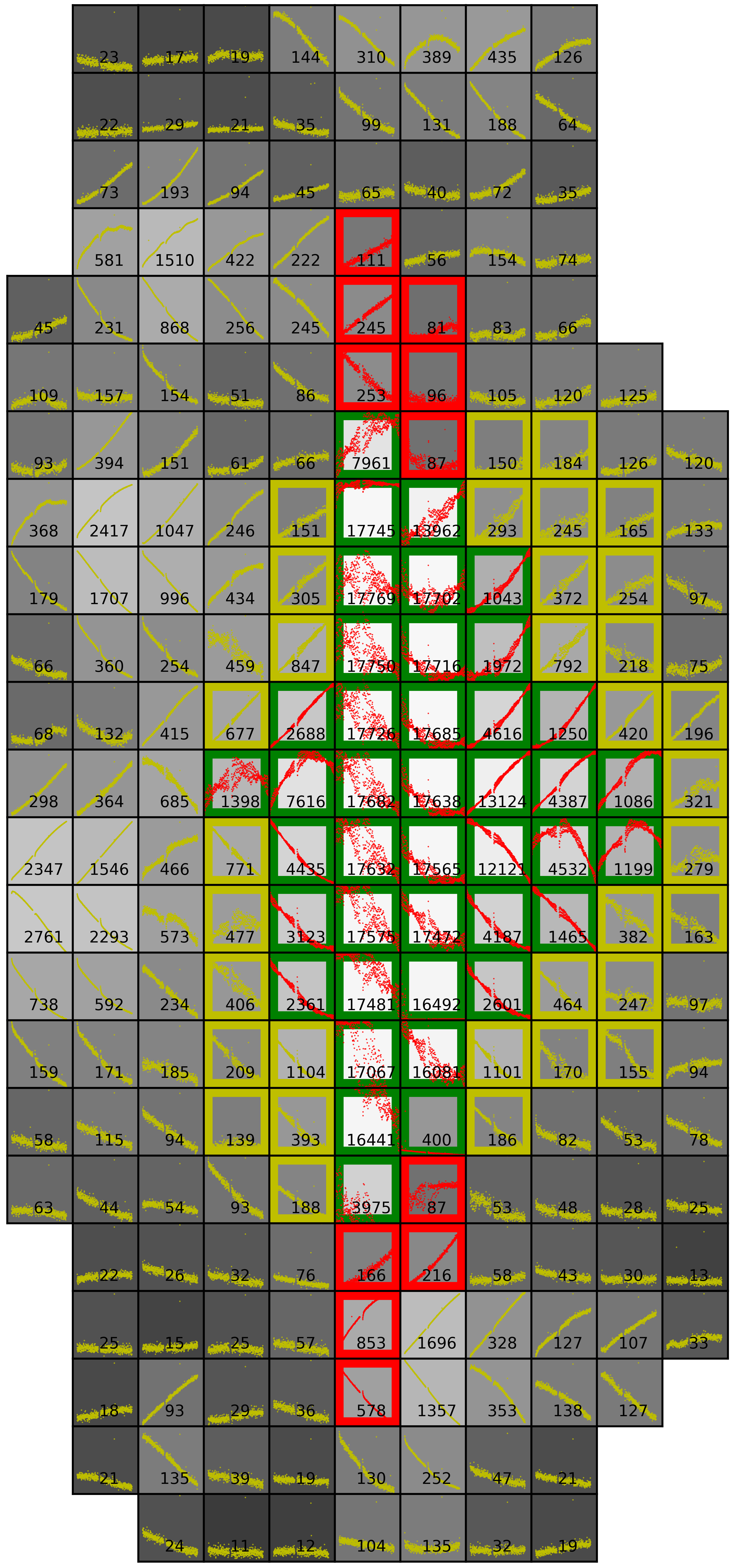}} 
\caption{Custom pixel mask for the Q8 data of KIC\,4930889. Individual pixel-level light curves are plotted for each pixel that was downloaded from the spacecraft. The numerical values and the background colour indicate the flux levels in each pixel. The pixels with green and red borders were used to extract the standard Kepler light curves. We added yellow pixels (significant signal is present) and removed red pixels (non-significant signal or contamination from nearby stars) in our custom mask for the light curve extraction. This results in a light curve with significantly less instrumental effects and/or contamination than the standard extraction.}
\label{KIC4930889pixelmask}
\end{figure}

\begin{figure*}
\resizebox{\hsize}{!}{\includegraphics{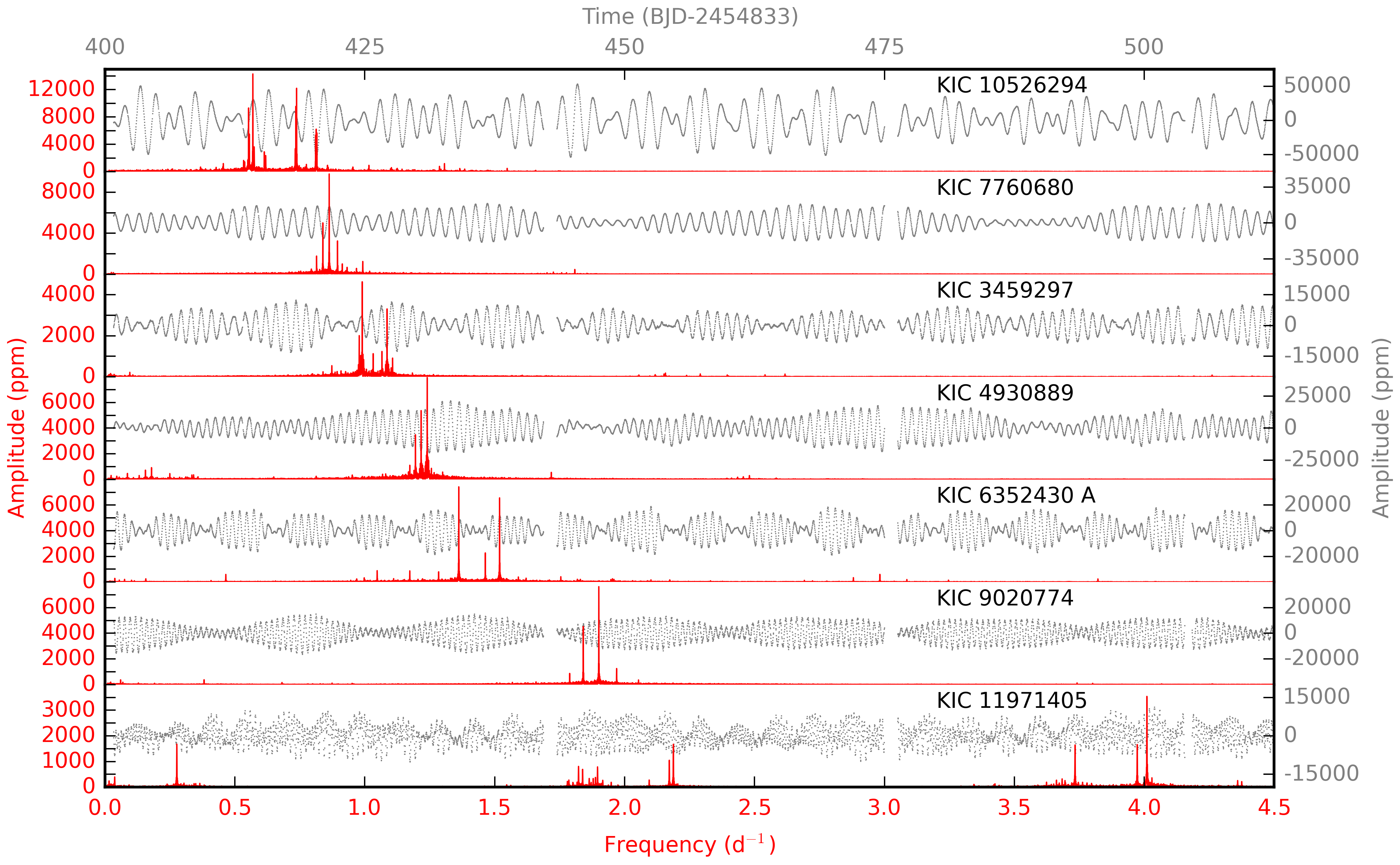}} 
\caption{Light curves (plotted using grey dots, showing a zoom-in to slightly less than 1/10th of the full data set) and the power spectra (plotted using red solid lines) of the five SPB stars studied in this paper along with the ones of KIC\,10526294 and KIC\,7760680 from \citet{2014A&A...570A...8P,2015ApJ...803L..25P} ordered according to the frequency of maximum power. For further information see Section\,\ref{spacingdiscussion}.}
\label{alllightcurvesandfouriers}
\end{figure*}

For the photometric analyses presented later in this paper we used all available \textit{Kepler} data from the nominal mission. These stars were observed in long cadence mode (LC) with a cadence of 29.43 minutes, for $\sim4\,\mathrm{years}$ between May 2009 and May 2013 (from the commissioning quarter Q0 to the end of the last science quarter Q17). There was no data taken of KIC\,6352430\,A and KIC\,9020774 during Q0, and because of the loss of module 3 of the CCD array, there was also no photometry gathered during Q6, Q10, and Q14 of KIC\,3459297 and KIC\,4930889. We used custom masks (see example in Fig.\,\ref{KIC4930889pixelmask}) to construct quarterly light curves from the target pixel files (Data Release 24) to minimise instrumental trends by adding pixels around the default masks with significant flux in them, and to avoid possible contamination from nearby objects by removing pixels containing light from sources other than the target. We cleaned all quarters from clear outliers, then detrended them using a division with a low-order (typically 2nd order) polynomial. In some specific cases quarters were broken up into pieces for this if the signal levels before and after an anomaly (e.g. the large coronal mass ejection during Q12) were visibly different. Afterwards the detrended quarters were merged to a continuous light curve and the counts were converted to ppm (see Fig.\,\ref{alllightcurvesandfouriers} for an overview of the final light curves).

\subsection{Frequency analysis}

To extract the pulsation modes (and describe any other periodic light variation), we followed a standard prewhitening procedure similar to that described in \citet{2014A&A...570A...8P}, with the small modifications that a) the prewhitening was carried out in the order of peak significance (instead of amplitude), and b) the prewhitening procedure was stopped when the signal-to-noise ratio (S/N; calculated in a $1\,\mathrm{d}^{-1}$ window) of the last prewhitened peak dropped below four (instead of using a $p$ value in hypothesis testing and filtering the frequency set later based on the measured S/N values). This resulted in the set of Fourier parameters ($A_j$ amplitudes, $f_j$ frequencies, and $\theta_j$ phases) that we used later for the analysis. An overview of the power spectra is shown in Fig.\,\ref{alllightcurvesandfouriers}.

\subsection{Searching for patterns}

There are two different kind of frequency patterns that we want to distinguish when dealing with SPB stars: combination peaks that are often found in groups and modes of the same degree $\ell$ forming a period series (of consecutive radial orders $n$). The former are a nuisance for the particular purpose of this paper because they complicate the identification of independent modes. However, combination frequencies could be used in the future to study nonlinear coupling of the modes, despite current uncertainties for their astrophysical implications in upper main-sequence  stars; see e.g. \citet{2015MNRAS.450.3015K}. Frequencies of modes with the same degree and consecutive radial order, on the other hand, depend explicitly on the internal structure and rotation frequency of the star, hence carry information about the physical properties inside the star (as already discussed in Section\,\ref{intro}); therefore it is very important to separate peaks that belong to these two very different groups \citep[see e.g.][]{2012AN....333.1053P}. Combination peaks are harmonics or linear combinations of the individual frequencies, but being able to explain a peak as a linear combination of other peaks does not imply that this peak cannot be an actual pulsation mode.

This remark gets especially important when searches for combinations are extended to higher orders \citep[$\mathcal{O}>2$ in][]{2012AN....333.1053P} allowing positive and negative combinations simultaneously (e.g. $c_i f_i + c_j f_j - c_k f_k$, where $c\in\mathbb{N}^{+}$, referred to as a mixed-type combination hereafter). Indeed, even with the frequency precision of the four years of \textit{Kepler} data, true individual modes might easily get matched as combinations in a blind search. The main reason behind this is that the frequency spectrum of SPB stars is very dense. For example, the slow rotator KIC\,10526294 shows a series of 19 $\ell=1$ modes (56 counting the rotationally split components) within $0.6\,\mathrm{d}^{-1}$ \citep{2014A&A...570A...8P}, and the moderate rotator KIC\,7760680 shows a series of 36 $\ell=1$ modes within $0.5\,\mathrm{d}^{-1}$ \citep{2015ApJ...803L..25P}; but, most importantly, these observed values are all in  agreement with numbers from excitation calculations \citep{2016MNRAS.455L..67M}. To complicate matters further, rotation introduces rotationally split components \citep[and distorts period spacing patterns; see e.g.][]{2013MNRAS.429.2500B} contributing to an even denser pulsation spectrum. Combination frequencies can most often explain the frequency groups found at higher frequencies \citep{2015MNRAS.450.3015K}, but one should pay special attention during a combination search because, with an ill-considered parameter selection, it is possible to match real frequencies in the main group of gravity modes as a mixed-type combination of a few selected peaks within the same group just by chance. That is why we advise against the use of mixed-type combinations during the search. In our experience, real combination peaks are in most cases simple harmonics ($c_i f_i$), or very simple positive or negative combinations ($c_i f_i \pm c_j f_j$) of only two strong parent frequencies. It is extremely rare that there is no better physical explanation for a peak than a mixed-type combination.

After we filter out the true combination peaks, we attempt to identify series of modes of the same degree $\ell$ among the remaining frequencies. Since internal mixing and rotation influence the shape of the period spacing function \citep{2008MNRAS.386.1487M,2013MNRAS.429.2500B}, we cannot use the autocorrelation function or histograms with success because the consecutive (in radial order $n$) modes are not expected to be equally spaced in period anymore. Instead, we have to look for characteristic patterns (rotation-induced slopes, and/or periodic waves) that resemble the expected period spacing function manually.

\subsection{Results for individual targets}

\subsubsection{KIC\,3459297}

The final processed light curve of KIC\,3459297 spans 1470.5\,days, contains $52\,411$ data points, and has a duty cycle of $72.8\%$. The frequency resolution of the periodogram (also known as the Rayleigh limit) is $1/T=0.00068\,\mathrm{d}^{-1}$. The prewhitening procedure resulted in 477 statistically significant model frequencies contributing to a variance reduction of $99.75\%$, and bringing down the average signal levels from 45.9--27.1--5.0--3.9--3.8 ppm to 4.3--2.3--1.3--1.1--1.0 ppm, measured in $2\,\mathrm{d}^{-1}$ windows centred on 1, 2, 5, 10, and $20\,\mathrm{d}^{-1}$, respectively.

The dominant modes, as expected for an SPB star, are in the $g$-mode frequency regime between 0.7 and $1.2\,\mathrm{d}^{-1}$, with the dominant pulsation mode at $0.990040(7)\,\mathrm{d}^{-1}$, having an amplitude of $4667(88)\,\mathrm{ppm}$. The amplitude weighted average of the model frequencies is $1.34\,\mathrm{d}^{-1}$, where the shift to higher frequencies is because of the presence of a large number of combination frequencies.

\begin{figure}
\resizebox{\hsize}{!}{\includegraphics{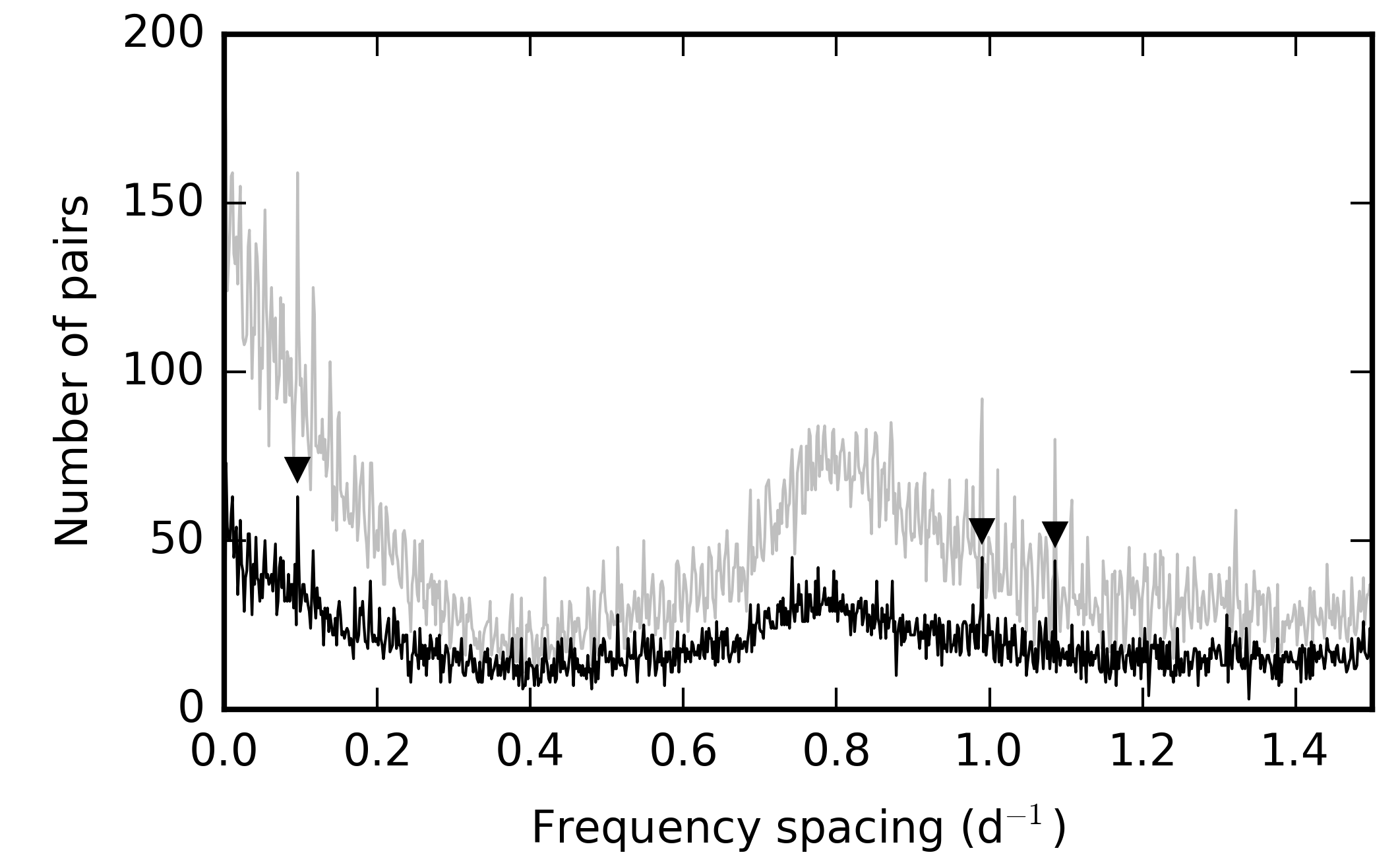}} 
\caption{Histogram of the distribution of frequency pairs in the frequency analysis of KIC\,3459297, calculated using a bin width of $2.5/T$ taking into account all statistically significant frequencies (grey solid line), or only a subset of frequencies where weaker signal from within $2.5/T$ of the stronger peaks was excluded (black solid line). The three values that are mentioned in the text are marked with black triangles.}
\label{frequencypairsKIC3459297}
\end{figure}

To confirm this, we follow the same combination search method as in \citet{2015ApJ...803L..25P}, restricting the search to combinations of maximum two parent frequencies. We find that all low frequency peaks are simple negative combinations of parent peaks in the main frequency regime, with $c\in [-1,1]$.  Almost all peaks towards higher frequencies, however, are simple positive combinations, with $c\in [0,8]$, except for a few peaks between $2.2$ and $2.8\,\mathrm{d}^{-1}$, some of which stay independent even when three parent frequencies and mixed-type combinations are allowed. Moreover, we find that looking at the histogram of frequency differences in Fig.\,\ref{frequencypairsKIC3459297} there are three bins that stand out: $0.990$, $1.086$, and $0.096\,\mathrm{d}^{-1}$. These match the frequency values of the two highest amplitude pulsation modes and their difference ($f_1$, $f_2$, and $f_2-f_1$), which implies that the high amplitude parent peaks contribute to a large number of combination frequencies. A similar behaviour has already been detected in KIC\,6352430\,A \citep{2013A&A...553A.127P}. Observing these differences in the histogram, for example $f_2-f_1$, does not require the presence of negative combination frequencies in the periodogram, as two combination frequencies that are formed as $f_1+f_i$ and $f_2+f_i$ also result in an observed separation of $f_2-f_1$.

While searching for characteristic patterns of a possible period series, we restrict ourselves to peaks that meet the \citet{1978Ap&SS..56..285L} criterion, which states that the minimal frequency separation that two close peaks must have to avoid influence on their apparent frequencies in the periodogram is $2.5\times$ the Rayleigh limit. In practice, given $f_i$ and $f_{i+1}$ frequencies of consecutive significant peaks, where $f_{i+1}-f_{i}<2.5/T$, we keep the frequency with the larger amplitude and discard the other frequency. We repeat this pairwise test until there are no more consecutive peaks in the periodogram that have a frequency difference below $2.5\times$ the Rayleigh limit. This lowers the total number of frequencies to 337. This number is significantly reduced further when we remove the simple combination frequencies. Then we take the $\sim5$ highest amplitude peaks and check whether they outline a recognisable period spacing pattern. Afterwards we try to add additional strong frequencies one-by-one, every time checking whether they fit in the period spacing pattern.

\begin{figure*}
\resizebox{\hsize}{!}{\includegraphics{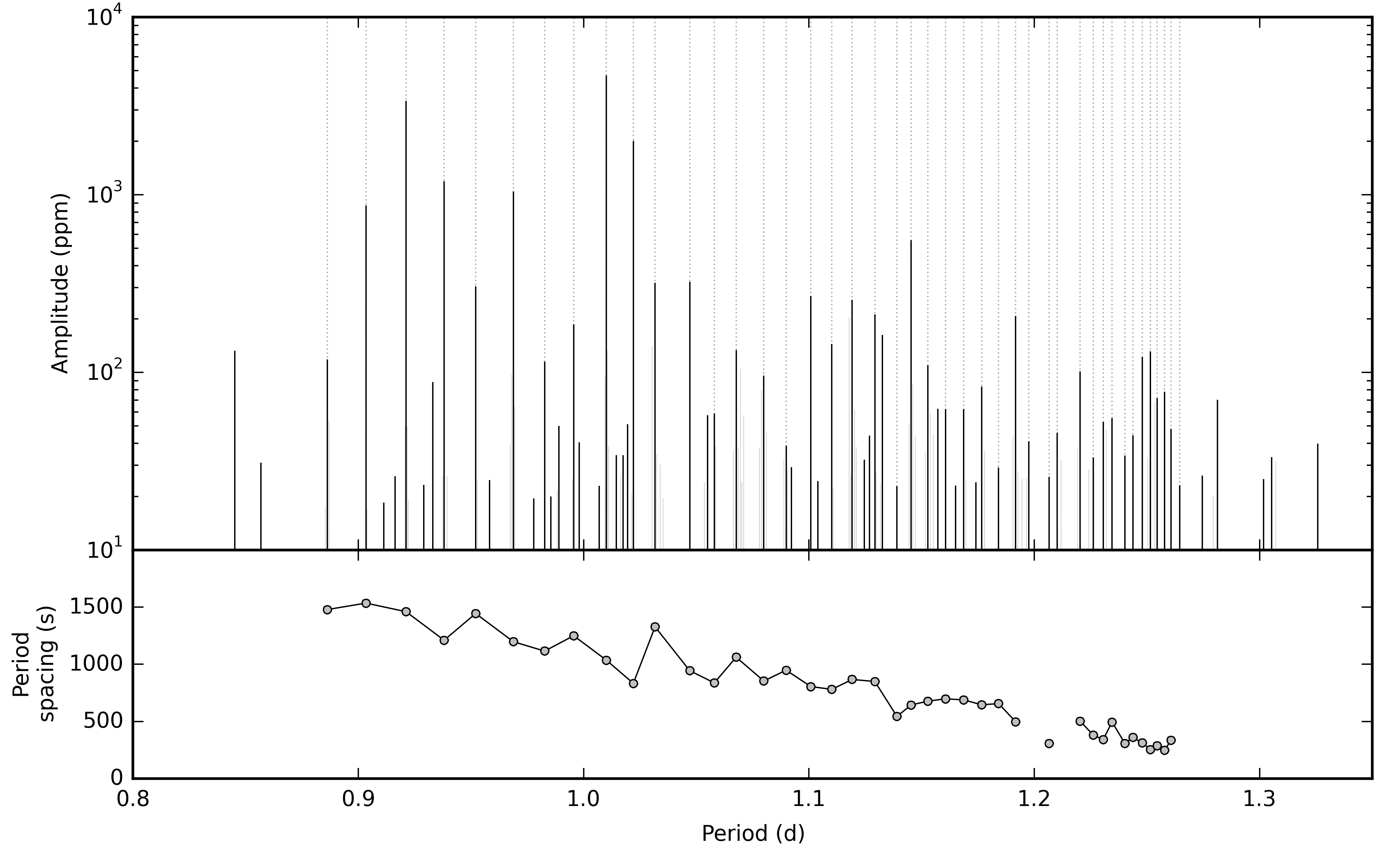}} 
\caption{Zoom-in of the period spectrum of KIC\,3459297 (top panel) with the members of the period spacing series indicated with vertical dashed lines. Peaks that are discarded during our pattern search are plotted using thin grey lines (for details, we refer to the text). The observed period spacing is shown in the bottom panel.}
\label{periodseriesKIC3459297}
\end{figure*}

\begin{figure}
\resizebox{\hsize}{!}{\includegraphics{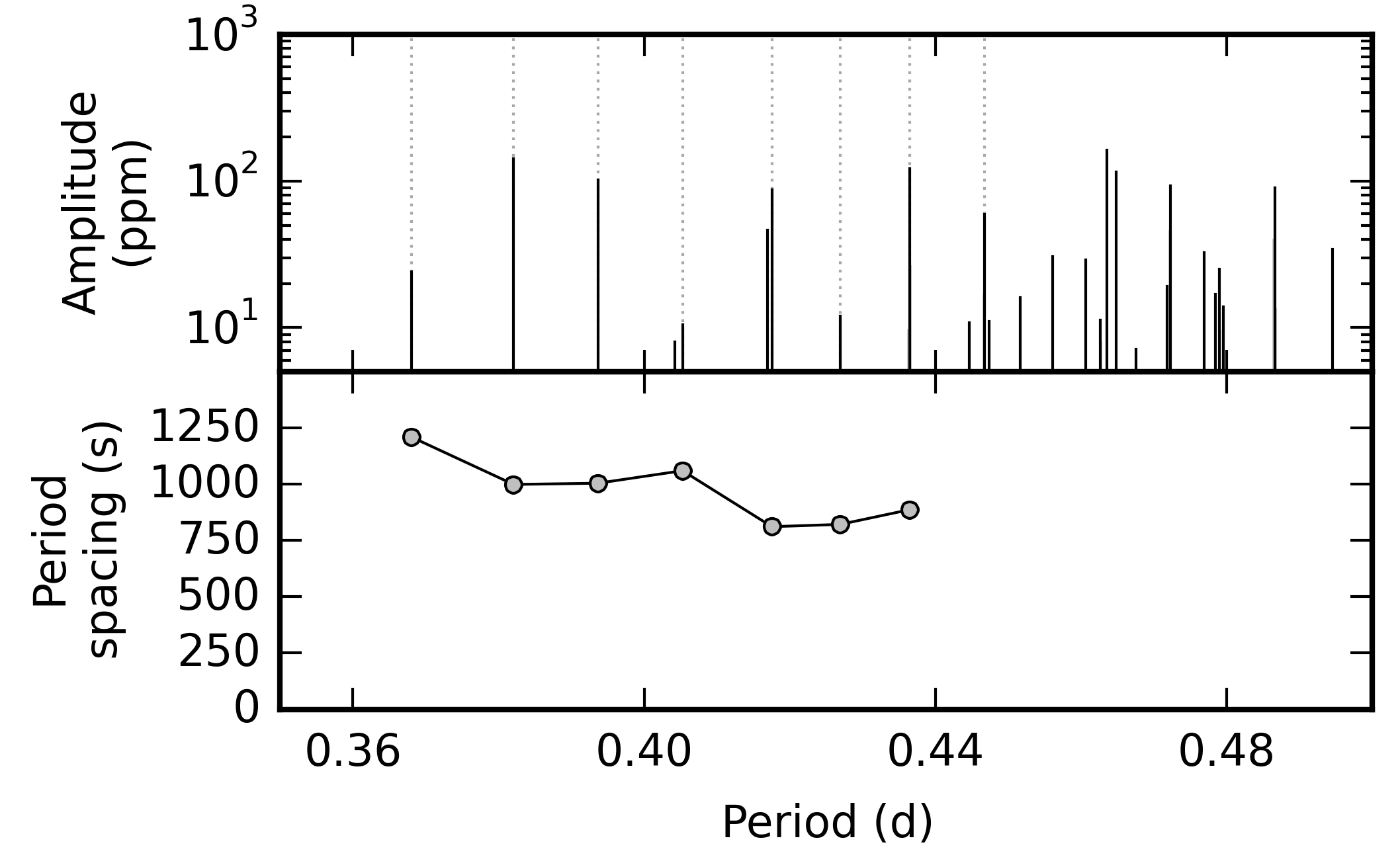}} 
\caption{Same as Fig.\,\ref{periodseriesKIC3459297}, but for the possible $l=2$ series in KIC\,3459297.}
\label{periodseriesKIC3459297alt}
\end{figure}

We find a long period series (consisting of 43 observed and 2 missing radial orders, which are listed in Table\,\ref{periodseriestableKIC3459297}) tilted by rotation, which exhibits short period variations on top of the general downwards trend (see Fig.\,\ref{periodseriesKIC3459297}). This points towards a star that is already relatively evolved on the main sequence (in good agreement with the observed $\log g$), and requires the absence of a large amount of diffusive mixing that would otherwise smear out the observed sawtooth-like features. Assuming that we observe a prograde $l=1$ series (based on the observed spacing, the tilt, and the spectroscopic fundamental parameters), it is possible that the peaks between $2.2$ and $2.8\,\mathrm{d}^{-1}$ (listed in Table\,\ref{periodseriestableKIC3459297alt}) that are not identified as simple combination frequencies might be members of a prograde ($m=1$ or $m=2$) $l=2$ series (see Fig.\,\ref{periodseriesKIC3459297alt}), but this could only be confirmed after carrying out modelling of the dipole series.

\begin{table}
\caption{Fourier parameters (periods, frequencies, amplitudes, and phases) of the modes of the main period series of KIC\,3459297. The displayed S/N values are calculated in a window of $1\,\mathrm{d}^{-1}$ centred on the given frequency. Numbers in parentheses are the formal errors of the last significant digit.}
\label{periodseriestableKIC3459297}
\centering
\scalebox{0.9}{
\begin{tabular}{l c c c c r}
\hline\hline
 \# & $p$ & $f$ & $A$ & $\theta$ & S/N\\
    & $\mathrm{d}$ & $\mathrm{d}^{-1}$ & $\mathrm{ppm}$ & $2\pi/\mathrm{rad}$ &  \\
\hline
  1 &      0.88627(2) &      1.12832(3) &          117(8) &   -0.40(7) & 11.3 \\
  2 &     0.903392(7) &     1.106940(9) &      8.7(2)$\times 10^{2}$ &    0.27(2) & 37.5 \\
  3 &     0.921169(5) &     1.085577(6) &     3.34(6)$\times 10^{3}$ &    0.03(2) & 62.2 \\
  4 &     0.938074(9) &      1.06601(1) &     1.18(3)$\times 10^{3}$ &    0.42(3) & 38.5 \\
  5 &      0.95209(2) &      1.05032(2) &      3.0(2)$\times 10^{2}$ &   -0.21(5) & 18.0 \\
  6 &     0.968813(9) &     1.032191(9) &     1.03(3)$\times 10^{3}$ &   -0.49(2) & 38.3 \\
  7 &      0.98268(3) &      1.01763(3) &          114(8) &   -0.18(7) & 10.9 \\
  8 &      0.99559(2) &      1.00443(2) &      1.9(1)$\times 10^{2}$ &    0.38(6) & 14.3 \\
  9 &     1.010060(7) &     0.990040(7) &     4.67(9)$\times 10^{3}$ &   -0.05(2) & 62.8 \\
 10 &     1.022058(8) &     0.978418(7) &     2.00(4)$\times 10^{3}$ &   -0.13(2) & 51.2 \\
 11 &      1.03170(2) &      0.96927(2) &      3.2(2)$\times 10^{2}$ &   -0.29(5) & 19.6 \\
 12 &      1.04708(2) &      0.95504(2) &      3.2(1)$\times 10^{2}$ &   -0.27(4) & 19.0 \\
 13 &      1.05800(4) &      0.94518(3) &           58(5) &    0.44(9) & 6.7 \\
 14 &      1.06769(3) &      0.93660(3) &      1.3(1)$\times 10^{2}$ &   -0.02(7) & 11.9 \\
 15 &      1.07999(3) &      0.92594(3) &           95(7) &    0.23(7) & 8.5 \\
 16 &      1.08987(6) &      0.91754(5) &           38(5) &    -0.4(1) & 6.3 \\
 17 &      1.10085(2) &      0.90839(2) &      2.7(1)$\times 10^{2}$ &    0.48(5) & 17.6 \\
 18 &      1.11016(3) &      0.90077(3) &      1.4(1)$\times 10^{2}$ &   -0.06(7) & 11.7 \\
 19 &      1.11920(2) &      0.89350(2) &      2.5(1)$\times 10^{2}$ &    0.02(5) & 17.4 \\
 20 &      1.12924(3) &      0.88555(2) &      2.1(1)$\times 10^{2}$ &   -0.28(6) & 17.9 \\
 21 &      1.13907(5) &      0.87791(4) &           23(3) &     0.0(1) & 4.3 \\
 22 &      1.14536(2) &      0.87309(1) &      5.5(2)$\times 10^{2}$ &    0.14(3) & 29.3 \\
 23 &      1.15281(4) &      0.86744(3) &          109(9) &    0.40(8) & 11.3 \\
 24 &      1.16066(5) &      0.86158(3) &           62(6) &   -0.44(9) & 6.8 \\
 25 &      1.16873(5) &      0.85563(3) &           62(6) &    0.30(9) & 6.7 \\
 26 &      1.17670(4) &      0.84984(3) &           83(7) &   -0.20(8) & 8.6 \\
 27 &      1.18417(6) &      0.84448(4) &           29(3) &    -0.4(1) & 4.5 \\
 28 &      1.19176(3) &      0.83910(2) &      2.1(1)$\times 10^{2}$ &   -0.45(5) & 14.2 \\
 29 &      1.19752(6) &      0.83506(4) &           41(5) &     0.1(1) & 6.1 \\
 30 &      1.20654(6) &      0.82881(4) &           26(3) &    -0.1(1) & 4.3 \\
 31 &      1.21010(6) &      0.82638(4) &           45(5) &    -0.3(1) & 6.4 \\
 32 &      1.22038(5) &      0.81942(3) &          100(9) &    0.30(9) & 11.2 \\
 33 &      1.22620(6) &      0.81553(4) &           33(3) &     0.0(1) & 4.4 \\
 34 &      1.23059(4) &      0.81262(3) &           52(4) &   -0.23(7) & 4.9 \\
 35 &      1.23453(5) &      0.81002(3) &           55(5) &    0.09(9) & 6.1 \\
 36 &      1.24026(5) &      0.80629(4) &           34(3) &   -0.41(9) & 4.4 \\
 37 &      1.24381(4) &      0.80398(3) &           44(3) &   -0.29(7) & 4.4 \\
 38 &      1.24797(4) &      0.80130(2) &          121(8) &    0.01(7) & 11.1 \\
 39 &      1.25158(4) &      0.79899(3) &          130(9) &    0.21(7) & 11.2 \\
 40 &      1.25451(5) &      0.79713(3) &           71(6) &   -0.34(9) & 7.4 \\
 41 &      1.25783(4) &      0.79502(3) &           77(6) &    0.09(7) & 6.9 \\
 42 &      1.26069(5) &      0.79322(3) &           48(4) &    0.12(9) & 5.5 \\
 43 &      1.26456(7) &      0.79079(4) &           23(3) &     0.3(1) & 4.2 \\
\hline
\end{tabular}}
\end{table}

\begin{table}
\caption{Fourier parameters (periods, frequencies, amplitudes, and phases) of the modes of the possible $l=2$ series in KIC\,3459297. The  S/N values shown are calculated in a window of $1\,\mathrm{d}^{-1}$ centred on the given frequency. Numbers in parentheses are the formal errors of the last significant digit.}
\label{periodseriestableKIC3459297alt}
\centering
\scalebox{0.9}{
\begin{tabular}{l c c c c r}
\hline\hline
 \# & $p$ & $f$ & $A$ & $\theta$ & S/N\\
    & $\mathrm{d}$ & $\mathrm{d}^{-1}$ & $\mathrm{ppm}$ & $2\pi/\mathrm{rad}$ &  \\
\hline
  1 &      0.36807(2) &       2.7169(2) &        2(1)$\times 10^{1}$ &    -0.4(4) & 11.7 \\
  2 &     0.382067(7) &      2.61734(4) &      1.4(2)$\times 10^{2}$ &     0.1(1) & 19.6 \\
  3 &      0.39363(1) &      2.54043(6) &      1.0(2)$\times 10^{2}$ &     0.1(2) & 19.6 \\
  4 &      0.40526(3) &       2.4675(2) &           10(5) &     0.2(4) & 5.8 \\
  5 &      0.41754(1) &      2.39499(7) &        9(2)$\times 10^{1}$ &     0.1(2) & 19.6 \\
  6 &      0.42693(3) &       2.3423(2) &           12(5) &     0.2(4) & 6.1 \\
  7 &      0.43644(1) &      2.29124(5) &      1.2(2)$\times 10^{2}$ &    -0.0(1) & 19.6 \\
  8 &      0.44671(2) &      2.23858(9) &        6(1)$\times 10^{1}$ &    -0.1(2) & 17.6 \\
\hline
\end{tabular}}
\end{table}

\subsubsection{KIC\,4930889}

The reduced light curve of KIC\,4930889 spans 1470.5\,days and consists of $52\,878$ data points. The data has a duty cycle of $73.5\%$. The Rayleigh limit of the periodogram is $1/T=0.00068\,\mathrm{d}^{-1}$. The prewhitening procedure delivered 380 statistically significant model frequencies, resulting in a variance reduction of $99.5\%$ and bringing down the average signal levels from 81.2--66.4--7.5--5.9--5.6 ppm to 12.2--7.1--1.3--0.8--0.5 ppm, measured in $2\,\mathrm{d}^{-1}$ windows centred on 1, 2, 5, 10, and $20\,\mathrm{d}^{-1}$, respectively.

\begin{figure}
\resizebox{\hsize}{!}{\includegraphics{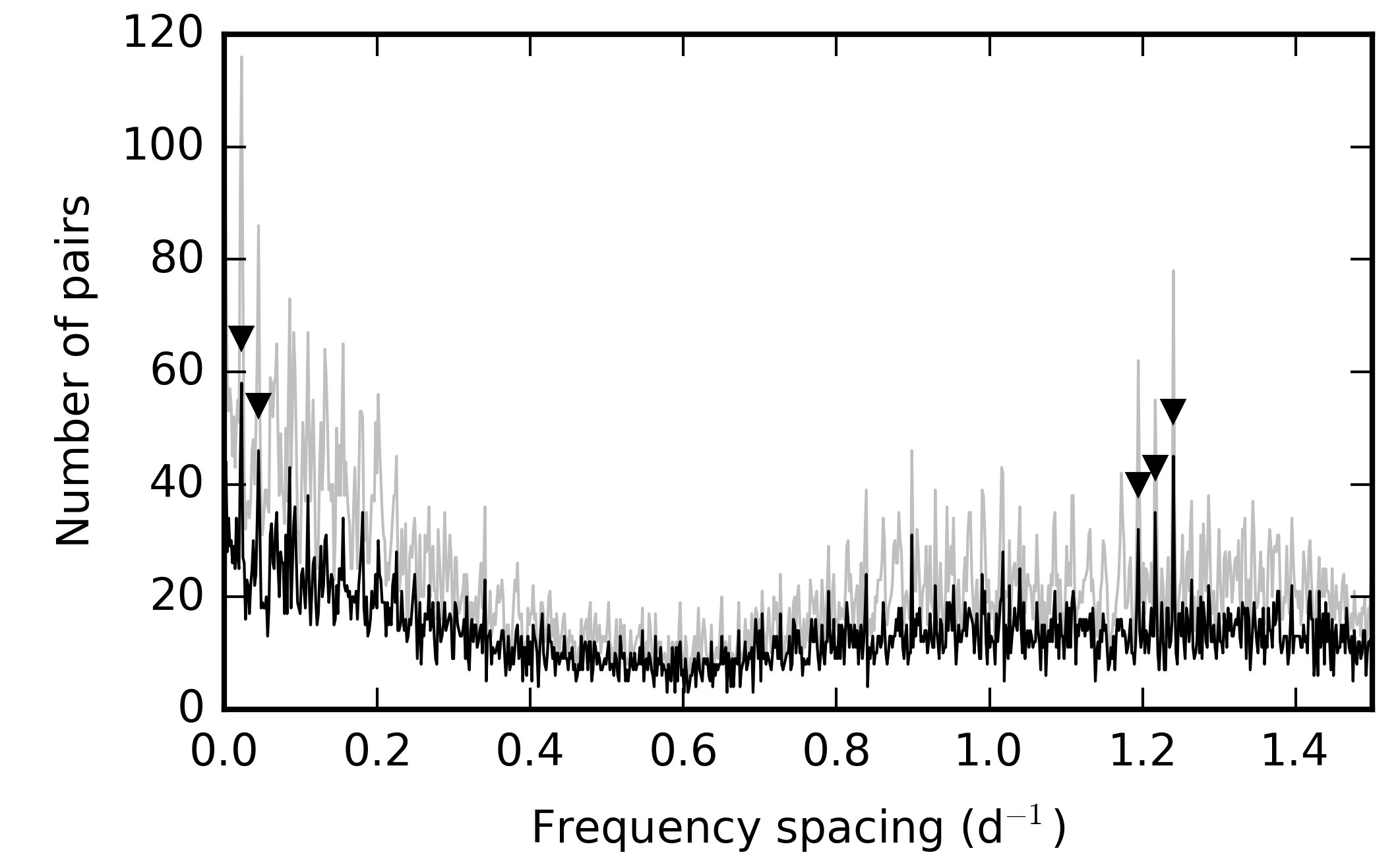}} 
\caption{Same as Fig.\,\ref{frequencypairsKIC3459297}, but for KIC\,4930889.}
\label{frequencypairsKIC4930889}
\end{figure}

The dominant modes are $g$ modes between 0.75 and $1.4\,\mathrm{d}^{-1}$. The strongest pulsation mode is at $1.240146(6)\,\mathrm{d}^{-1}$ with an amplitude of $8093(136)\,\mathrm{ppm}$. The amplitude weighted average of the model frequencies is $1.28\,\mathrm{d}^{-1}$. Almost all power above this frequency range comes from combination frequencies (with a possible exception of the relatively strong peak at $1.71770(4)\,\mathrm{d}^{-1}$), while a number of low frequency peaks cannot be accounted for as simple negative combinations. Again, as illustrated by the five marked bins (at $0.023$, $0.045$, $1.194$, $1.216$, and $1.240\,\mathrm{d}^{-1}$) in Fig.\,\ref{frequencypairsKIC4930889}, the three dominant modes contribute to a large number of combination frequencies (with the bins corresponding to $f_1-f_2$,$f_1-f_3$,$f_3$,$f_2$, and $f_1$, respectively).

\begin{figure*}
\resizebox{\hsize}{!}{\includegraphics{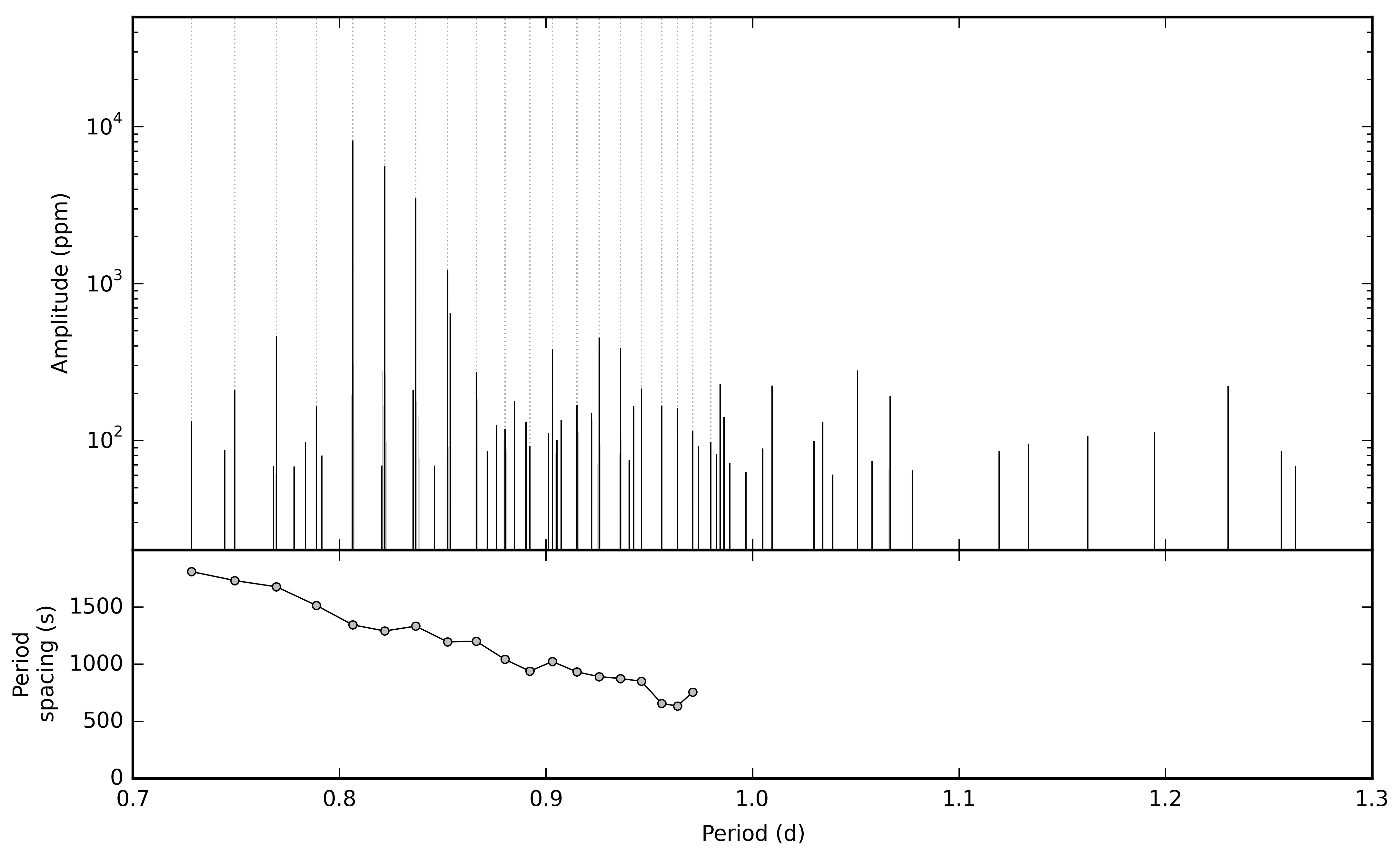}} 
\caption{Same as Fig.\,\ref{periodseriesKIC3459297}, but for KIC\,4930889.}
\label{periodseriesKIC4930889}
\end{figure*}

Our manual search for possible period series (after filtering for close frequencies and leaving only 297 model peaks) gives a positive result again. We identify a tilted period series consisting of 20 consecutive radial orders (see Table\,\ref{periodseriestableKIC4930889} and Fig.\,\ref{periodseriesKIC4930889}). This period series is much smoother than in the case of KIC\,3459297, showing only three shallow dips (around 0.82, 0.89, and 0.96 day). These properties point towards an earlier evolutionary stage and possibly a higher diffusive mixing.

\begin{table}
\caption{Fourier parameters (periods, frequencies, amplitudes, and phases) of the modes of the main period series of KIC\,4930889. The S/N values shown are calculated in a window of $1\,\mathrm{d}^{-1}$ centred on the given frequency. Numbers in parentheses are the formal errors of the last significant digit.}
\label{periodseriestableKIC4930889}
\centering
\scalebox{0.9}{
\begin{tabular}{l c c c c r}
\hline\hline
 \# & $p$ & $f$ & $A$ & $\theta$ & S/N\\
    & $\mathrm{d}$ & $\mathrm{d}^{-1}$ & $\mathrm{ppm}$ & $2\pi/\mathrm{rad}$ &  \\
\hline
  1 &      0.72836(3) &      1.37295(6) &      1.3(2)$\times 10^{2}$ &    -0.1(2) & 7.3 \\
  2 &      0.74932(3) &      1.33454(5) &      2.1(3)$\times 10^{2}$ &     0.1(1) & 8.4 \\
  3 &      0.76938(2) &      1.29974(3) &      4.6(4)$\times 10^{2}$ &    0.11(9) & 17.0 \\
  4 &      0.78882(3) &      1.26772(5) &      1.6(2)$\times 10^{2}$ &    -0.2(1) & 8.3 \\
  5 &     0.806357(4) &     1.240146(6) &      8.1(1)$\times 10^{3}$ &   -0.03(2) & 64.9 \\
  6 &     0.821928(4) &     1.216652(6) &     5.58(9)$\times 10^{3}$ &    0.04(2) & 68.6 \\
  7 &     0.836879(4) &     1.194916(6) &     3.47(6)$\times 10^{3}$ &   -0.48(2) & 65.7 \\
  8 &      0.85232(1) &      1.17327(2) &     1.22(5)$\times 10^{3}$ &    0.31(4) & 32.1 \\
  9 &      0.86617(3) &      1.15451(4) &      2.7(3)$\times 10^{2}$ &     0.3(1) & 11.4 \\
 10 &      0.88008(4) &      1.13626(5) &      1.2(2)$\times 10^{2}$ &     0.2(1) & 5.8 \\
 11 &      0.89215(5) &      1.12089(6) &        9(2)$\times 10^{1}$ &     0.1(2) & 5.1 \\
 12 &      0.90302(3) &      1.10739(3) &      3.8(4)$\times 10^{2}$ &    0.35(9) & 13.3 \\
 13 &      0.91488(4) &      1.09304(4) &      1.7(2)$\times 10^{2}$ &    -0.3(1) & 7.4 \\
 14 &      0.92569(3) &      1.08028(3) &      4.5(4)$\times 10^{2}$ &    0.10(9) & 17.0 \\
 15 &      0.93601(3) &      1.06837(3) &      3.9(3)$\times 10^{2}$ &    0.39(9) & 12.2 \\
 16 &      0.94614(4) &      1.05692(5) &      2.1(3)$\times 10^{2}$ &    -0.2(1) & 8.2 \\
 17 &      0.95600(5) &      1.04602(5) &      1.7(2)$\times 10^{2}$ &     0.2(1) & 8.3 \\
 18 &      0.96360(4) &      1.03778(5) &      1.6(2)$\times 10^{2}$ &    -0.3(1) & 7.5 \\
 19 &      0.97095(6) &      1.02992(6) &      1.1(2)$\times 10^{2}$ &    -0.4(2) & 6.3 \\
 20 &      0.97970(6) &      1.02072(7) &      1.0(2)$\times 10^{2}$ &     0.0(2) & 5.6 \\
\hline
\end{tabular}}
\end{table}

\begin{figure}
\resizebox{\hsize}{!}{\includegraphics{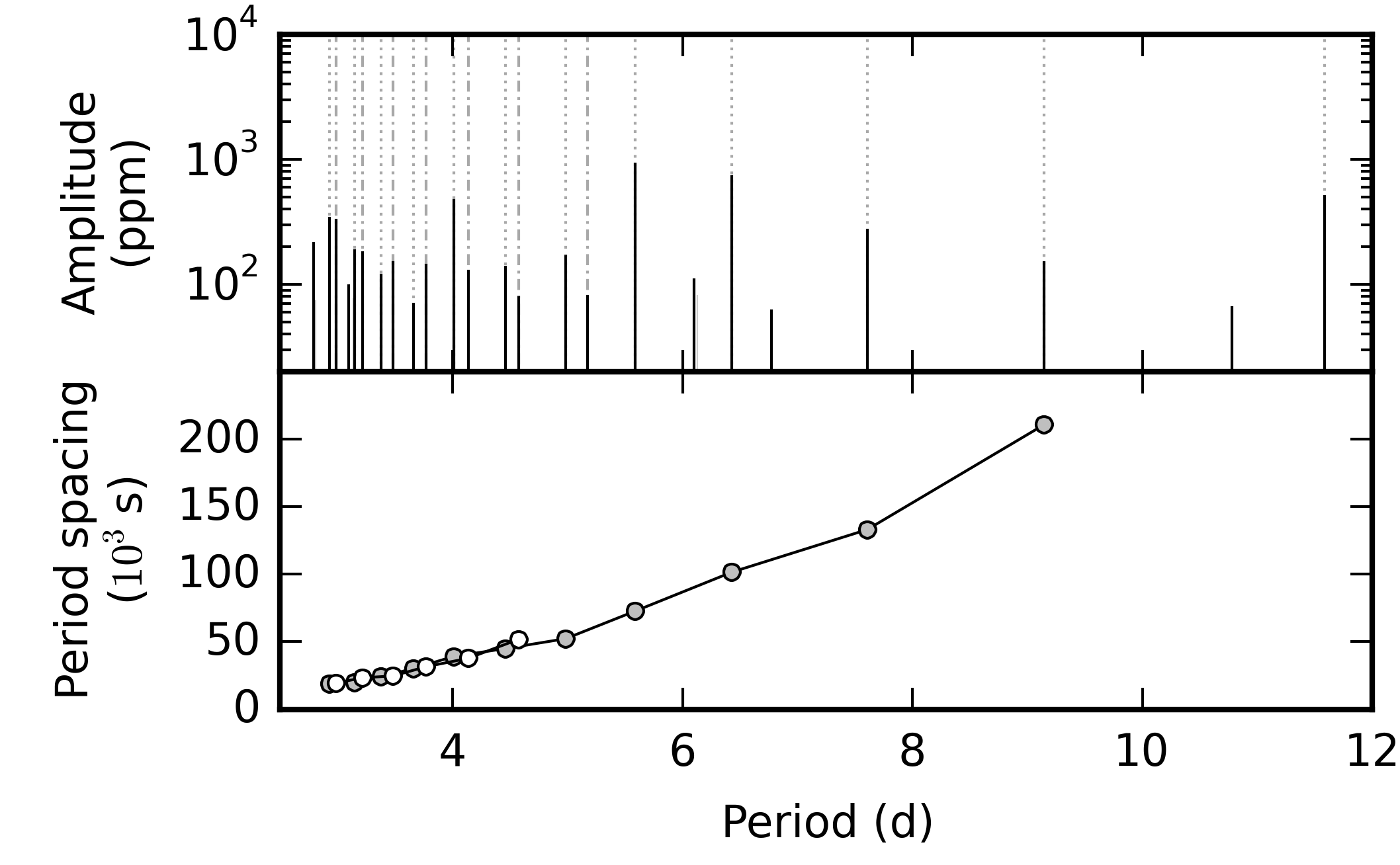}} 
\caption{Same as Fig.\,\ref{periodseriesKIC3459297}, but for the proposed additional series of the low frequency peaks in KIC\,4930889.}
\label{periodseriesKIC4930889alt}
\end{figure}

Since we find an unusually high amount of modes at lower frequencies that cannot be explained as simple combinations, we also searched for period series in this region. This results in two very similar period series, both showing an upwards tilt typical for retrograde modes (see Table\,\ref{periodseriestableKIC4930889alt} and Fig.\,\ref{periodseriesKIC4930889alt}). During the construction of these series we disregarded the results from the combination search, as some peaks might very easily be identified as combination frequencies simply by chance since the region of possible parent frequencies is very dense, and because the fact that none of the higher amplitude peaks in the low frequency region were identified as combinations to start with. 

\begin{table}
\caption{Fourier parameters (periods, frequencies, amplitudes, and phases) of the two extra possible period series of KIC\,4930889. The S/N values shown are calculated in a window of $1\,\mathrm{d}^{-1}$ centred on the given frequency. Numbers in parentheses are the formal errors of the last significant digit.}
\label{periodseriestableKIC4930889alt}
\centering
\scalebox{0.9}{
\begin{tabular}{l c c c c r}
\hline\hline
 \# & $p$ & $f$ & $A$ & $\theta$ & S/N\\
    & $\mathrm{d}$ & $\mathrm{d}^{-1}$ & $\mathrm{ppm}$ & $2\pi/\mathrm{rad}$ &  \\
\hline
  1 &       2.9324(3) &      0.34102(4) &      3.4(4)$\times 10^{2}$ &     0.5(1) & 13.2 \\
  2 &       3.1502(5) &      0.31744(5) &      1.9(3)$\times 10^{2}$ &    -0.4(1) & 8.3 \\
  3 &       3.3806(7) &      0.29580(6) &      1.2(2)$\times 10^{2}$ &     0.0(2) & 7.0 \\
  4 &        3.660(1) &      0.27325(7) &        7(1)$\times 10^{1}$ &     0.2(2) & 4.4 \\
  5 &       4.0090(5) &      0.24944(3) &      4.7(4)$\times 10^{2}$ &    0.09(8) & 17.0 \\
  6 &        4.462(1) &      0.22413(5) &      1.4(2)$\times 10^{2}$ &    -0.5(1) & 7.1 \\
  7 &        4.983(1) &      0.20069(5) &      1.7(2)$\times 10^{2}$ &    -0.4(1) & 8.3 \\
  8 &       5.5877(6) &      0.17897(2) &      9.2(5)$\times 10^{2}$ &   -0.45(5) & 27.5 \\
  9 &       6.4294(9) &      0.15554(2) &      7.3(4)$\times 10^{2}$ &    0.02(6) & 22.5 \\
 10 &        7.605(2) &      0.13149(4) &      2.7(3)$\times 10^{2}$ &     0.2(1) & 11.2 \\
 11 &        9.146(4) &      0.10934(5) &      1.5(2)$\times 10^{2}$ &     0.4(1) & 7.4 \\
 12 &       11.587(4) &      0.08631(3) &      5.1(4)$\times 10^{2}$ &   -0.45(7) & 17.9 \\
\hline
  1 &       2.9896(3) &      0.33450(4) &      3.3(3)$\times 10^{2}$ &    -0.4(1) & 12.2 \\
  2 &       3.2150(5) &      0.31104(5) &      1.8(3)$\times 10^{2}$ &     0.1(1) & 8.3 \\
  3 &       3.4847(6) &      0.28697(5) &      1.5(2)$\times 10^{2}$ &     0.3(1) & 7.5 \\
  4 &       3.7710(8) &      0.26518(6) &      1.4(2)$\times 10^{2}$ &     0.2(1) & 7.5 \\
  5 &        4.136(1) &      0.24175(6) &      1.3(2)$\times 10^{2}$ &    -0.3(2) & 7.0 \\
  6 &        4.573(1) &      0.21866(7) &        8(1)$\times 10^{1}$ &     0.5(2) & 4.8 \\
  7 &        5.173(2) &      0.19330(7) &        8(1)$\times 10^{1}$ &     0.1(2) & 4.9 \\
\hline
\end{tabular}}
\end{table}

\subsubsection{KIC\,6352430\,A}

The full reduced \textit{Kepler} photometry of KIC\,6352430\,A covers 1459.5\,days \citep[compared to the 1141.5 days that were available in][]{2013A&A...553A.127P} and contains $65\,478$ data points (with a duty cycle of 91.7\%). The Rayleigh limit is  $1/T=0.000685\,\mathrm{d}^{-1}$. The iterative prewhitening was stopped after the removal of 858 significant frequencies, which resulted in a variance reduction of 99.9\%. The average signal levels in the original and residual light curve are 55.8--52.6--6.2--5.1--4.8 ppm and 5.4--3.6--0.9--0.6--0.3 ppm, measured in $2\,\mathrm{d}^{-1}$ windows centred on 1, 2, 5, 10, and $20\,\mathrm{d}^{-1}$, respectively.

\begin{figure}
\resizebox{\hsize}{!}{\includegraphics{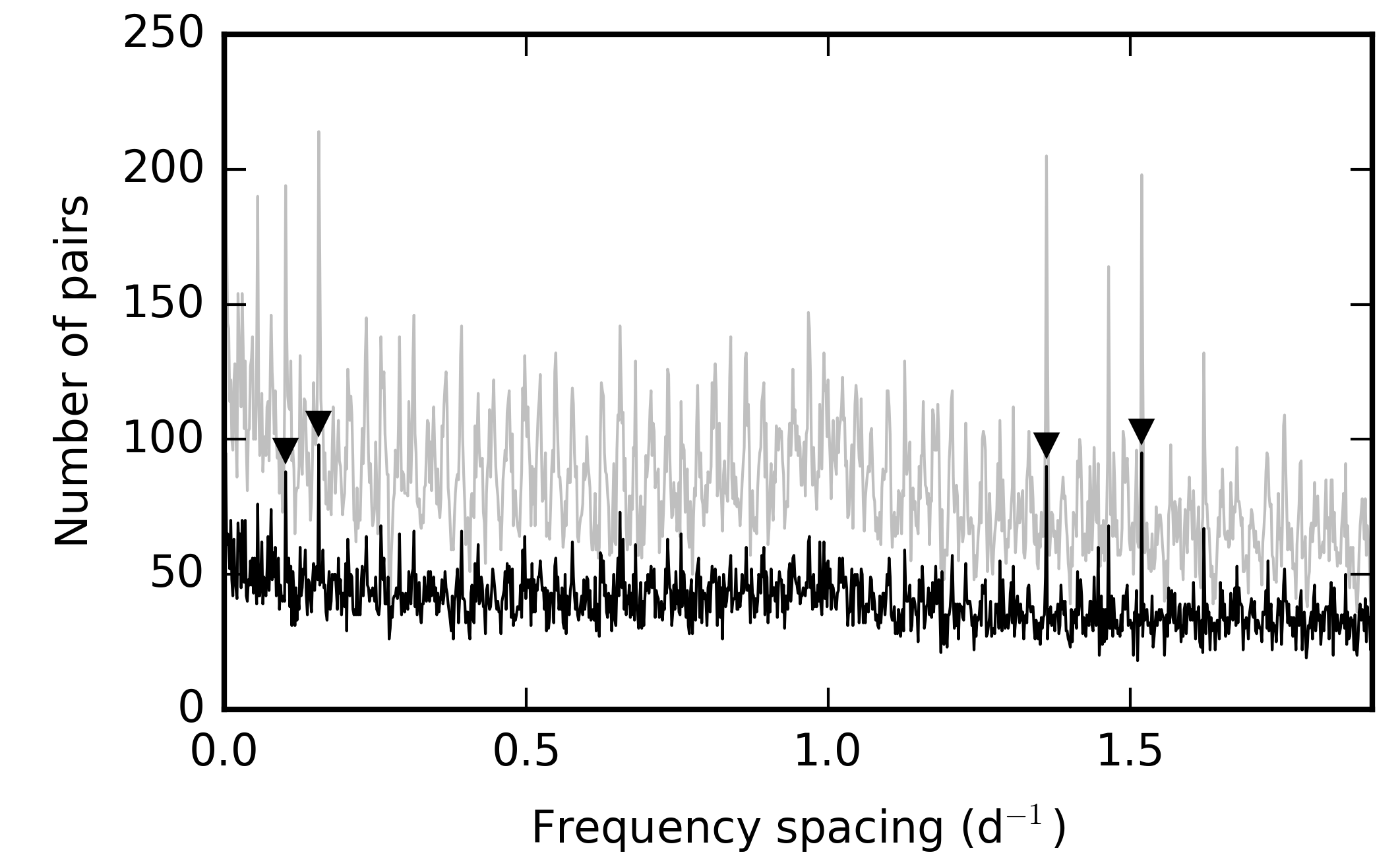}} 
\caption{Same as Fig.\,\ref{frequencypairsKIC3459297}, but for KIC\,6352430\,A.}
\label{frequencypairsKIC6352430}
\end{figure}

The dominant modes are in the $g$ mode regime between 0.9 and $1.8\,\mathrm{d}^{-1}$ with the strongest mode at $1.361716(6)\,\mathrm{d}^{-1}$, having an amplitude of $7406(111)\,\mathrm{ppm}$. The amplitude weighted average of the model frequencies is $1.84\,\mathrm{d}^{-1}$, where the shift towards higher frequencies is again a clear sign of a large number of combination frequencies. Unlike the previous stars, the transition between the dominant region and the region of combination frequencies is continuous; there are no apparent gaps in the power spectrum where no signal could be found. All peaks below $0.7\,\mathrm{d}^{-1}$ can be explained as simple negative combinations, except for the frequency at $0.03768(3)\,\mathrm{d}^{-1}$, which can be identified as the spectroscopically derived binary orbital frequency of $f_\mathrm{SB2}=0.03766(3)\,\mathrm{d}^{-1}$ from \citet{2013A&A...553A.127P}. We note that the fact that one can identify a frequency as a simple combination does not directly mean that there is no better explanation for the peak in question. In fact, we also observe four harmonics of the orbital frequency up to $5\times f_\mathrm{SB2}$, but some of these get matched as simple negative combinations thanks to the high power density in the $g$-mode region; however a much more logical explanation, especially given their sequentially smaller amplitudes towards higher harmonics, is that these are orbital harmonics. Above $1.9\,\mathrm{d}^{-1}$ a quickly growing percentage of peaks are identified as combination frequencies, and beyond $2.5\,\mathrm{d}^{-1}$ basically all power can be traced back to parent frequencies in the dominant $g$-mode region (with a few relatively large amplitude exceptions below $4\,\mathrm{d}^{-1}$). Again, large amplitude modes have a major contribution to combination frequencies, as the strongest bins in Fig.\,\ref{frequencypairsKIC6352430} fall at frequencies of the strongest modes and their differences (with the marked bins at 0.102, 0.157, 1.361, and $1.519\,\mathrm{d}^{-1}$ corresponding to $f_3-f_1$, $f_2-f_1$, $f_1$, and $f_2$, respectively).

\begin{figure*}
\resizebox{\hsize}{!}{\includegraphics{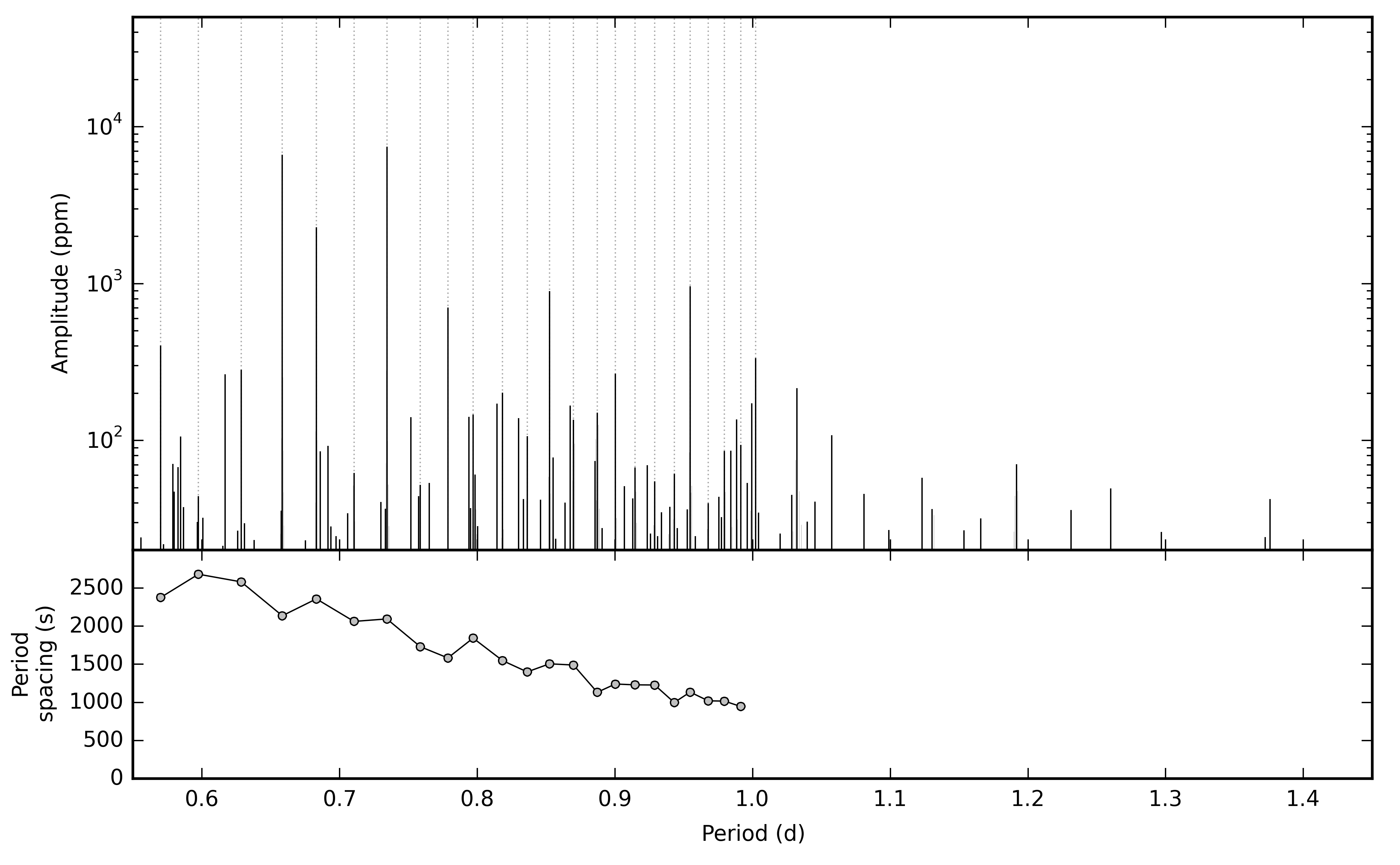}} 
\caption{Same as Fig.\,\ref{periodseriesKIC3459297}, but for KIC\,6352430\,A.}
\label{periodseriesKIC6352430}
\end{figure*}

The purpose of revisiting KIC\,6352430\,A is to look for tilted period series, and (after filtering for close peaks that left us with 584 model peaks) screening through the independent frequencies in the region of interest, we manage to detect one (see Table\,\ref{periodseriestableKIC6352430} and Fig.\,\ref{periodseriesKIC6352430}) that is very similar in topography to that of KIC\,3459297.

\begin{table}
\caption{Fourier parameters (periods, frequencies, amplitudes, and phases) of the modes of the period series detected in KIC\,6352430\,A. The S/N values shown are calculated in a window of $1\,\mathrm{d}^{-1}$ centred on the given frequency. Numbers in parentheses are the formal errors of the last significant digit.}
\label{periodseriestableKIC6352430}
\centering
\scalebox{0.9}{
\begin{tabular}{l c c c c r}
\hline\hline
 \# & $p$ & $f$ & $A$ & $\theta$ & S/N\\
    & $\mathrm{d}$ & $\mathrm{d}^{-1}$ & $\mathrm{ppm}$ & $2\pi/\mathrm{rad}$ &  \\
\hline
  1 &     0.569977(7) &      1.75446(2) &      4.0(2)$\times 10^{2}$ &    0.26(6) & 21.4 \\
  2 &      0.59749(3) &      1.67367(7) &           44(8) &    -0.4(2) & 6.5 \\
  3 &      0.62855(1) &      1.59097(3) &      2.8(2)$\times 10^{2}$ &    0.20(7) & 18.1 \\
  4 &     0.658438(1) &     1.518747(3) &     6.57(5)$\times 10^{3}$ &   0.037(7) & 108.4 \\
  5 &     0.683179(3) &     1.463745(6) &     2.26(3)$\times 10^{3}$ &   -0.19(2) & 69.3 \\
  6 &      0.71049(3) &      1.40749(6) &           61(9) &     0.2(1) & 7.8 \\
  7 &     0.734367(3) &     1.361716(6) &      7.4(1)$\times 10^{3}$ &   -0.09(1) & 83.9 \\
  8 &      0.75863(3) &      1.31817(6) &           52(8) &     0.3(2) & 6.4 \\
  9 &     0.778667(7) &      1.28425(1) &      7.0(2)$\times 10^{2}$ &    0.23(3) & 34.8 \\
 10 &      0.79700(2) &      1.25471(3) &      1.4(1)$\times 10^{2}$ &    0.32(8) & 11.3 \\
 11 &      0.81834(2) &      1.22199(3) &      2.0(1)$\times 10^{2}$ &    0.10(7) & 14.0 \\
 12 &      0.83627(3) &      1.19579(4) &      1.1(1)$\times 10^{2}$ &     0.2(1) & 9.6 \\
 13 &     0.852473(8) &      1.17306(1) &      8.9(3)$\times 10^{2}$ &    0.37(3) & 41.3 \\
 14 &      0.86991(2) &      1.14955(3) &      1.3(1)$\times 10^{2}$ &    0.26(8) & 11.6 \\
 15 &      0.88716(3) &      1.12720(3) &      1.5(1)$\times 10^{2}$ &   -0.39(8) & 12.2 \\
 16 &      0.90026(2) &      1.11079(2) &      2.6(2)$\times 10^{2}$ &   -0.29(6) & 17.8 \\
 17 &      0.91462(4) &      1.09334(5) &           67(9) &     0.1(1) & 7.2 \\
 18 &      0.92886(5) &      1.07659(6) &           54(8) &     0.4(1) & 6.1 \\
 19 &      0.94308(5) &      1.06036(5) &           61(9) &    -0.1(1) & 7.2 \\
 20 &      0.95463(1) &      1.04753(1) &      9.5(3)$\times 10^{2}$ &    0.29(3) & 40.9 \\
 21 &      0.96777(7) &      1.03331(7) &           39(7) &    -0.0(2) & 5.9 \\
 22 &      0.97957(4) &      1.02085(4) &           85(9) &    -0.3(1) & 7.8 \\
 23 &      0.99134(4) &      1.00874(4) &        9(1)$\times 10^{1}$ &    -0.2(1) & 8.9 \\
 24 &      1.00229(2) &      0.99772(2) &      3.3(2)$\times 10^{2}$ &    0.10(6) & 22.5 \\
\hline
\end{tabular}}
\end{table}

\subsubsection{KIC\,9020774}

The reduced light curve of KIC\,9020774 contains $64\,863$ data points over 1450.1\,days (with a duty cycle of 91.4\%). The Rayleigh limit of the data set is $1/T=0.00069\,\mathrm{d}^{-1}$. The iterative prewhitening procedure resulted in a model with 281 significant frequencies and a variance reduction of 99.1\%. The average signal levels before and after the removal of the model are 41.7--45.3--7.6--6.3--6.0 ppm and 7.0--4.5--3.9--3.8--3.7 ppm, measured in $2\,\mathrm{d}^{-1}$ windows centred on 1, 2, 5, 10, and $20\,\mathrm{d}^{-1}$, respectively.

\begin{figure}
\resizebox{\hsize}{!}{\includegraphics{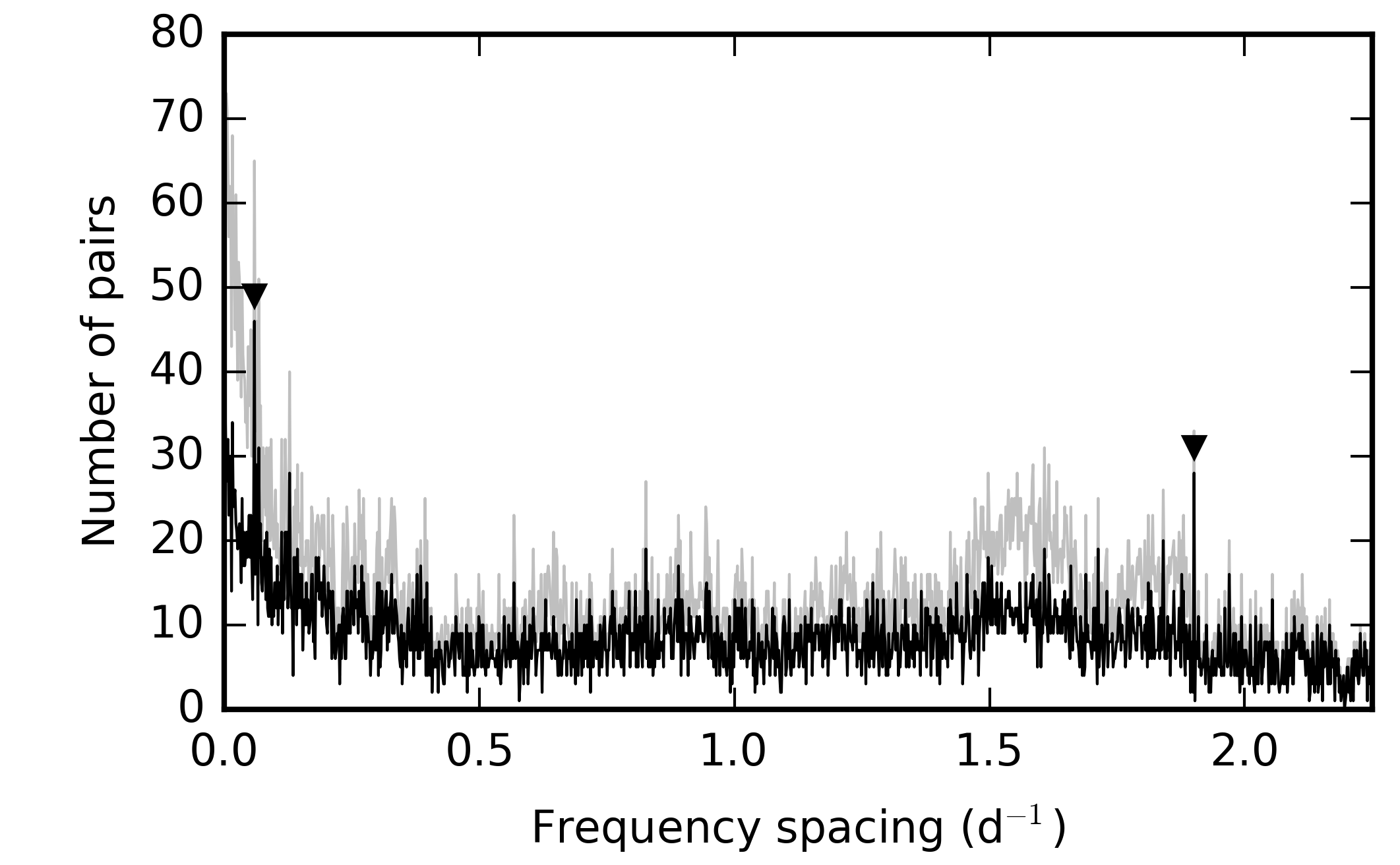}} 
\caption{Same as Fig.\,\ref{frequencypairsKIC3459297}, but for KIC\,9020774.}
\label{frequencypairsKIC9020774}
\end{figure}

The dominant modes in the periodogram are $g$ modes between 1.2 and $2.3\,\mathrm{d}^{-1}$. The strongest modes are localised in a tight cluster, with the dominant frequency at $1.900723(3)\,\mathrm{d}^{-1}$ with an amplitude of $7672(65)\,\mathrm{ppm}$. The amplitude weighted average of all model frequencies is $2.08\,\mathrm{d}^{-1}$. This star also shows the combination frequencies that we have seen in the previous cases, meaning that nearly all peaks below $0.75\,\mathrm{d}^{-1}$ are simple negative combinations (with a possible exception of a stronger peak at $0.381316(7)\,\mathrm{d}^{-1}$), and that peaks beyond the $g$-mode region are simple harmonics and positive combinations of the independent $g$ modes. Not surprisingly anymore, the strongest modes contribute to a large number of these frequencies, as illustrated by the strong bins at 0.059 and $1.901\,\mathrm{d}^{-1}$ in Fig.\,\ref{frequencypairsKIC9020774}, corresponding to $f_1-f_2$ and $f_1$, respectively.

\begin{figure*}
\resizebox{\hsize}{!}{\includegraphics{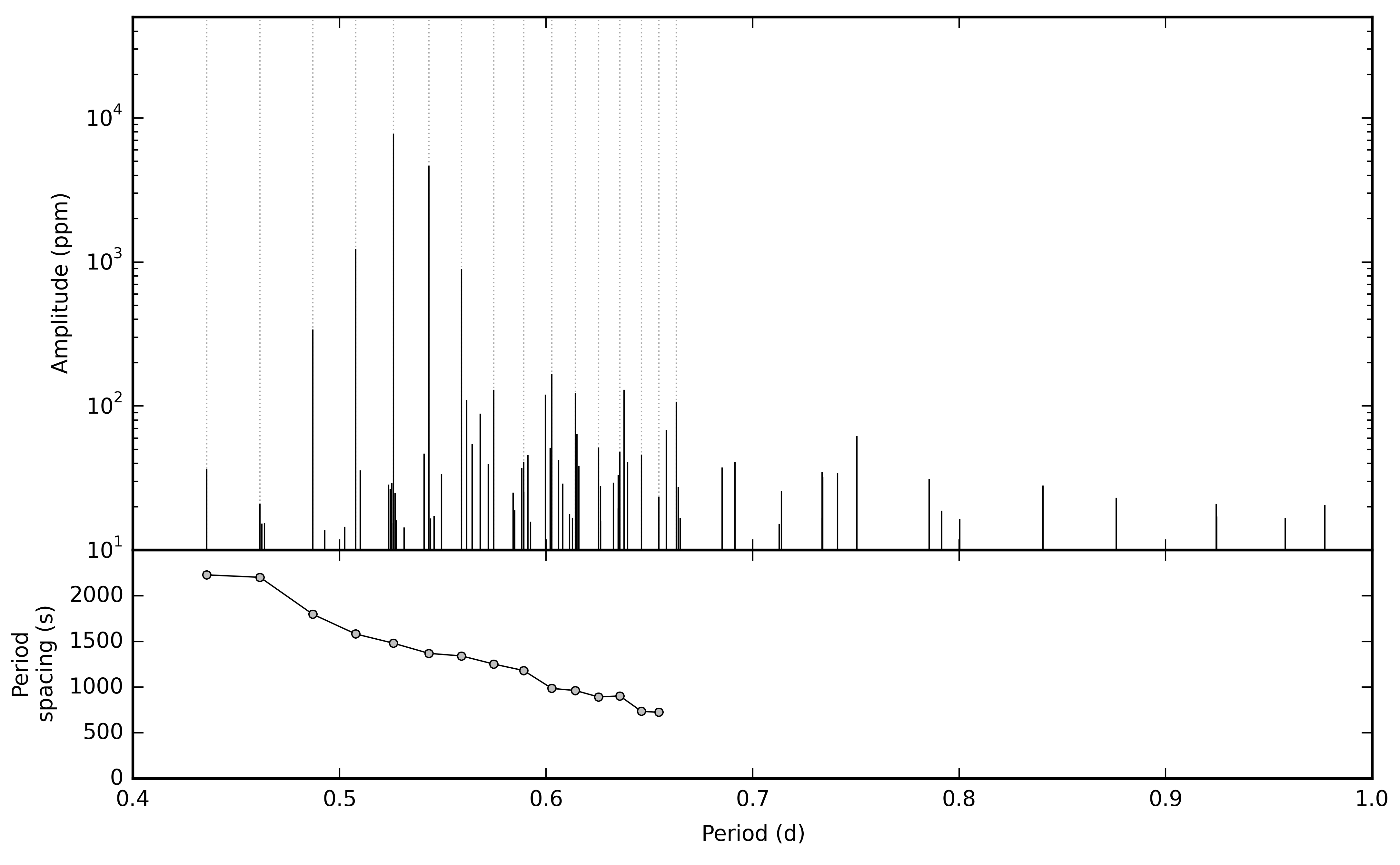}} 
\caption{Same as Fig.\,\ref{periodseriesKIC3459297}, but for KIC\,9020774.}
\label{periodseriesKIC9020774}
\end{figure*}

Screening through the 239 peaks that remain after filtering the close peaks out, we manage to find another tilted period series (see Table\,\ref{periodseriestableKIC9020774} and Fig.\,\ref{periodseriesKIC9020774}). The series is steep and smooth, which is in good agreement with the faster rotation rate of the star and its position in the Kiel diagram in Fig.\,\ref{hrd}.

\begin{table}
\caption{Fourier parameters (periods, frequencies, amplitudes, and phases) of the modes of the period series of KIC\,9020774. The S/N values shown are calculated in a window of $1\,\mathrm{d}^{-1}$ centred on the given frequency. Numbers in parentheses are the formal errors of the last significant digit.}
\label{periodseriestableKIC9020774}
\centering
\scalebox{0.9}{
\begin{tabular}{l c c c c r}
\hline\hline
 \# & $p$ & $f$ & $A$ & $\theta$ & S/N\\
    & $\mathrm{d}$ & $\mathrm{d}^{-1}$ & $\mathrm{ppm}$ & $2\pi/\mathrm{rad}$ &  \\
\hline
  1 &      0.43570(1) &      2.29514(5) &           36(5) &     0.3(1) & 10.6 \\
  2 &      0.46150(2) &      2.16685(8) &           21(4) &    -0.4(2) & 6.2 \\
  3 &     0.486991(2) &     2.053427(9) &          336(8) &    0.13(2) & 26.2 \\
  4 &     0.507805(1) &     1.969261(4) &     1.21(1)$\times 10^{3}$ &   -0.36(1) & 77.1 \\
  5 &    0.5261156(9) &     1.900723(3) &     7.67(7)$\times 10^{3}$ &   0.388(9) & 104.5 \\
  6 &    0.5432530(4) &     1.840763(1) &     4.61(2)$\times 10^{3}$ &   0.487(4) & 118.5 \\
  7 &     0.559106(1) &     1.788570(4) &          882(8) &    0.37(1) & 75.6 \\
  8 &     0.574634(6) &      1.74024(2) &          128(6) &    0.35(5) & 14.5 \\
  9 &      0.58913(2) &      1.69741(5) &           41(5) &    -0.3(1) & 9.5 \\
 10 &     0.602811(7) &      1.65889(2) &          165(8) &   -0.43(5) & 26.3 \\
 11 &     0.614221(7) &      1.62808(2) &          122(6) &   -0.08(5) & 14.5 \\
 12 &      0.62537(2) &      1.59906(4) &           51(5) &    -0.1(1) & 10.6 \\
 13 &      0.63570(2) &      1.57307(4) &           48(5) &     0.2(1) & 9.4 \\
 14 &      0.64616(2) &      1.54761(4) &           46(5) &    -0.1(1) & 9.5 \\
 15 &      0.65467(3) &      1.52749(7) &           23(4) &     0.0(2) & 5.5 \\
 16 &     0.663052(9) &      1.50818(2) &          106(6) &   -0.41(6) & 14.5 \\
\hline
\end{tabular}}
\end{table}

\subsubsection{KIC\,11971405}\label{KIC11971405photometry}

From the viewpoint of its high projected rotational velocity of $242\pm14\,\mathrm{km\,s}^{-1}$ and peculiar, exceptionally weak Be spectra, KIC\,11971405 is  the most extreme pulsator in our sample. As \citet{2015MNRAS.450.3015K} have already pointed out while studying the occurrence of linear combination frequencies in a sample of gravity mode pulsators, KIC\,11971405 shows several outburst recurring on an irregular timescale. The authors also hypothesised the Be nature of the star, which we have now confirmed in Sect.\,\ref{KIC11971405spectroscopy}. This is also in good agreement with the high $v \sin i$ value.

The light curve looks like a typical SPB light curve (but with complex beating patterns and a shorter than typical period) until the middle of Q11, when the first large outburst occurs at TBJD ($\mathrm{BJD}-2454833$) 1053. The outburst can be described as a general growth of both the average brightness and amplitude of the light variability. The outburst phase has a faster rising and a longer decaying part; the total length of the first outburst is $\sim15$\,days and the maximum brightness is reached $\sim5$\,days after the start. The second outburst, which is extremely similar to the first one both in morphology and temporal evolution, starts at TBJD 1129. These are followed by much shorter and shallower mini-outbursts centred at TBJD 1304, 1424, 1517, and 1547 (with a few possible smaller outbursts in between). Then just at the beginning of Q17, starting at TBJD 1559, there is another larger outburst, but since it starts exactly at the beginning of a quarter, it might be slightly influencing our detrending and vice versa.

Similar behaviour has been found in the hybrid B0.5\,IVe CoRoT target HD\,49330 by \citet{2009A&A...506...95H}. Likely because the CoRoT observation was 10 times shorter in time frame, only one outburst was recorded; but the magnitude of that event was very similar to the brightenings observed here, while its duration was 3 to 4 times longer. Furthermore, although not to the same extent, HD\,49330 also exhibited enhanced pulsation amplitudes during the outburst. However, in contrast to the weak Be signatures observed in KIC\,11971405, the CoRoT target showed a very strong (and varying) emission component \citep{2009A&A...506..103F}.

We employ a short-time Fourier transformation (STFT) of the full data set to illustrate the long-term changes in the amplitude spectrum. Afterwards, we apply our normal iterative prewhitening procedure on the Q0-Q11.1 data (until the first monthly downlink within Q11) to derive model frequencies that we can use in our combination and period series search. We do not use the full data set for this procedure, as our initial attempts have shown that it is not possible to adequately model the full light curve with a limited set of frequencies in a meaningful way (which is not surprising when trying to reproduce aperiodic changes with a sum of sines).

The full light curve itself spans 1470.5\,days and contains $65\,968$ data points ($91.7\%$ duty cycle and $0.00068\,\mathrm{d}^{-1}$ Rayleigh limit). Because of the complex beating patterns originating in many closely spaced frequencies and the limited duration of the outbursts, it is impossible to use a window width for the STFT that is long enough to avoid visual artefacts, but short enough so that it recovers some of the temporal changes that occur during an outburst. When the temporal resolution is chosen to be too small, such as in the wavelet analysis performed in \citet{2016arXiv160802872R}, dense frequency groups remain unresolved. This implies that the observed changes in amplitude are simply a result of the beating of the unresolved modes. Furthermore, because of the merging or smearing effect introduced by the degraded resolution in frequency, the beating among frequencies can lead to artificial changes in amplitudes and/or frequencies, turning a dense group of modes that are otherwise stable in frequency into an unresolved, periodically varying pseudo-mode. In  \citet[][Fig.\,6. and 7.]{2016arXiv160802872R}, $g_2$, $g_3$, $g_4$, and $g_5$ are suffering from this artefact, which explains the observed short-term correlations between the amplitudes
of neighbouring frequency groups and hides the actual long-term changes. Here, we take a careful approach to avoid such artefacts.

\begin{figure*}
\resizebox{\hsize}{!}{\includegraphics{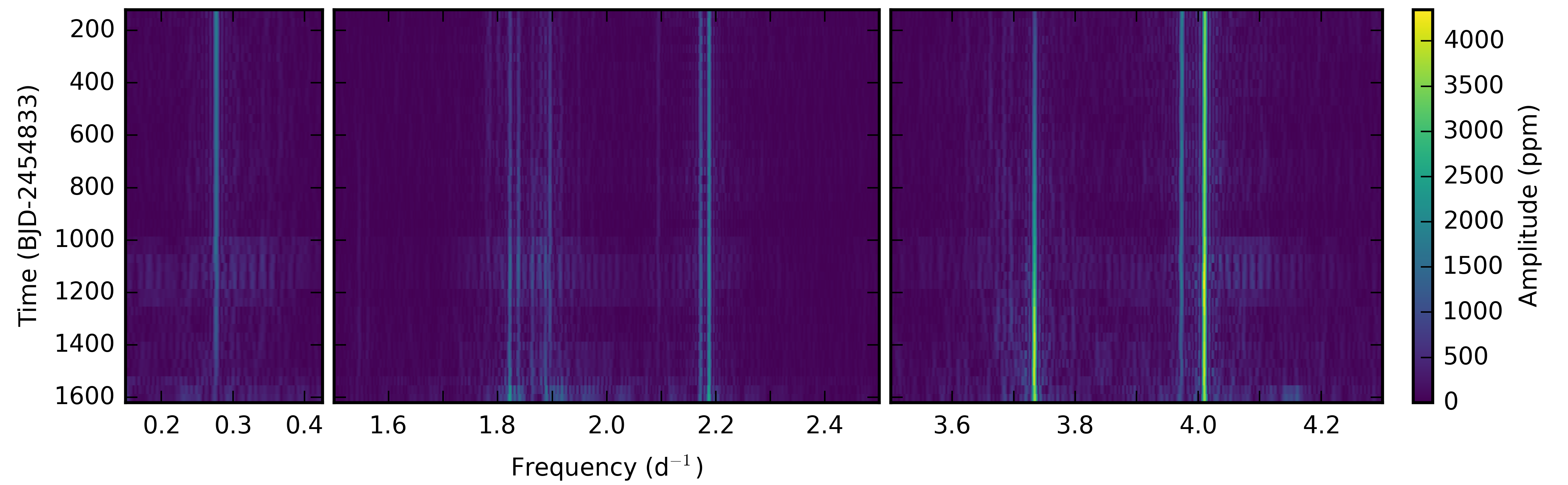}} 
\caption{Short-time Fourier transform of the full light curve of KIC\,11971405, showing the regions of dominant pulsation power.}
\label{stft11971405}
\end{figure*}

\begin{figure}
\resizebox{\hsize}{!}{\includegraphics{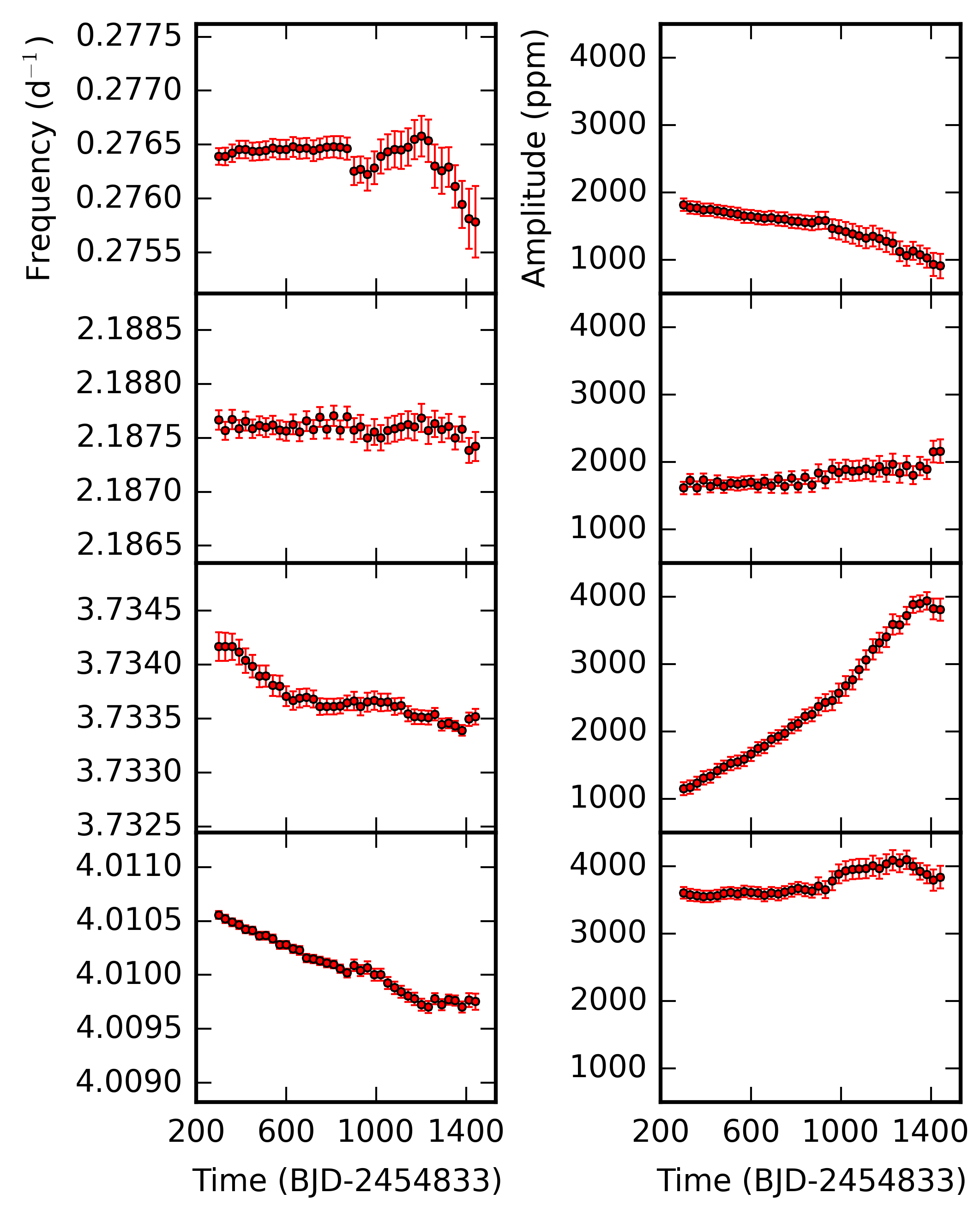}} 
\caption{Frequency (left panels) and amplitude (right panels) variations of the four strongest pulsation modes traced from the STFT of KIC\,11971405, which was calculated using 360\,day wide windows that were stepped with 30 days.}
\label{stftchanges11971405}
\end{figure}

From our tests it seems that all dominant frequencies get a boost in amplitude during the outburst events (although the one at $3.73380(1)\,\mathrm{d}^{-1}$ to a lesser extent). This explains the visual broadening of the light curve around these events without an apparent change in the pattern of the light variation, which one would expect if the outburst were connected only to a given frequency. What is more interesting is that there is a clear change in the amplitude of two of the most dominant modes on a much longer timescale. The STFT of the three dominant frequency regions of KIC\,11971405 are shown in Fig.\,\ref{stft11971405} (calculated using 180\,day wide windows that were slid with 30 days). Tracing the changes in amplitude and frequency (Fig.\,\ref{stftchanges11971405}), we find that while the peak at $3.73380(1)\,\mathrm{d}^{-1}$ continuously grows in time (from around 1000 to 4000\,ppm), the peak at $0.27644(2)\,\mathrm{d}^{-1}$ loses power (from around 1800 to 900\,ppm). The smaller modes are much more difficult to track in the STFT, but is seems that there are some other peaks with smaller, long-term amplitude changes. Moreover, looking at the frequency values, the dominant peaks between 3.6 and $4.2\,\mathrm{d}^{-1}$ (in the 3rd panel of Fig.\,\ref{stft11971405}) show a small long-term decreasing trend in frequency, while the dominant modes in the other two frequency regions are stable over time; the low frequency peak at $0.27644(2)\,\mathrm{d}^{-1}$ is strongly influenced by the presence of outbursts, which is why there are apparent variations after the first outburst. These changes in frequency are marginally larger than the Rayleigh limit of the full data set, but the trends are very clear and free of scatter, especially for the peak at $4.01030(1)\,\mathrm{d}^{-1}$, which is the dominant frequency of the Q0-Q11.1 data (with an apparent amplitude of $3507(56)$\,ppm).

The Q0-Q11.1 data set that we use later on consists of $41\,175$ data points and covers 911.7\,days ($92.3\%$ duty cycle and $0.0011\,\mathrm{d}^{-1}$ Rayleigh limit). The iterative prewhitening procedure resulted in 1541 model frequencies; the amplitude weighted average of all model frequencies is $3.14\,\mathrm{d}^{-1}$. While the data set is significantly shorter than those of the four other stars that we have analysed above, the number of frequencies is much higher, which raises some concerns. Although the frequency spectrum seems to be very rich in combination frequencies even beyond the three dominant frequency regions of 0 to $0.7\,\mathrm{d}^{-1}$, 1.3 to $2.3\,\mathrm{d}^{-1}$, and 3.1 to $4.5\,\mathrm{d}^{-1}$, there is no easy way to explain the presence of this many stable modes from a theoretical point of view. Moreover, there is also an excessive amount of peaks in this set of model frequencies that are so closely spaced that they must influence each other during the prewhitening, which is known to create spurious model frequencies. This is why we always filter these close frequencies before searching for period series, but in this star, the filtering removes half the frequencies (leaving 765 of them). Even though a rich and dense frequency spectrum in a fast rotator SPB is not at all impossible given the predicted presence of the numerous rotationally split components of various series of modes of different degrees and radial orders, it is more likely that a large amount of the very closely spaced frequency groups (where a larger central peak is surrounded by a few peaks that have lower amplitudes as we go further from the larger peak) are simply spurious peaks caused by the long-term frequency shifts observed in the STFT of the data. These structures are not to be confused with the narrowly spaced rotationally split triplets or quintuplets, which can have very similar appearance, since those only occur in extremely slow rotator stars.

While this model provides a variance reduction of $99.8\%$ in the Q0-Q11.1 data, it fails miserably in reproducing the outbursts or the light curve after the outbursts. At this point we have no ways to model these features in the light curve, so we proceed with the analysis of the frequency spectrum of the Q0-Q11.1 data.

\begin{figure}
\resizebox{\hsize}{!}{\includegraphics{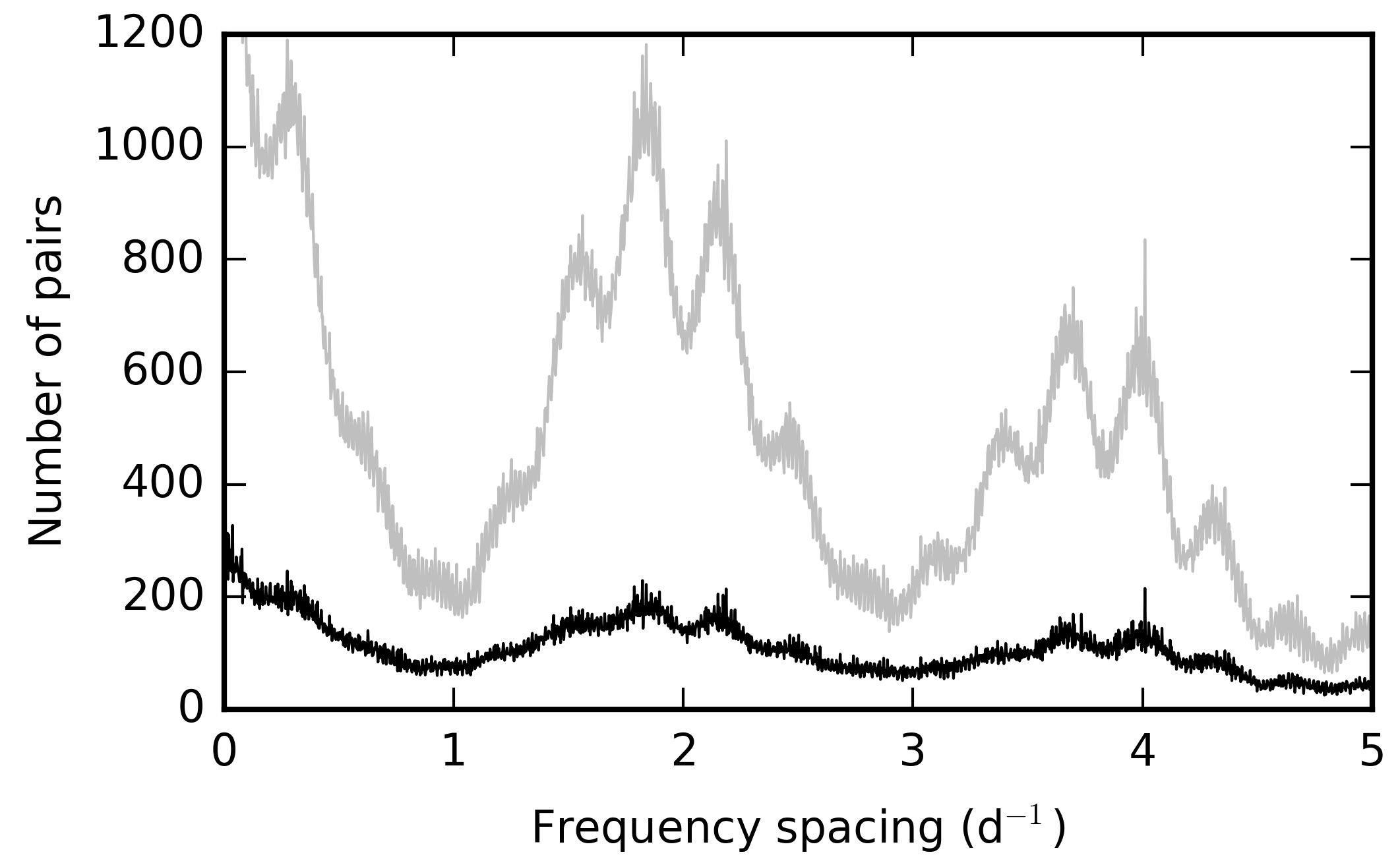}} 
\caption{Same as Fig.\,\ref{frequencypairsKIC3459297}, but for KIC\,11971405.}
\label{frequencypairsKIC11971405}
\end{figure}

The frequency spectrum of KIC\,11971405 is extremely structured. The histogram of the frequency pairs in Fig.\,\ref{frequencypairsKIC11971405} also reflects this (as frequency groups in the periodogram result in groups in the histogram too), but unlike the previous cases, there is not one bin or a few bins that would strongly rise above the underlying general structure. The bin corresponding to the dominant mode is stronger, but not as significant as in the other stars. There is a clear presence of possible combination frequencies, but owing to the complexity of the spectrum, at first sight it is much less clear which peaks could be parent modes and which are the combinations. While in our experience parent peaks are always stronger than the combinations, which is well illustrated by the previous four stars in this paper, it has been shown recently that it is also possible to have combination frequencies that are stronger than their parent peaks \citep{2015MNRAS.450.3015K}. Our standard combination search shows that the independent modes are situated in the 1.3 to $2.3\,\mathrm{d}^{-1}$ region, while as usual, the low frequency peaks are simple negative combinations and the peaks towards higher frequencies are low order positive combinations, even without omitting our standard amplitude criterion \citep{2012AN....333.1053P}. There are some large amplitude modes (basically all strong frequencies -- including the dominant one -- outside the main 1.3 to $2.3\,\mathrm{d}^{-1}$ region), which cannot be explained as combinations without looking over this criterion. But when we only look at frequency values, they are also explained simply, as already shown in \citet{2015MNRAS.450.3015K}. Except for these few large amplitude peaks, the amplitudes in the 1.3 to $2.3\,\mathrm{d}^{-1}$ region are in general larger than in the other groups.

\begin{figure*}
\resizebox{\hsize}{!}{\includegraphics{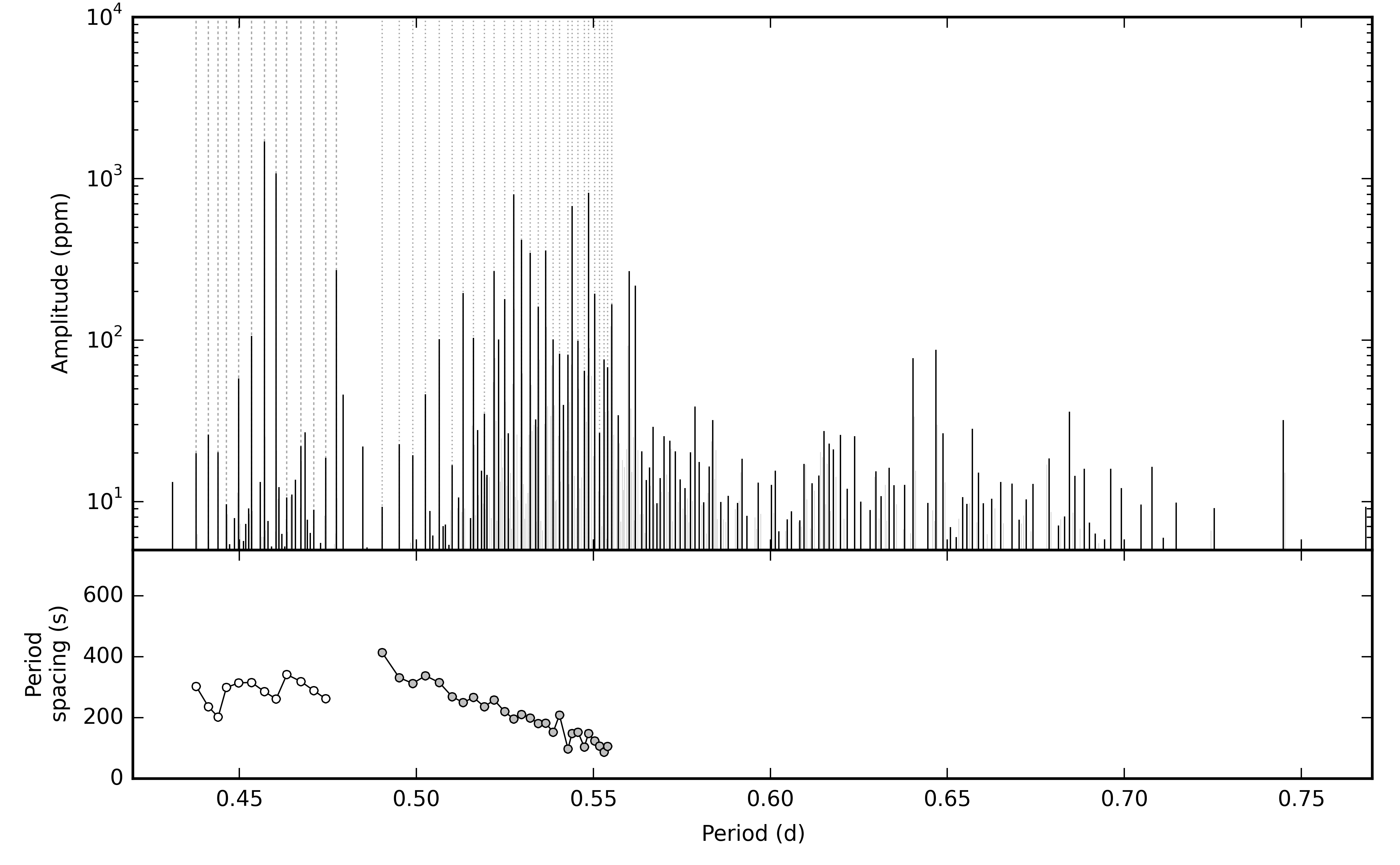}} 
\caption{Same as Fig.\,\ref{periodseriesKIC3459297}, but for KIC\,11971405.}
\label{periodseriesKIC11971405}
\end{figure*}

There is an additional observation that strengthens our opinion of why the 1.3 to $2.3\,\mathrm{d}^{-1}$ region must contain the independent pulsation modes, and that is the presence of possible period series. We find two possible series at period spacing values, which are compatible with the observed high projected rotational velocity and they are both in this region (see Table\,\ref{periodseriestableKIC11971405} and Fig.\,\ref{periodseriesKIC11971405}). The second series gets slightly more noisy towards higher periods, likely because of the density of the frequency spectrum and the low value of the period spacing, causing the peaks to influence each other in that region. The detection of $g$-mode period series along with the observed weak Be spectrum confirms that Be stars are simply fast rotator SPB stars.

\begin{table}
\caption{Fourier parameters (periods, frequencies, amplitudes, and phases) of the modes of the two possible period series of KIC\,11971405. The S/N values shown are calculated in a window of $1\,\mathrm{d}^{-1}$ centred on the given frequency. Numbers in parentheses are the formal errors of the last significant digit.}
\label{periodseriestableKIC11971405}
\centering
\scalebox{0.9}{
\begin{tabular}{l c c c c r}
\hline\hline
 \# & $p$ & $f$ & $A$ & $\theta$ & S/N\\
    & $\mathrm{d}$ & $\mathrm{d}^{-1}$ & $\mathrm{ppm}$ & $2\pi/\mathrm{rad}$ &  \\
\hline
  1 &      0.43781(4) &       2.2841(2) &           20(7) &    -0.4(4) & 7.8 \\
  2 &      0.44132(4) &       2.2660(2) &           26(8) &    -0.2(3) & 8.8 \\
  3 &      0.44404(5) &       2.2521(2) &           20(8) &     0.2(4) & 8.5 \\
  4 &      0.44638(6) &       2.2403(3) &           10(5) &    -0.3(5) & 6.0 \\
  5 &      0.44984(2) &       2.2230(1) &        6(1)$\times 10^{1}$ &    -0.0(2) & 11.8 \\
  6 &      0.45347(2) &      2.20523(9) &      1.0(1)$\times 10^{2}$ &    -0.3(1) & 18.0 \\
  7 &     0.457121(3) &      2.18761(2) &     1.68(4)$\times 10^{3}$ &    0.48(3) & 47.2 \\
  8 &     0.460422(4) &      2.17192(2) &     1.07(3)$\times 10^{3}$ &    0.19(3) & 38.0 \\
  9 &      0.46345(5) &       2.1577(3) &           10(4) &     0.1(4) & 5.7 \\
 10 &      0.46740(4) &       2.1395(2) &           22(6) &    -0.3(3) & 7.2 \\
 11 &      0.47109(5) &       2.1228(2) &            9(3) &    -0.0(4) & 4.6 \\
 12 &      0.47442(4) &       2.1078(2) &           18(5) &     0.4(3) & 6.6 \\
 13 &     0.477452(9) &      2.09445(4) &      2.7(2)$\times 10^{2}$ &    0.23(7) & 17.5 \\
\hline
  1 &      0.49037(5) &       2.0393(2) &            9(3) &     0.2(3) & 4.6 \\
  2 &      0.49516(4) &       2.0196(2) &           22(6) &     0.2(3) & 7.2 \\
  3 &      0.49899(4) &       2.0040(2) &           19(5) &    -0.4(3) & 6.6 \\
  4 &      0.50261(3) &       1.9896(1) &        5(1)$\times 10^{1}$ &     0.3(2) & 9.7 \\
  5 &      0.50651(2) &      1.97428(8) &      1.0(1)$\times 10^{2}$ &    -0.1(1) & 10.2 \\
  6 &      0.51016(4) &       1.9602(2) &           17(5) &     0.4(3) & 6.0 \\
  7 &      0.51326(1) &      1.94832(5) &      1.9(2)$\times 10^{2}$ &    0.19(9) & 14.5 \\
  8 &      0.51615(2) &      1.93741(8) &      1.0(1)$\times 10^{2}$ &    -0.1(1) & 10.1 \\
  9 &      0.51923(4) &       1.9259(1) &           35(8) &    -0.2(2) & 8.5 \\
 10 &      0.52196(1) &      1.91585(4) &      2.7(2)$\times 10^{2}$ &   -0.35(7) & 17.5 \\
 11 &      0.52495(2) &      1.90495(5) &      1.8(2)$\times 10^{2}$ &   -0.44(9) & 13.4 \\
 12 &     0.527495(6) &      1.89575(2) &      7.9(3)$\times 10^{2}$ &   -0.49(3) & 35.4 \\
 13 &      0.52975(1) &      1.88767(3) &      4.1(2)$\times 10^{2}$ &    0.38(6) & 22.3 \\
 14 &      0.53218(1) &      1.87905(4) &      3.4(2)$\times 10^{2}$ &    0.48(6) & 20.1 \\
 15 &      0.53448(2) &      1.87096(6) &      1.6(2)$\times 10^{2}$ &    -0.1(1) & 17.9 \\
 16 &      0.53657(1) &      1.86368(4) &      3.6(2)$\times 10^{2}$ &   -0.31(6) & 20.1 \\
 17 &      0.53867(2) &      1.85641(7) &      1.0(1)$\times 10^{2}$ &     0.2(1) & 9.0 \\
 18 &      0.54044(3) &       1.8504(1) &        8(1)$\times 10^{1}$ &     0.3(2) & 10.1 \\
 19 &      0.54285(3) &      1.84213(9) &        8(1)$\times 10^{1}$ &     0.5(1) & 12.0 \\
 20 &     0.543978(6) &      1.83831(2) &      6.7(2)$\times 10^{2}$ &   -0.45(4) & 34.9 \\
 21 &      0.54569(2) &      1.83254(8) &      1.0(1)$\times 10^{2}$ &     0.4(1) & 9.0 \\
 22 &      0.54745(3) &       1.8266(1) &        6(1)$\times 10^{1}$ &     0.3(2) & 8.9 \\
 23 &     0.548645(7) &      1.82267(2) &      8.1(3)$\times 10^{2}$ &    0.25(4) & 32.4 \\
 24 &      0.55036(2) &      1.81700(5) &      1.9(2)$\times 10^{2}$ &   -0.23(9) & 14.4 \\
 25 &      0.55179(4) &       1.8123(1) &           26(5) &     0.1(2) & 6.6 \\
 26 &      0.55302(3) &       1.8083(1) &        7(1)$\times 10^{1}$ &     0.5(2) & 8.9 \\
 27 &      0.55402(3) &       1.8050(1) &        7(1)$\times 10^{1}$ &     0.2(2) & 12.0 \\
 28 &      0.55525(2) &      1.80099(6) &      1.7(2)$\times 10^{2}$ &    0.38(9) & 18.0 \\
\hline
\end{tabular}}
\end{table}

There is a possible scenario that could simultaneously explain the formation of unexpectedly large amplitude combination peaks an the observed changes in the amplitude and frequency of these modes (Fig.\,\ref{stftchanges11971405}). Take the peak at $4.01030(1)\,\mathrm{d}^{-1}$, which is explained by \citeauthor{2015MNRAS.450.3015K} as a combination of the peaks $1.82267(2)\,\mathrm{d}^{-1}$ and $2.18761(2)\,\mathrm{d}^{-1}$. This peak can also be explained as a combination of, at least, two other strong peaks (with a match better than the Rayleigh limit), namely $1.83831(2)\,\mathrm{d}^{-1}$ and $2.17192(2)\,\mathrm{d}^{-1}$. The difference between the two possible combination peaks is $0.000044\,\mathrm{d}^{-1}$ or 1/25th of the Rayleigh limit. Assuming that both these combination frequencies are actually produced by the four\ strong parent peaks, we get two peaks that cannot be resolved by 3-4 years of \textit{Kepler} observations. Thus we detect these as one peak that shows amplitude and frequency variations on the timescale of the beat period of the two unresolved peaks, which is of the order of 60 years. This perfectly explains  the excessive amplitude at a given phase of the beat period and the change in amplitude, although the observed change in frequency is larger than what this model predicts. In general, faster rotator SPB stars have denser $g$-mode period series, thus the formation of similar unresolved combination peaks becomes easier. 

\begin{figure}
\resizebox{\hsize}{!}{\includegraphics{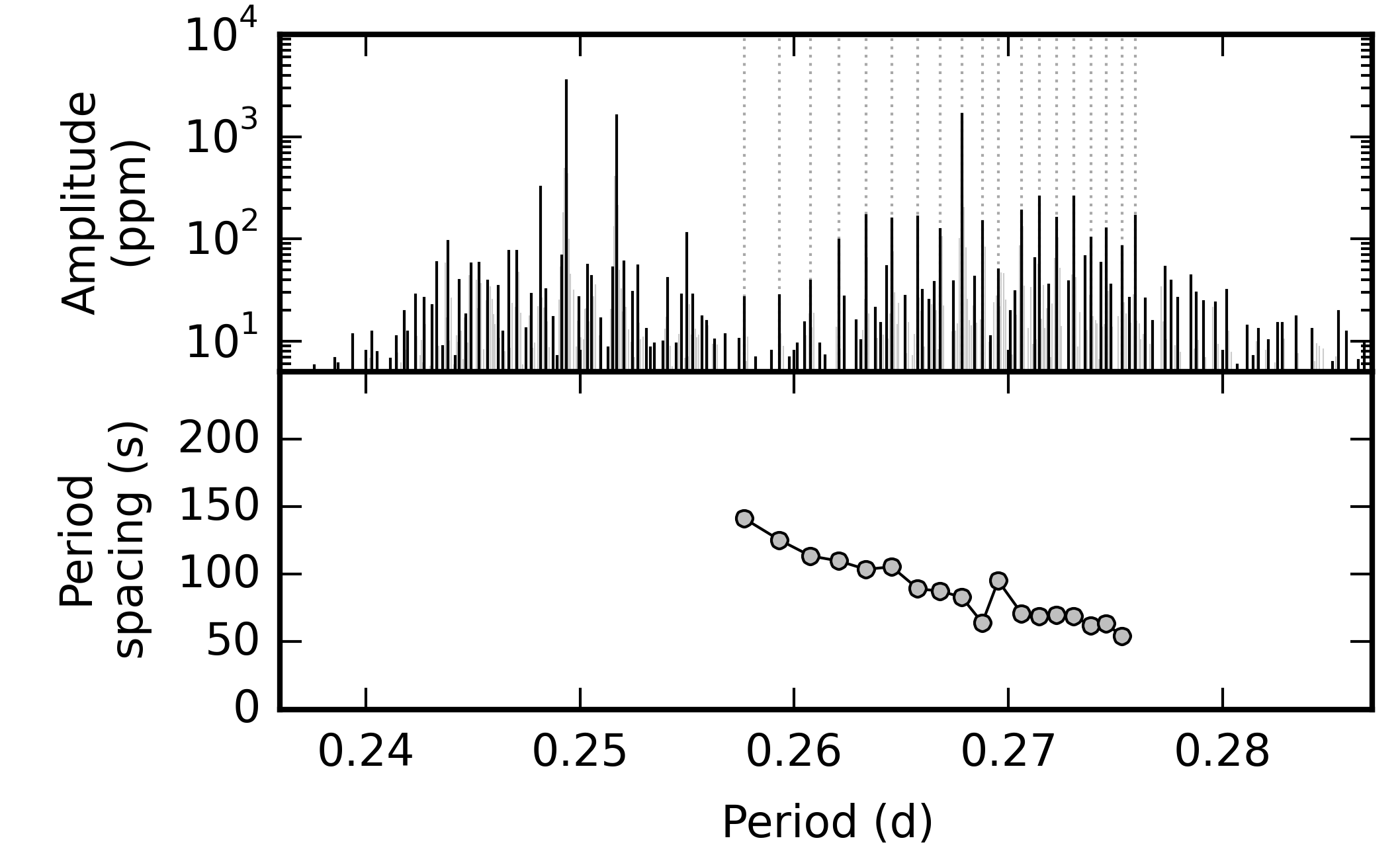}} 
\caption{Same as Fig.\,\ref{periodseriesKIC3459297}, but for the possible series among the high frequency peaks in KIC\,11971405.}
\label{periodseriesKIC11971405alt}
\end{figure}

For sufficiently fast rotation, the frequency ranges of prograde quadrupole modes and of combination frequencies of dipole modes, may overlap. Given the large number of frequencies, combination frequencies of dipole modes matching observed quadrupole modes can be expected. On the other hand, a series of pure combination frequencies is unlikely to be identifiable as quadrupole modes. In order to avoid an incorrect interpretation, any seismic modelling of such a series has to be carried out with caution. In this case, we actually manage to detect a period spacing pattern (see Table\,\ref{periodseriestableKIC11971405alt} and Fig.\,\ref{periodseriesKIC11971405alt}) among the possible combination frequencies around $3.75\,\mathrm{d}^{-1}$. In Section\,\ref{spacingdiscussion} we show that this series is consistent with the rotation rate that we derived from the prograde dipole series, assuming $(\ell,m) = (2,2)$. This indicates that this series could be genuine, and we might be observing a complex mix of beating combination frequencies and actual independent pulsation modes.

\begin{table}
\caption{Fourier parameters (periods, frequencies, amplitudes, and phases) of the modes of the possible series among the high frequency peaks in KIC\,11971405. The S/N values shown are calculated in a window of $1\,\mathrm{d}^{-1}$ centred on the given frequency. Numbers in parentheses are the formal errors of the last significant digit.}
\label{periodseriestableKIC11971405alt}
\centering
\scalebox{0.9}{
\begin{tabular}{l c c c c r}
\hline\hline
 \# & $p$ & $f$ & $A$ & $\theta$ & S/N\\
    & $\mathrm{d}$ & $\mathrm{d}^{-1}$ & $\mathrm{ppm}$ & $2\pi/\mathrm{rad}$ &  \\
\hline
  1 &     0.257683(7) &       3.8807(1) &           27(5) &     0.2(2) & 5.8 \\
  2 &     0.259317(6) &      3.85628(9) &           28(4) &    -0.0(2) & 5.6 \\
  3 &     0.260769(7) &       3.8348(1) &           38(6) &     0.3(2) & 7.5 \\
  4 &     0.262084(5) &      3.81558(7) &      1.0(1)$\times 10^{2}$ &     0.4(1) & 12.0 \\
  5 &     0.263355(3) &      3.79716(5) &      1.7(1)$\times 10^{2}$ &    0.36(8) & 9.7 \\
  6 &     0.264555(3) &      3.77993(5) &      1.6(1)$\times 10^{2}$ &    0.19(8) & 9.9 \\
  7 &     0.265777(3) &      3.76256(5) &      1.6(1)$\times 10^{2}$ &   -0.03(8) & 9.7 \\
  8 &     0.266810(4) &      3.74799(6) &      1.2(1)$\times 10^{2}$ &     0.2(1) & 9.0 \\
  9 &     0.267824(1) &      3.73380(1) &     1.65(4)$\times 10^{3}$ &   -0.42(2) & 45.2 \\
 10 &     0.268785(3) &      3.72044(5) &      1.5(1)$\times 10^{2}$ &   -0.13(8) & 8.9 \\
 11 &     0.269527(7) &       3.7102(1) &           50(8) &     0.4(2) & 8.8 \\
 12 &     0.270628(3) &      3.69510(5) &      1.9(1)$\times 10^{2}$ &    0.18(8) & 18.0 \\
 13 &     0.271448(3) &      3.68394(4) &      2.6(2)$\times 10^{2}$ &    0.13(6) & 14.5 \\
 14 &     0.272244(4) &      3.67318(5) &      1.6(1)$\times 10^{2}$ &    0.26(8) & 10.0 \\
 15 &     0.273053(3) &      3.66229(4) &      2.6(2)$\times 10^{2}$ &    0.04(6) & 13.4 \\
 16 &     0.273849(5) &      3.65165(6) &      1.0(1)$\times 10^{2}$ &     0.2(1) & 12.0 \\
 17 &     0.274568(4) &      3.64208(5) &      1.2(1)$\times 10^{2}$ &    0.12(9) & 12.0 \\
 18 &     0.275301(5) &      3.63239(7) &        8(1)$\times 10^{1}$ &     0.1(1) & 9.8 \\
 19 &     0.275929(4) &      3.62412(5) &      1.7(1)$\times 10^{2}$ &   -0.36(8) & 10.1 \\
\hline
\end{tabular}}
\end{table}

\begin{figure}
\resizebox{\hsize}{!}{\includegraphics{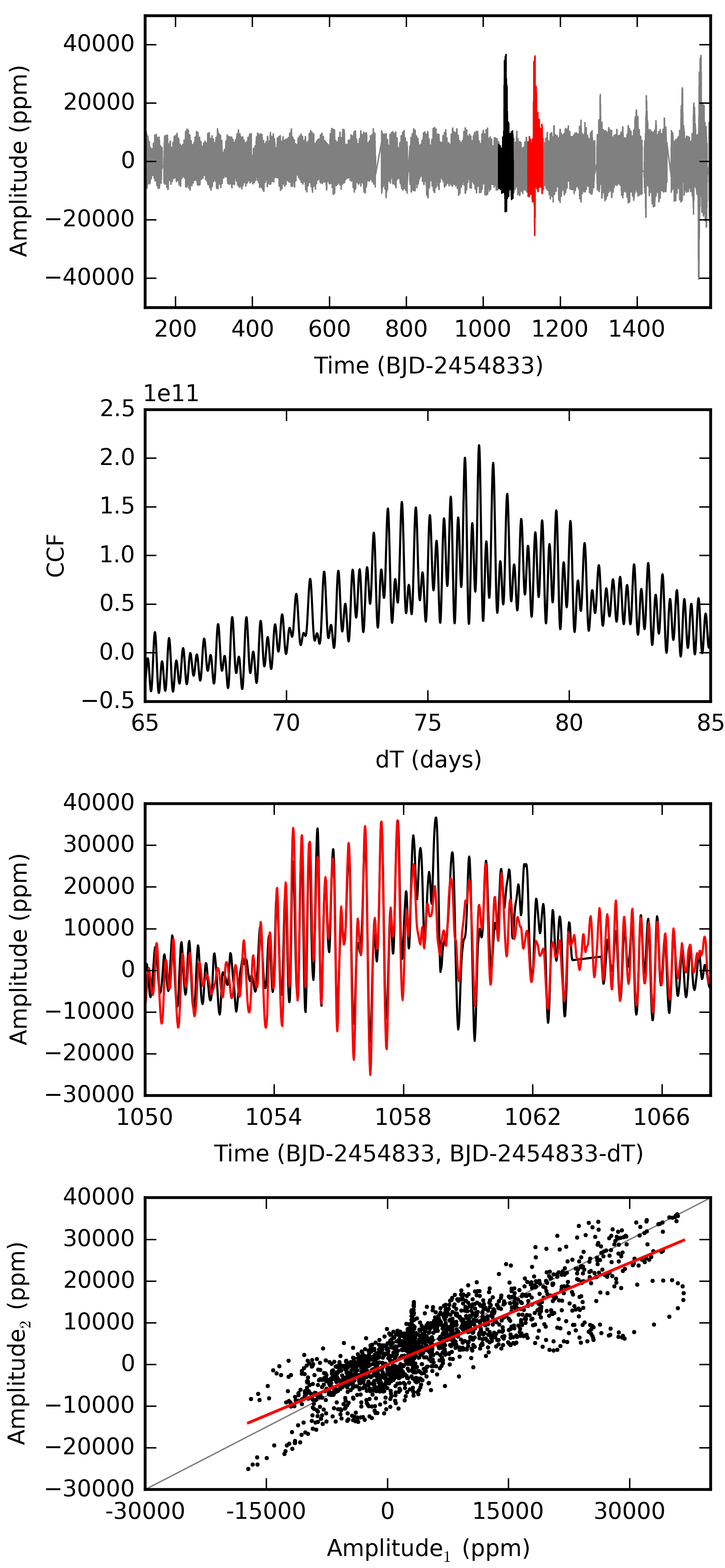}} 
\caption{(\textit{Upper panel}) Full detrended light curve of KIC\,11971405 (in grey) with the regions of the two most prominent outburst indicated in black and red. (Second panel) The cross-correlation function of the two outbursts. (Third panel) The second outburst (red solid line) shifted back in time with $\mathrm{d}T=76.81\,\mathrm{day}$ (the value of the highest peak in the CCF function on the second panel) and plotted over the first outburst (black solid line). The match between the two light curves, especially around the maximum amplitude phase, is impressive. (\textit{Lower panel}). A correlation diagram between the observed photometric signal of the two outbursts in the range shown on the third panel. A $1:1$ line is plotted in the background as a grey solid line, and the best-fit linear through the scatter plot is plotted using a red solid line ($y = 0.81\pm0.01 \times x$, $R^2 = 0.79$).}
\label{outburstcorrelationsKIC11971405}
\end{figure}

\paragraph{Photometric outbursts:}\label{KIC11971405outbursts} Inspecting the first two large outbursts in the light curve (see Fig.\,\ref{outburstcorrelationsKIC11971405}, first panel), we find that there is a very noticeable similarity between the two, as already mentioned above. To illustrate this, first we used a cross-correlation function (CCF; see Fig.\,\ref{outburstcorrelationsKIC11971405}, second panel) to calculate the time shift between the two events, then plotted them on top of each other (to illustrate the similarity;\ see Fig.\,\ref{outburstcorrelationsKIC11971405}, third panel) and measured whether there is a good linear correlation between these outbursts (see Fig.\,\ref{outburstcorrelationsKIC11971405}, last panel). We find that the two outbursts are indeed very similar (with a correlation coefficient $R^2 = 0.79$), and the second follows the first outburst with a $\mathrm{d}T=76.81\,\mathrm{day}$ delay. The fact that there is such a striking similarity between the topography of the pulsation signal during these outbursts must mean that there is a connection between the outbursts and pulsations. Since the topography is related to the phases of the pulsation modes and the different apparent beatings, the observed strong correlation means that the outbursts happen at the same phase (or constellation of phases) regarding the stellar oscillations. It has been predicted that $g$ modes could indeed be responsible for the outburst events in Be stars, and there is growing observational evidence supporting this idea \citep[see e.g.][]{1998ASPC..135..343R,2014IAUS..301..173S}.

\subsection{Period spacing patterns across the SPB instability strip:\ Initial pattern fitting}\label{spacingdiscussion}

\begin{figure*}
\resizebox{\hsize}{!}{\includegraphics{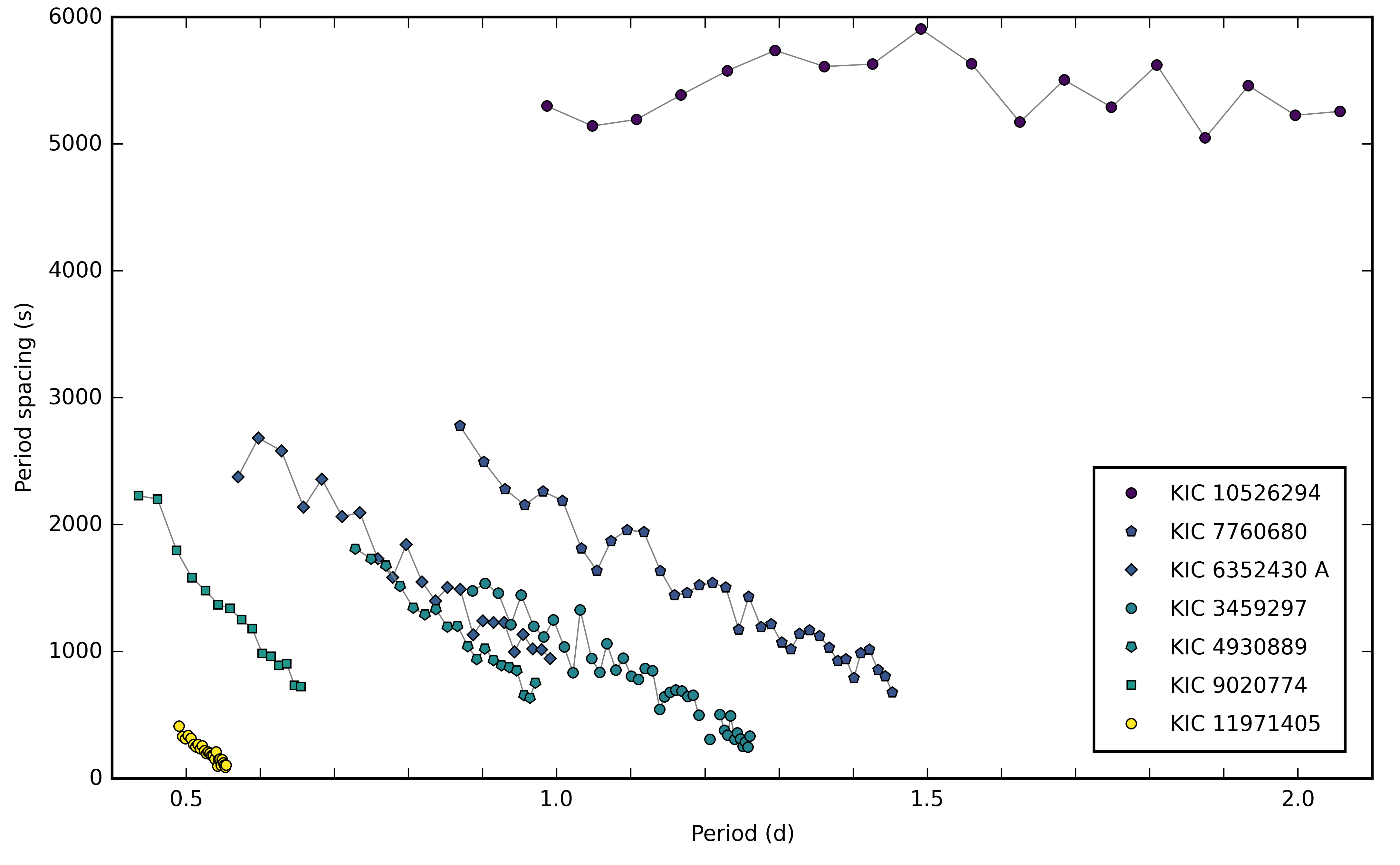}} 
\caption{Main observed period spacing patterns of the five SPB stars studied in this paper along with those of KIC\,10526294 and KIC\,7760680 from \citet{2014A&A...570A...8P,2015ApJ...803L..25P}. The colours represent the projected rotation velocity of the star; slow rotators are plotted using dark symbols and fast rotators using bright symbols.}
\label{allperiodspacings}
\end{figure*}

To give an overview of the observed period spacing patterns across the SPB instability to date, we show each of the longest main series detected in SPB stars observed by \textit{Kepler} in Fig.\,\ref{allperiodspacings}. Just by looking at this figure, along with Fig.\,\ref{alllightcurvesandfouriers}, it is clear that stars that rotate faster (or -- to be more precise -- have a higher projected rotational velocity), show their strongest excited modes and their main period series towards shorter periods (higher frequencies), and have smaller average spacing values due to a larger tilt.

The detected period spacing patterns can then be analysed to carry out a mode identification and derive the approximate near-core rotation rates by fitting model patterns to these series. Because the interior chemical gradients do not affect the global slope of the observed period spacing patterns, we can ignore their effect for now and start from an asymptotic period spacing pattern, as described by \citet{1980ApJS...43..469T}. Realistic values for the asymptotic period spacing of SPB stars range from around 4000\,s to 11000\,s. The influence of rotation is then introduced using the traditional approximation, where we assumed that, first, the star is spherically symmetric and, second, is rigidly rotating. To this end, we applied the traditional approximation module from the stellar pulsation code GYRE v4.3 \citep{2013MNRAS.435.3406T}. We then fit the resulting model patterns to the observed series. The used analysis methods are described in more detail in \citet{2016arXiv160700820V}.

\begin{figure}
\resizebox{\hsize}{!}{\includegraphics{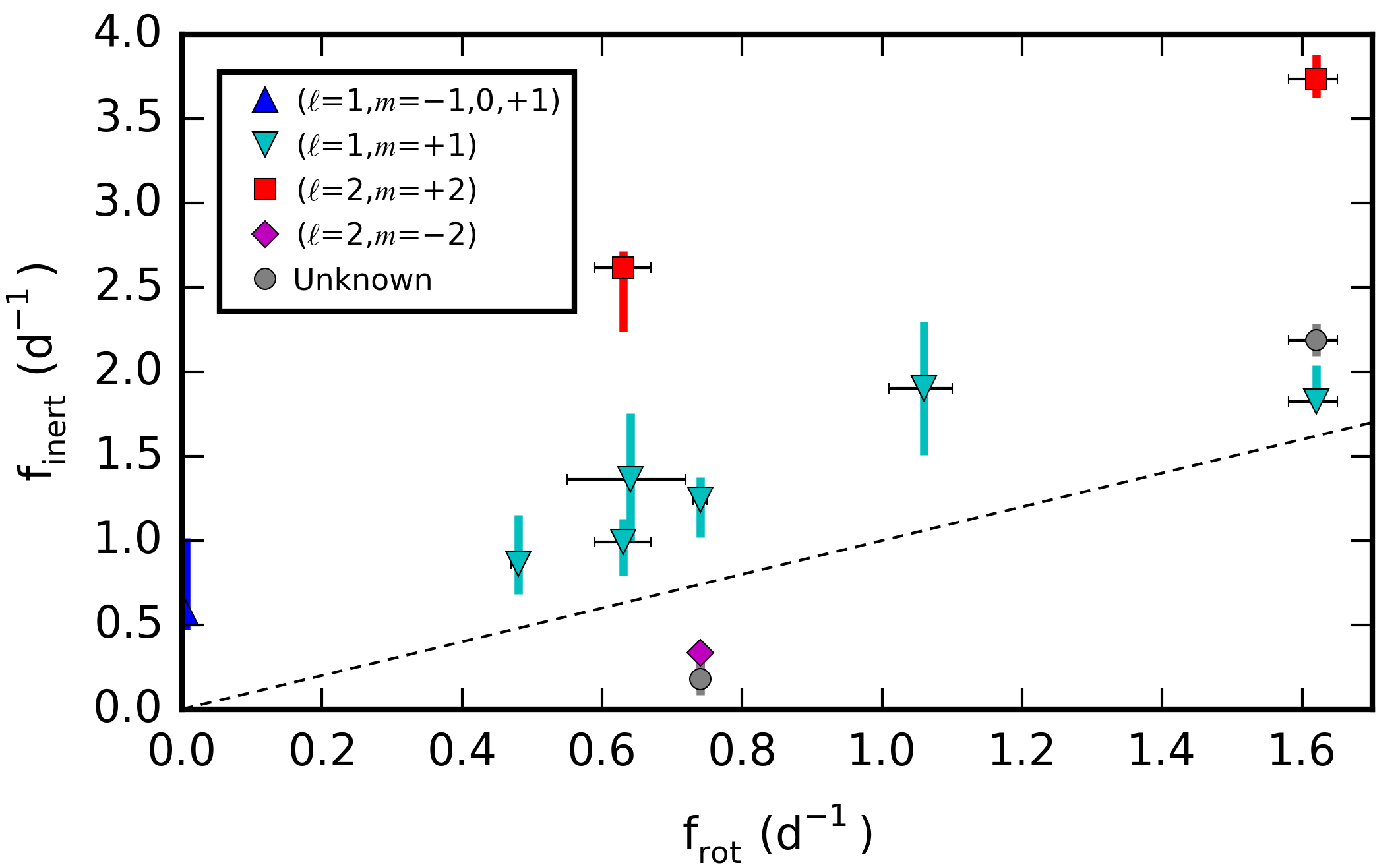}} 
\caption{Dominant pulsation frequency in the inertial frame as a function of the derived rotation frequency $f_\mathrm{rot}$ for the detected period series \citep[along with the previously known series of KIC\,10526294 and KIC\,7760680 from][]{2014A&A...570A...8P,2015ApJ...803L..25P}. The thick vertical lines indicate the full extent of the observed period series. The dashed black line indicates where $f_\mathrm{inert}$ equals to $f_\mathrm{rot}$.}
\label{periodseries_finert_vs_frot}
\end{figure}

\begin{figure}
\resizebox{\hsize}{!}{\includegraphics{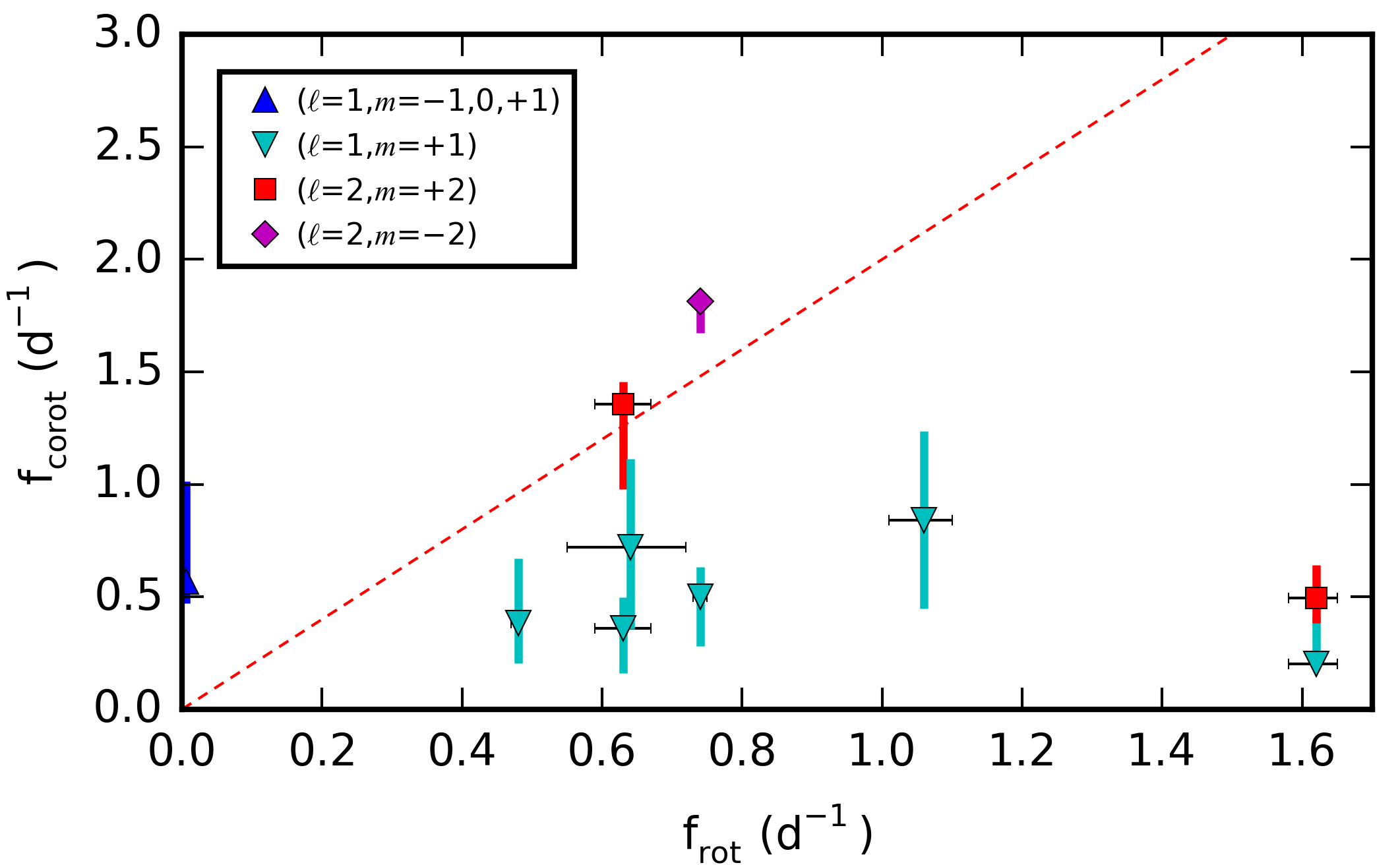}} 
\caption{Same as Fig.\,\ref{periodseries_finert_vs_frot}, but in the corotating frame. The dashed red line shows where the pulsations pass from the superinertial regime (above the line) into the subinertial regime (below the line).}
\label{periodseries_fcorot_vs_frot}
\end{figure}

The dominant period spacing pattern (on Fig.\,\ref{allperiodspacings}) of each of the stars was found to be a series of $(\ell,m)$ = $(1,+1)$ prograde dipole modes. This is consistent with the results obtained by \citet{2016arXiv160700820V} for $\gamma$\,Dor stars. Other derived mode identifications include $(\ell,m)$ = $(2,-2)$ for one of the rising series in KIC\,4930889, and $(2,+2)$ for the extra series in KIC\,3459297 and for the possible series in the range of combination frequencies in KIC\,11971405. However, contrary to the study of the lower mass $\gamma$\,Dor stars, none of the observed series were identified as Rossby waves. A possible explanation for this non-detection may be related to the absence of a thin convective envelope in SPBs or different properties in the interior rotation between these two groups of gravity-mode pulsators; a proper interpretation of this different behaviour requires future theoretical research. The results of this exercise are illustrated in Fig.\,\ref{periodseries_finert_vs_frot} and Fig.\,\ref{periodseries_fcorot_vs_frot}, while the derived rotation frequencies and asymptotic period spacings are listed in Table,\,\ref{patternfitparameters}. Finally, the derived asymptotic spacing for KIC\,4930889 is more compatible with the fundamental parameters of the primary component of this system than of the secondary.

\begin{table}
\caption{Results from the initial pattern fitting of the observed period series.}
\label{patternfitparameters}
\centering
\renewcommand{\arraystretch}{1.25}
\scalebox{1}{
\begin{tabular}{l l l l}
\hline\hline
    & $f_\mathrm{rot}\,(\mathrm{d}^{-1})$ & $dP_\mathrm{asym}\,(\mathrm{s})$ & Mode ID $(\ell,m)$\\
\hline
KIC\,3459297            & $0.63^{+0.04}_{-0.04}$ & $5840^{+950}_{-860}$         & $(1,+1)$, $(2,+2)$\\
KIC\,6352430\,A         & $0.64^{+0.08}_{-0.09}$ & $4910^{+1000}_{-800}$        & $(1,+1)$\\
KIC\,4930889\,(A?)      & $0.74^{+0.01}_{-0.01}$ & $6160^{+320}_{-320}$         & $(1,+1)$, $(2,-2)$\\
KIC\,9020774            & $1.06^{+0.04}_{-0.05}$ & $5610^{+640}_{-600}$         & $(1,+1)$\\
KIC\,11971405           & $1.62^{+0.03}_{-0.04}$ & $6890^{+1700}_{-1600}$       & $(1,+1)$, $(2,+2)$\\
\hline
\end{tabular}}
\tablefoot{For comparison, we recall the rotation frequencies derived -- using different methods -- for the two slower rotator stars that are used for comparison in this paper: KIC\,10526294 has a radius averaged rotation frequency of around $0.005\,\mathrm{d}^{-1}$ \citep{2015ApJ...810...16T} and KIC\,7760680 rotates at $0.48\,\mathrm{d}^{-1}$ \citep{2016arXiv160402680M}.}
\end{table}

%%%%%%%%%%%%%%%%%
%%%Conclusions%%%
%%%%%%%%%%%%%%%%%

\section{Conclusions}\label{conclusions}

By presenting four new SPBs and revisiting one of the previously published binary components, we  completed the observational analyses of the stars in our \textit{Kepler} Guest Observer Programme. We raised the number of in-depth analysed SPB stars in the initial \textit{Kepler} field from four to eight, and the number of SPB stars with an observed period spacing from four to nine; seven of these, including the five new SPB stars in this paper, are from \textit{Kepler}. Simultaneously we have shown that period series are not only common, but they -- and their combination peaks -- dominate the frequency spectrum of SPB stars. We can find a wealth of information about the internal physics of these objects imprinted in the topology of these series, thus they serve as foundations on which we can build an improved theory.

While this investigation along with previous studies \citep{2013A&A...553A.127P,2014A&A...570A...8P,2015ApJ...803L..25P} have already provided previously unseen constraints for theoretical modelling efforts \citep{2015A&A...580A..27M,2015ApJ...810...16T,2016arXiv160402680M}, the exploitation of \textit{Kepler} data needs to continue. There must be still a few dozen SPB stars in the \textit{Kepler} data archives. These light curves are ten times better than anything we had pre-\textit{Kepler}, and there is no upcoming mission the coming decade that would deliver data with the same or superior properties given that PLATO \citep{2014ExA....38..249R} will only get launched in 2025. Although TESS \citep[Transiting Exoplanet Survey Satellite;][launch in 2017]{2014SPIE.9143E..20R} will be an all-sky survey, it will at best provide only one-fourth of the frequency resolution of \textit{Kepler,} and only in the very limited continuous viewing zones around the ecliptic poles. Therefore in the meantime we must continue with screening the \textit{Kepler} data for similar period series in SPB stars covering various regions of the instability strip to contribute to a statistically meaningful sample of these intriguing objects via in-depth modelling efforts based on their seismic barcodes. 

%%%%%%%%%%%%%%%%%%%%%%
%%%Acknowledgements%%%
%%%%%%%%%%%%%%%%%%%%%%

\begin{acknowledgements}
The research leading to these results has received funding from The Research Foundation -- Flanders (FWO), Belgium, from the European Research Council (ERC) under the European Union's Horizon 2020 research and innovation programme (grant agreement N$^\circ$670519: MAMSIE), and from the People Programme (Marie Curie Actions) of the European Union's Seventh Framework Programme FP7/2007-2013 under REA grant agreement No.\,623303 (ASAMBA). TVR acknowledges financial support from the Fund for Scientific Research of Flanders (FWO), Belgium, under grant agreement G.0B69.13. We are grateful to Bill Paxton and Richard Townsend for their valuable work on the stellar evolution code MESA and stellar pulsation code GYRE. EN acknowledges support from the NCN grant No.\,2014/13/B/ST9/00902. Funding for the \textit{Kepler} mission is provided by the NASA Science Mission directorate. Some of the data presented in this paper were obtained from the Multimission Archive at the Space Telescope Science Institute (MAST). STScI is operated by the Association of Universities for Research in Astronomy, Inc., under NASA contract NAS5-26555. Support for MAST for non-HST data is provided by the NASA Office of Space Science via grant NNX09AF08G and by other grants and contracts. This work is partly based on observations made with the William Herschel Telescope, which is operated on the island of La Palma by the Isaac Newton Group in the Spanish Observatorio del Roque de los Muchachos (ORM) of the Instituto de Astrof\'{i}sica de Canarias (IAC). We made use of the matplotlib colormaps by Nathaniel J. Smith, Stefan van der Walt, and Eric Firing.
\end{acknowledgements}

\bibliographystyle{aa}
\bibliography{KeplerBstars_IV}

\end{document}